\documentclass[a4paper,11pt]{article}
\pdfoutput=1 

\usepackage{jheppub} 

\usepackage{hyperref}
\usepackage{pdfpages}
\numberwithin{figure}{section}

\usepackage{epstopdf}

\usepackage[colorinlistoftodos]{todonotes}
\usepackage{epsfig}
\usepackage{graphicx,color}

\newcommand{\be}{\begin{equation}}
\newcommand{\ee}{\end{equation}}

\newcommand{\A}{{\cal A}}

\newcommand{\tr}{{\rm tr}}

\newcommand{\rmq}{{\rm \bf q}}
\newcommand{\rmg}{{\rm \bf g}}
\newcommand{\rmp}{{ \boldsymbol{\gamma}}}

\newcommand{\beq}{\begin{eqnarray}}
\newcommand{\eeq}{\end{eqnarray}}

\newcommand{\eg}{{\it e.g. }}

\newcommand{\bq}{{\bar q}}

\definecolor{red}{rgb}{1,0,0}
\definecolor{gray}{rgb}{0.5,0.5,0.5}

\newcommand{\uq}{{\underline q}}

\usepackage[normalem]{ulem} 
\renewcommand\sout{\bgroup  \ULdepth=-.5ex \ULset}

\usepackage{hyperref}

\def\bea{\begin{eqnarray}}
\def\eea{\end{eqnarray}}

\title{TMD factorization for dijets + photon production from the dilute-dense CGC framework}

\author[a]{Tolga Altinoluk,}
\author[b,c]{Renaud Boussarie,}
\author[d]{Cyrille Marquet,}
\author[e]{and Pieter Taels }

\affiliation[a]{National Centre for Nuclear Research, 00-681 Warsaw, Poland}
\affiliation[b]{Institute of Nuclear Physics, Polish Academy of Sciences, Radzikowskiego 152, PL-31-342 Krak\'ow, Poland}
\affiliation[c]{Physics Department, Brookhaven National Laboratory, Upton, NY 11973, USA}
\affiliation[d]{Centre de Physique Th\'eorique, \'Ecole Polytechnique, CNRS, Universit\'e Paris-Saclay, F-91128 Palaiseau, France}
\affiliation[e]{INFN Sezione di Pavia, via Bassi 6, I-27100 Pavia, Italy}

\emailAdd{tolga.altinoluk@ncbj.gov.pl}
\emailAdd{renaud.boussarie@ifj.edu.pl}
\emailAdd{cyrille.marquet@polytechnique.edu}
\emailAdd{pieter.taels@ca.infn.it}

\date{\today}

\preprint{???}

\abstract{
We calculate the production of a photon and two jets at forward rapidity in proton-nucleus collisions, within the hybrid dilute-dense framework in the Color Glass Condensate (CGC) formalism. After obtaining the cross section for both the quark- and gluon-initiated channels, we consider the correlation limit, in which the vector sum of the transverse momenta of the three outgoing particles is small with respect to the individual transverse momenta. In this limit, the cross section simplifies considerably and can be written in a factorized form, sensitive to various unpolarized and linearly-polarized transverse-momentum-dependent gluon distribution functions (gluon TMDs). Thus, we demonstrate for the first time that the emergence of a TMD factorization formula in the correlation limit, from CGC expressions, holds beyond the previously-considered simpler $2\to2$ processes.
}

\begin{document}
\maketitle
\flushbottom
\allowdisplaybreaks

\section{Introduction}
One of the most important principles in QCD is factorization, the
systematic separation of short- and long-distance physics for a
sufficiently inclusive cross section. The former involves the scattering
of quarks and gluons, and is perturbatively calculable in the presence
of a hard scale. Long-distance physics is non-perturbative and pertains,
amongst others, to the structure of the hadron in terms of these quarks
and gluons, encoded in the parton distribution functions (PDFs) and
fragmentation functions (FFs) which need to be extracted from experiments,
or from lattice QCD calculations. Radiative corrections are resummed
with the help of evolution equations in the hard scale, and absorbed
into a renormalization of these PDFs/FFs.

In the case of collinear factorization, there is a single hard scale,
and the PDFs and FFs are only a function of this scale and of the
longitudinal momentum fraction $x$ of the parton. However, in less
inclusive processes that depend on a second, smaller scale, these
distributions need to be generalized to include the dependence
on the transverse momentum of the parton, yielding the so-called transverse-momentum-dependent (TMD) PDFs and FFs. 
These TMD PDFs/FFs are of great interest
since their measurement offers insight in the three-dimensional structure
of the proton in terms of the QCD degrees of freedom. However, in
contrast to their collinear counterparts, which are believed to be
universal, they depend on the hard process under consideration and are therefore more
complicated.

The radiative corrections that are considered in collinear factorization involve logarithms $\ln Q^{2}/\mu^{2}$
in the scale, and can be resummed with the help of the Dokshitzer-Gribov-Lipatov-Altarelli-Parisi
(DGLAP) \cite{dglap} evolution equations. In the TMD factorization framework \cite{tmd factorization}, additional logarithms of the form $\ln Q^{2}/k_t^{2}$ need to be resummed, where $k_t$ is the small transverse momentum scale in the problem. 

At very high energies, however, soft logarithms
in the center-of-mass energy $\ln s$ are expected to be dominant,
necessitating a resummation and subsequently an evolution equation
in rapidity instead of the scale, provided by the (linear) Balitsky-Fadin-Kuraev-Lipatov (BFKL)
equation \cite{bfkl}. BFKL involves a universal unintegrated gluon PDF, which just like the TMDs depends
on transverse momentum. One of the most fascinating aspects of
QCD at high energy or small $x$, is the phenomenon
of saturation \cite{GLRMQ}: the damping of the steep rise of the gluon density predicted by BFKL, caused by gluon recombinations which become important due to the high density. The low-$x$ evolution of the gluon density in the presence of these nonlinear interactions
is described by an effective theory known as the Color Glass Condensate
(CGC), which leads to the nonlinear BK-JIMWLK evolution equations \cite{JIMWLK,cgc,balitsky,kovchegov}. 

Within the CGC framework, the hybrid formalism \cite{Dumitru:2005gt} is used to study single inclusive particle production at next-to-leading order \cite{Elastic_vs_Inelastic}-\cite{Ducloue:2017mpb} and heavy-quark production \cite{Altinoluk:2015vax}, at forward rapitidy. In this set-up, the wave function of the projectile proton is treated in the spirit of collinear factorization\footnote{Multiple collinear scattering effects from the projectile protons have been studied in \cite{Boussarie:2018zwg}.}. Perturbative corrections to this wave function are provided by the usual QCD splitting processes. On the other hand, the dense target is treated in the CGC, i.e. it is defined as a distribution of strong color fields, which during the scattering transfer transverse momentum to the propagating partonic configuration.  

Recently, a lot of effort has been dedicated to the understanding of gluon TMDs in the CGC.
In particular, it was shown in \cite{Marquet:2016cgx} that nonlinear corrections are intimately connected to the process dependence
of the gluon TMDs, and that only in the linear regime can one speak of a universal BFKL unintegrated gluon distribution.
Various processes at small $x$ which involve two scales have been studied, such as dijet, heavy-quark pair or photon-jet
production in electron-nucleus or proton-nucleus collisions \cite{firstlowxTMDs,Akcakaya:2012si,Dumitru:2015gaa,Kotko:2015ura,Boer:2016fqd,Boer:2017xpy,Marquet:2017xwy}, see \cite{Petreska:2018cbf} for a recent review.
The usual approach is to calculate the cross section in the CGC, after
which one takes the small dipole or correlation limit which corresponds
to a leading-power expansion in the ratio $p_{t}^{2}/Q^{2}$ of the hard scales,
with $\Lambda_{\mathrm{QCD}}^{2}\ll p_{t}^{2}\ll Q^{2}$. On the one
hand, the analysis of the CGC in terms of gluon TMDs allows for a
better understanding of QCD dynamics at high energy. On the other
hand, once a CGC expression is obtained for a gluon TMD in the small
$x$ limit, its nonlinear evolution in rapidity can be computed with
the help of the JIMWLK \cite{Marquet:2016cgx} or BK \cite{vanHameren:2016ftb}
equations. In addition, non-perturbative models such as the McLerran-Venugopalan
(MV) \cite{MV} or Golec-Biernat-W{\"u}sthoff (GBW) model \cite{gbw}
can be used to write down analytical expressions.

In this paper, we study the process $pA\to\mathrm{jet}+\mathrm{jet}+\gamma$
at forward rapidity in a hybrid setup
%
%
\cite{Dumitru:2005gt}.
The derivation of that cross section is given in section 2. We then take, in section 3,
the correlation limit, by making a small dipole expansion, which corresponds
to a leading-power expansion in the small ratio $p_{t}^{2}/Q^{2}$,
where $p_{t}=q_{1}+q_{2}+q_{3}$ is the vector sum of the transverse
momenta of the outgoing particles. The result can be cast in an explicit
hybrid TMD factorized formula, in which both unpolarized ($\mathcal{F}^{\left(i\right)}$)
and linearly-polarized ($\mathcal{H}^{\left(i\right)}$) gluon TMDs
play a role, appearing in several types $i$, distinct by their gauge-link structure.
Finally, section 4 is devoted to our conclusions.

\section{Production cross section}

Dijet+photon production in $pA$ collisions can be initiated from two different channels. The first one that we will consider is the quark initiated channel. In this process, the incoming quark with vanishing transverse momenta in the projectile wave function, is dressed by emitting a photon and a gluon at $O(g_eg_s)$, where $g_e$ is the QED and $g_s$ is the strong coupling constant. The photon can be emitted either before or after the emission of the gluon. The dressed quark state, which takes into account both orderings of the photon and gluon emissions, then scatters off the target via eikonal interactions and produces a quark, a gluon and a photon in the final state. In general, the transverse momenta of the produced jets are much larger than the saturation scale of the target $Q_s$, and we shall thus consider them to be hard. According to the hybrid factorization ansatz \cite{Dumitru:2005gt}, the total production cross section for this process is written as a convolution of the partonic cross section and the quark distribution function $f_q(x_p,\mu^2)$ inside the proton:
\beq
\label{Q_in_full_X_section}
\frac{d\sigma^{pA\to \gamma gq+X}}{d^3\uq_1d^3\uq_2d^3\uq_3}=\int dx_p \, f_q(x_p,\mu^2) \, \frac{d\sigma^{qA\to \gamma gq+X}}{d^3\uq_1d^3\uq_2d^3\uq_3}\;,
\eeq 
where $x_p$ is the longitudinal momentum fraction carried by the incoming quark, $\mu^2$ is the factorization scale, and the three-momenta $\uq_i=(q_i^+,q_i)$ are the longitudinal and transverse momenta of the produced jets. 

Recently, this process was studied in \cite{Altinoluk:2018uax} focusing on specific kinematics, in which the dominant contribution to the cross section comes from the configuration in which the photon is radiated collinearly to the incoming quark, after which the latter splits into a quark-gluon pair with large relative transverse momenta. In these specific kinematics, the case in which the photon is radiated \textit{after} the quark-gluon splitting was neglected.
In this paper, we have in mind to study a different kinematical regime, and for that we need study the dijet+photon production in more generality. We shall include all the contributions that were neglected in \cite{Altinoluk:2018uax}, with the exception of the instantaneous diagrams of light-cone perturbation theory, which do not contribute to the TMD regime we want to study.

The second channel that we will consider is the gluon initiated channel. In this process, the incoming gluon, with vanishing transverse momenta, splits into a quark-antiquark pair with large relative transverse momenta in the projectile wave function at $O(g_s)$. Then, at $O(g_eg_s)$, the incoming gluon is dressed with a photon which is emitted either from the quark or from the antiquark. The dressed gluon state eikonally scatters off the target and produces the final hard jets and the hard photon. Similarly to the quark initiated channel, the total production cross section for this process can be written as the convolution of the partonic cross section with the gluon distribution function $f_g (x_p,\mu^2)$ of the incoming proton:
\beq
\label{G_in_full_X_section}
\frac{d\sigma^{pA\to q\gamma\bq+X}}{d^3\uq_1d^3\uq_2d^3\uq_3}=\int dx_p \, f_g (x_p,\mu^2) \, \frac{d\sigma^{gA\to q\gamma\bq+X}}{d^3\uq_1d^3\uq_2d^3\uq_3}\;,
\eeq 
with $x_p$ being the longitudinal momentum fraction carried by the gluon. Recently, this gluon initiated channel was also studied in \cite{Benic:2016uku,Benic:2018hvb} for central production  \footnote{See \cite{Roy:2018jxq} for the eA scattering.}, and subsequently in \cite{Benic:2017znu} in a regime where the quark (or antiquark) ends up with much less transverse momentum than the photon it has emitted .

\subsection{Quark initiated channel: $qA\to qg\gamma+X$}  

We start our analysis with the quark initiated channel. The partonic level cross section is formally defined as the expectation value of the number operator in the outgoing state: 
\beq
\label{Partonic_q_chan}
&&
(2\pi)^9\frac{d\sigma^{qA\to \gamma gq+X}}{d^3\uq_1d^3\uq_2d^3\uq_3}(2\pi)\delta(p^+-q_1^+-q_2^+-q_3^+) \nonumber\\
&&= \frac{1}{2N_c}\sum_{s,\alpha}{}_{\rm out}\big\langle(\rmq)[p^+,0]_s^\alpha\big| O(\uq_1,\uq_2,\uq_3) \big|(\rmq)[p^+,0]^\alpha_{s}\big\rangle_{\rm out}\;,
\eeq 
where the normalization $1/2N_c$ comes from averaging over the spin and color indices of the outgoing state in the amplitude and in the complex conjugate amplitude. The number operator is built from creation and annihilation operators of the final state particles. Therefore, in the quark initiated channel one needs to select the Fock state containing a quark, a gluon and a photon from the outgoing wave function, for which the number operator is given by
\beq
\label{Num_Op_Quark_Channel}
O(\uq_1,\uq_2,\uq_3)=\gamma_\lambda^\dagger(\uq_1)\gamma_\lambda(\uq_1)  a_i^{\dagger b}(\uq_2)a^b_i(\uq_2)  b_t^{\dagger \beta}(\uq_3)b^\beta_t(\uq_3)\;.
\eeq
In the above expression, $\gamma_\lambda(\uq_1)$  is the annihilation operator of a dressed photon with polarization $\lambda$ and three-momentum $\uq_1$,  $a^b_i(\uq_2)$ is the one for a dressed gluon with color $b$, polarization $i$, and three-momentum $\uq_2$, and finally $b^\beta_t(\uq_3)$ is the annihilation operator of dressed quark with color $\beta$, spin $t$, and three-momentum $\uq_3$. It is convenient to express the expectation value of the number operator in the mixed Fourier basis, which gives:
\beq
\label{Num_Op_Q_Mixed}
&&
{}_{\rm out}\big\langle(\rmq)[p^+,0]_s^\alpha\big| O(\uq_1,\uq_2,\uq_3) \big|(\rmq)[p^+,0]^\alpha_{s}\big\rangle_{\rm out}=\int_{y_1z_1\, y_2z_2\, y_3z_3}e^{iq_1\cdot(y_1-z_1)+iq_2\cdot(y_2-z_2)+iq_3\cdot(y_3-z_3)}\nonumber\\
&&
\hspace{3.3cm}
\; {}_{\rm out}\big\langle(\rmq)[p^+,0]_s^\alpha\big| \gamma_\lambda^\dagger(q_1^+, y_1)\gamma_{\lambda}(q_1^+,z_1) \, a^{\dagger b}_i(q_2^+,y_2)a^b_i(q_2^+,z_2) \,
\nonumber\\
&&
\hspace{7cm}
\times\; 
 b^{\dagger\beta}_t(q_3^+,y_3)b^\beta_t(q_3^+,z_3)
\big|(\rmq)[p^+,0]^\alpha_{s}\big\rangle_{\rm out}\;.
\eeq
The action of the creation and annihilation operators on the dressed states is defined in the standard way, for example in the case of a gluon: 
\beq
\label{action_ann}
a_i^b(q_2^+,z_2)\big| (\rmg)[k_2^+,x_2]_\eta^c\big\rangle_{D}&=&2\pi \, \delta^{bd}\delta_{i\eta}\, \delta(k_2^+-q_2^+)\delta^{(2)}(x_2-z_2)\;,\\
\label{action_cre}
a^{\dagger b}_i(q_2^+,y_2)|0\rangle&=&\big|(\rmg)[q^+_2,y_2]_i^b\big\rangle_D\;.
\eeq

In order to evaluate the cross section Eq. \ref{Partonic_q_chan}, we need the expression for the outgoing state in the quark initiated channel, which was derived in \cite{Altinoluk:2018uax}. Here, we will not repeat the derivation but rather give a sketch of it. The strategy we employ is the following (see \eg\cite{Marquet:2007vb}): we start with the perturbative expression of the dressed quark state in full momentum space, given in terms of the bare states. Then, we Fourier transform the state into mixed (longitudinal momentum and transverse coordinate) space, and use this expression to calculate the eikonal interaction with the target, which will give us the outgoing state in terms of the bare states. Finally, to obtain the final expression we express the outgoing state in terms of the dressed states. This is all done in the $A^+=0$ gauge.

In momentum space, the dressed quark state with longitudinal momentum $p^+$, vanishing transverse momenta, color $\alpha$ and spin $s$ is given in terms of the bare states as 
\beq
&&\left| (\rmq) [p^+,0]^\alpha_s\right\rangle_D= 
Z^q\left| (\rmq) [p^+,0]_s^\alpha\right\rangle_0\nonumber \\ 
&+&
Z^{q\gamma} \; g_e\sum_{s',\lambda} \int \frac{dk_1^+}{2\pi} \frac{d^2k_1}{(2\pi)^2} \; 
F_{(\rmq\rmp)}^{(1)}\left[ (\rmp)[k_1^+,k_1]^\lambda, (\rmq)[p^+-k_1^+,-k_1]_{ss'}\right]\nonumber\\
&&
\hspace{6cm}
\times
\left| (\rmq) \left[p^+-k_1^+,-k_1\right]_{s'}^\alpha ; (\rmp) [k_1^+,k_1]_\lambda \right\rangle_0 \nonumber\\
&+&Z^{qg} \; g_s\sum_{s',\eta} \int \frac{dk_2^+}{2\pi}\frac{d^2k_2}{(2\pi)^2} \, t^c_{\alpha\beta} \; 
F^{(1)}_{(\rmq\rmg)}\left[ (\rmg)[k_2^+,k_2]^\eta, (\rmq)[p^+-k_2^+,-k_2]_{ss'}\right] \nonumber\\
&&
\hspace{6cm}
\times
\left| (\rmq)\left[p^+-k_2^+,-k_2 \right]_{s'}^\beta; (\rmg)\left[k_2^+,k_2\right]_{\eta}^c \right\rangle_0 \nonumber\\
&+&
Z^{qg\gamma}\; g_sg_e\sum_{s's''}\sum_{\lambda\eta} \int \frac{dk_1^+}{2\pi}\frac{d^2k_1}{(2\pi)^2}\frac{dk_2^+}{2\pi}\frac{d^2k_2}{(2\pi)^2} \; t^c_{\alpha\beta}\nonumber\\
&&
\times
\Bigg\{ 
F^{(2)}_{(\rmq\rmp-\rmq\rmg)}\left[ (\rmp)[k_1^+,k_1]^\lambda , (\rmg)[k_2^+,k_2]^\eta , (\rmq)[p^+-k_1^+-k_2^+,-k_1-k_2]_{ss''} \right]
\nonumber\\
&&
\hspace{0.2cm}
+ \; 
F^{(2)}_{(\rmq\rmg-\rmq\rmp)}\left[ (\rmg)[k_2^+,k_2]^\eta , (\rmp)[k_1^+,k_1]^\lambda , (\rmq)[p^+-k_2^+-k_1^+,-k_2-k_1]_{ss''}\right]
%
\Bigg\}\nonumber\\
&&
\hspace{1.5cm}
\times
\left| (\rmq) \left[ p^+-k_1^+-k_2^+, -k_1-k_2\right]_{s''}^\beta , (\rmg) \left[k_2^+,k_2 \right]^c_\eta , (\rmp) \left[k_1^+,k_1\right]^\lambda \right\rangle_0 \;.
\label{dwf1}
\eeq
Here $Z^q$, $Z^{q\gamma}$, $Z^{qg}$ and $Z^{qg\gamma}$ are the normalization functions. In the wave function approach, they provide the virtual contributions to the production process, and are determined by using the orthogonality conditions of the states \cite{Mueller:1993rr}. Since we are focused on the tree level production of a hard photon and two hard jets, the explicit  expressions for these normalization functions are not relevant for our purposes, and they can be set to identity. At this point, we would like to emphasize that there is also an instantaneous quark contribution in the dressed state. One should include this contribution to calculate the complete expression for the dressed quark at $O(g_eg_s)$. However, this instantaneous contribution is suppressed by two powers of the hard momenta of the produced jets, and we can therefore safely neglect it for our purposes.

The functions $F^{(1)}_{(\rmq\rmp)}$ and $F^{(1)}_{(\rmq\rmg)}$ define the momentum structure of the quark-photon and and quark-gluon splittings. The quark-photon splitting function reads (see for example \cite{Altinoluk:2018uax}):
\beq
\label{Splitting_quark_photon}
F^{(1)}_{(\rmq\rmp)}\left[ (\rmp)[k_1^+,k_1]^\lambda, (\rmq)[p^+-k_1^+, p-k_1]_{ss'}\right]=\left[\frac{-1}{\sqrt{2\xi_1p^+}}\right]
\phi^{\lambda\bar\lambda}_{ss'}(\xi_1)\frac{(\xi_1p-k_1)^{\bar\lambda}}{(\xi_1p-k_1)^2}\;,
\eeq
with 
\beq
\label{splt_Amp}
\phi^{\lambda\bar\lambda}_{ss'}(\xi_1)=\left[ (2-\xi_1)\delta^{\lambda\bar\lambda}\delta_{ss'}-i\epsilon^{\lambda\bar\lambda}\sigma^3_{ss'}\xi_1\right]\;,
\eeq
where $\sigma^3$ is the third Pauli matrix and where we have defined the longitudinal momentum ratio $\xi_1\equiv k_1^+/p^+$. The function $F^{(1)}_{(\rmq\rmg)}$ has the same structure as  $F^{(1)}_{(\rmq\rmp)}$, and its explicit expression can be read off from \eqref{Splitting_quark_photon} by exchanging $(1\to 2)$. 

\begin{figure}
\begin{centering}
\includegraphics[trim=7.3cm 21cm 6.1cm 4.5cm, clip=true,scale=1]{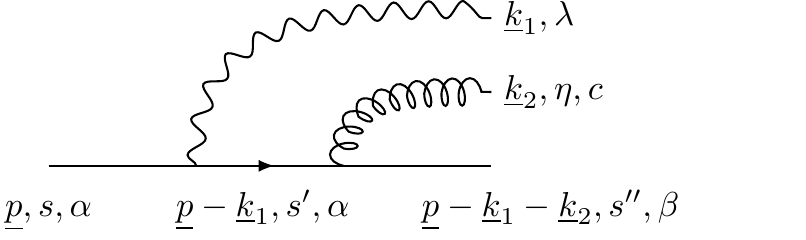}\includegraphics[trim=7.3cm 21cm 7.3cm 4.5cm, clip=true,scale=1]{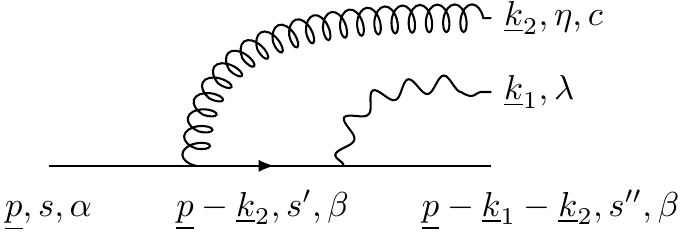}
\par\end{centering}
\caption{The dressed quark state to order $O(g_eg_s)$, with the two possible orderings of the photon resp. gluon emission by the quark.}
\label{fig:quark_channel_notations}  
\end{figure}

The functions $F^{(2)}_{(\rmq\rmp-\rmq\rmg)}$ and $F^{(2)}_{(\rmq\rmg-\rmq\rmp)}$ in Eq. \eqref{dwf1} define the momentum structure of successive quark-photon and quark-gluon splittings, and differ in the sequence of the emissions (see Fig. \ref{fig:quark_channel_notations}). The explicit expression for $F^{(2)}_{(\rmq\rmp-\rmq\rmg)}$, in which the photon is emitted before the gluon, reads (see for example \cite{Altinoluk:2018uax})
\beq
\label{Splitting_photon_gluon}
&&
F^{(2)}_{(\rmq \rmp-\rmq\rmg)}\left[(\rmp)[k_1^+,k_1]^\lambda, (\rmg)[k_2^+,k_2]^\eta, (\rmq)[p^+-k_1^+-k_2^+,p-k_1-k_2]_{ss''}\right]
\\
&&
=\sum_{s'}
\left[\frac{-1}{\sqrt{2\xi_1p^+}}\phi^{\lambda\bar\lambda}_{ss'}(\xi_1)\right]
\left[\frac{-1}{\sqrt{2\xi_2p^+}}\tilde\phi^{\eta\bar\eta}_{s's''}(\xi_1,\xi_2)\right]
\frac{(\xi_1p-k_1)^{\bar\lambda}}{(\xi_1p-k_1)^2}
\nonumber\\
&&
\hspace{0.3cm}
\times\; 
\frac{[\xi_2(p-k_1)-\bar\xi_1k_2]^{\bar\eta}}{\xi_2(\xi_1p-k_1)^2+\xi_1(\xi_2p-k_2)^2-(\xi_2k_1-\xi_1k_2)^2}\;,\nonumber
\eeq
with 
\beq
\tilde\phi^{\eta\bar\eta}_{s's''}(\xi_1,\xi_2)=\frac{\xi_1}{\bar\xi_1}\left[ (2\bar\xi_1-\xi_2)\delta^{\eta\bar\eta}\delta_{s's''}-i\epsilon^{\eta\bar\eta}\sigma^3_{s's''}\xi_2\right]\;,
\eeq
where the ratios of longitudinal momenta are defined as
\begin{equation}
\xi_1\equiv k_1^+/p^+\ ,\quad\xi_2\equiv k_2^+/p^+\ ,\quad\bar\xi_1 \equiv 1-\xi_1\ ,\quad\mbox{and}\quad\bar\xi_2 \equiv 1-\xi_2\ .
\end{equation}

As mentioned earlier, our next step is to perform a two dimensional Fourier transform on Eq. \eqref{dwf1}, in order to write the dressed quark state in mixed (longitudinal momenta and transverse coordinate) space. The details of this calculation can be found in \cite{Altinoluk:2018uax} and the final result reads
\beq
\label{Dressed_mixed_space}
&&
\hspace{-0.5cm}
\left| (\rmq) [p^+,0]^\alpha_s\right\rangle_D= 
\int_{\omega}\left| (\rmq) [p^+,\omega]_s^\alpha\right\rangle_0
\\ 
&+&
g_e\sum_{s'\lambda} \int \frac{dk_1^+}{2\pi} \int_{\omega, v, x_1} 
\bigg[\frac{(-i)}{\sqrt{2\xi_1p^+}}\,  \phi^{\lambda\bar\lambda}_{ss'}(\xi_1) \bigg]\, A^{\bar\lambda}(v-x_1) \, \delta^{(2)}\left[\omega-(\bar\xi_1v+\xi_1x_1)\right]
\nonumber\\
&&
\hspace{7cm}
\times
\left| (\rmq) \left[p^+-k_1^+,v\right]_{s'}^\alpha ; (\rmp) [k_1^+,{x_1}]_\lambda \right\rangle_0 \nonumber\\
&+&
g_s\sum_{s'\eta} \int \frac{dk_2^+}{2\pi}\int_{\omega, v, {x_2}} \, t^c_{\alpha\beta}\, 
\bigg[\frac{(-i)}{\sqrt{2\xi_2p^+}}\, \phi^{\eta\bar\eta}_{ss'}(\xi_2)\bigg] \, A^{\bar\eta}(v-x_2) \, \delta^{(2)}\left[\omega-(\bar\xi_2v+\xi_2x_2)\right]
\nonumber\\
&&
\hspace{7cm}
\times
\left| (\rmq)\left[p^+-k_2^+,v\right]_{s'}^\beta; (\rmg)\left[k_2^+,{x_2}\right]_{\eta}^c \right\rangle_0 
\nonumber\\
&+&g_sg_e\sum_{s's''}\sum_{\lambda\eta} \int \frac{dk_1^+}{2\pi}\frac{dk_2^+}{2\pi}t^c_{\alpha\beta}
\int_{\omega,\,  v, \,x_1, \, x_2, \, x_3}\nonumber\\
&&
\times
\Bigg\{
\delta^{(2)}\left[v- \left\{\left(1-\frac{\xi_2}{\bar\xi_1}\right)x_3+\frac{\xi_2}{\bar\xi_1}x_2\right\}\right]
\delta^{(2)}\left[ \omega-\left(\xi_1x_1+\bar\xi_1v\right)\right]
\nonumber\\
&&
\hspace{0.2cm}
\times
\bigg[\frac{(-i)}{\sqrt{2\xi_1p^+}}\phi^{\lambda\bar\lambda}_{ss'}(\xi_1)\bigg]
\bigg[\frac{(-i)}{\sqrt{2\xi_2p^+}}\phi^{\eta\bar\eta}_{s's''}\left(\frac{\xi_2}{\bar\xi_1}\right)\bigg]
A^{\bar\eta}(x_3-x_2){\cal A}^{\bar\lambda}\bigg(\xi_1,v-x_1; \frac{\xi_2}{\bar\xi_1}, x_3-x_2\bigg)
\nonumber\\
&&
\hspace{0.5cm}
+\, 
\delta^{(2)}\left[v- \left\{\left(1-\frac{\xi_1}{\bar\xi_2}\right)x_3+\frac{\xi_1}{\bar\xi_2}x_1\right\}\right]
\delta^{(2)}\left[ \omega-\left(\xi_2x_2+\bar\xi_2v\right)\right]
\nonumber\\
&&
\hspace{0.2cm}
\times
\bigg[\frac{(-i)}{\sqrt{2\xi_2p^+}}\phi^{\eta\bar\eta}_{ss'}(\xi_2)\bigg]
\bigg[\frac{(-i)}{\sqrt{2\xi_1p^+}}\phi^{\lambda\bar\lambda}_{s's''}\left(\frac{\xi_1}{\bar\xi_2}\right)\bigg]
A^{\bar\lambda}(x_3-x_1){\cal A}^{\bar\eta}\bigg(\xi_2,v-x_2; \frac{\xi_1}{\bar\xi_2}, x_3-x_1\bigg)\Bigg\}
\nonumber\\
&&
\hspace{3.5cm}
\times
\left| (\rmq)\left[ p^+-k_1^+-k_2^+,{x_3}\right]_{s''}^\beta; (\rmg)\left[k_2^+,{x_2}\right]^c_\eta; (\rmp)\left[k_1^+,{x_1}\right]^\lambda \right\rangle_0\nonumber\,.
\eeq
Here, we have introduced the standard Weizs\"acker-Williams field, defined as 
\beq
\label{Standard_WW}
A^\lambda(x-y)=-\frac{1}{2\pi}\frac{(x-y)^\lambda}{(x-y)^2}\; , 
\eeq
as well as the modified Weizs\"acker-Williams field for two successive splittings, which we define as
\beq
\label{Modified_WW}
{\cal A}^\lambda\bigg(\xi_1,v-x_1; \frac{\xi_2}{\bar\xi_1},x_3-x_2\bigg)=-\frac{1}{2\pi}\frac{\xi_1(v-x_1)^\lambda}{\xi_1(v-x_1)^2+\frac{\xi_2}{\bar\xi_1}\Big(1-\frac{\xi_2}{\bar \xi_1}\Big)(x_3-x_2)^2}\, .
\eeq
We should remark that, in the original derivation of the dressed quark state Eq. (\ref{Dressed_mixed_space}), the Ioffe time constraint \cite{Gribov:1965hf,Ioffe:1969kf} on the lifetime of emitted pairs in the quark wave function was taken into account \cite{Altinoluk:2018uax}. This constraint provides a consistent description of the coherent scattering of the incoming wave function on the target, which is crucial for next-to-leading order calculations \cite{nlohybridpart2} since it has a direct impact on the factorization scale. However, since our calculation is essentially a tree-level one, the Ioffe time restriction can be neglected for our purposes.

The incoming dressed quark state scatters off the target via multiple eikonal interactions, which preserve the transverse position and the spin of its components. However, the quark and the gluon inside the dressed quark undergo a rotation in color space, which can be encoded in a Wilson line in the fundamental resp. adjoint representation, at the transverse position of the quark or gluon. The Wilson lines are defined in the standard way in terms of the color fields $\alpha^-$ of the target as 
\beq
\label{Wilson_def}
S_{F,A}(x)={\cal P} \, e^{ig\int dx^+\tau^a\alpha^-_a(x^+,x)}\;,
\eeq
where $\tau^a$ is the $SU(N_c)$ generator either in the fundamental or the adjoint representation, as indicated in the subscript $F$ or $A$. Multiplying the kets inside Eq. (\ref{Dressed_mixed_space}) with the appropriate Wilson lines to account for the scattering, we obtain the outgoing wave function in terms of bare states. However, the outgoing wave function needs to be expressed in terms of the dressed states, for which the procedure and its calculation is presented in detail in \cite{Altinoluk:2018uax}. The result is:
\beq
\label{out_full}
&&
\left| (\rmq)[p^+,0]_s^{\alpha}\right\rangle_{\rm out}= \int_\omega S_F^{\alpha\beta}(\omega) \left|(\rmq)[p^+,\omega]_s^{\beta}\right\rangle_D \\
&&
+g_e\sum_{s'\lambda}\int\frac{dk_1^+}{2\pi}\int_{\omega v x_1}
\left[S_F^{\alpha\beta}(v)-S_F^{\alpha\beta}(\omega)\right]
\bigg[\frac{(-i)}{\sqrt{2\xi_1p^+}}\phi^{\lambda\bar\lambda}_{ss'}(\xi_1)\bigg]
A^{\bar\lambda}(v-x_1)\nonumber\\
&&
\hspace{5cm}
\times
\delta^{(2)}\left[\omega-(\bar\xi_1v+\xi_1x_1)\right]
\left| (\rmq)[p^+-k_1^+,v]^\beta_{s}; (\rmp)[k_1^+,x_1]^\lambda\right\rangle_D\nonumber\\
&&
+g_s\sum_{s'\eta}\int\frac{dk_2^+}{2\pi}\int_{\omega v x_2}
\left[ t^c_{\alpha\beta}S_F^{\beta\sigma}(v)S^{cd}_A(x_2) - S_F^{\alpha\beta}(\omega) t^d_{\beta\sigma}\right] 
\bigg[\frac{(-i)}{\sqrt{2\xi_2p^+}} \phi^{\eta\bar\eta}_{ss'}(\xi_2)\bigg]
A^{\bar\eta}(v-x_2) \nonumber\\
&&
\hspace{5cm}
\times
\delta^{(2)}\left[ \omega-(\bar\xi_2v+\xi_2x_2)\right]
\left| (\rmq)[p^+-k_2^+,v]^\sigma_{s'}; (\rmg)[k_2^+,x_2]^c_{\eta}\right\rangle_D\nonumber\\
&&
+g_sg_e \sum_{s's''}\sum_{\lambda\eta}\int \frac{dk_1^+}{2\pi}\frac{dk_2^+}{2\pi} \int_{wvx_1x_2x_3} 
\delta^{(2)}\Big[\omega-(\xi_1x_1+\bar\xi_1v)\Big]
\nonumber\\
&&
\times
\delta^{(2)}\bigg[v-\left\{\left(1-\frac{\xi_2}{\bar\xi_1}\right)x_3+\frac{\xi_2}{\bar\xi_1}x_2\right\}\bigg]
\bigg[\frac{(-i)}{\sqrt{2\xi_1p^+}} \phi^{\lambda\bar\lambda}_{ss'}(\xi_1)\bigg]
\bigg[\frac{(-i)}{\sqrt{2\xi_2p^+}} \phi^{\eta\bar\eta}_{s's''}\left(\frac{\xi_2}{\bar\xi_1}\right)\bigg]A^{\bar\eta}(x_3-x_2)
\nonumber\\
&&
\times
\bigg\{
\left[ t^c_{\alpha\beta}S_F^{\beta\sigma}(x_3)S_A^{cd}(x_2) -S_F^{\alpha\beta}(\omega)t^d_{\beta\sigma}\right]
{\cal A}^{\bar\lambda}\bigg(\xi_1,v-x_1; \frac{\xi_2}{\bar\xi_1}, x_3-x_2\bigg)\nonumber\\
&&
\hspace{2.5cm}
-
\left[S_F^{\alpha\beta}(v)-S_F^{\alpha\beta}(\omega)\right]t^d_{\beta\sigma} \, 
A^{\bar\lambda}(v-x_1)\bigg\}
\nonumber\\
&&
\hspace{5cm}
\times
\left| (\rmq)[p^+-k_1^+-k_2^+,x_3]^{\sigma}_{s''}, (\rmg)[k_2^+,x_2]^d_{\eta}, (\rmp) [k_1^+,x_1]^{\lambda}\right\rangle_D
\nonumber\\
&&
+g_sg_e \sum_{s's''}\sum_{\lambda\eta}\int \frac{dk_1^+}{2\pi}\frac{dk_2^+}{2\pi} \int_{wvx_1x_2x_3} 
\delta^{(2)}\Big[\omega-(\xi_2x_2+\bar\xi_2v)\Big]
\nonumber\\
&&
\times
\delta^{(2)}\bigg[v-\left\{\left(1-\frac{\xi_1}{\bar\xi_2}\right)x_3+\frac{\xi_1}{\bar\xi_2}x_1\right\}\bigg]
\bigg[\frac{(-i)}{\sqrt{2\xi_2p^+}} \phi^{\eta\bar\eta}_{ss'}(\xi_2)\bigg]
\bigg[\frac{(-i)}{\sqrt{2\xi_1p^+}} \phi^{\lambda\bar\lambda}_{s's''}\left(\frac{\xi_1}{\bar\xi_2}\right)\bigg]
A^{\bar\lambda}(x_3-x_1)
\nonumber\\
&&
\times
\bigg\{
\left[ t^c_{\alpha\beta}S_F^{\beta\sigma}(x_3)S_A^{cd}(x_2) -S_F^{\alpha\beta}(\omega)t^d_{\beta\sigma}\right]
{\cal A}^{\bar\eta} \bigg(\xi_2,v-x_2; \frac{\xi_1}{\bar\xi_2}, x_3-x_1\bigg)\nonumber\\
&&
\hspace{2.5cm}
-
\left[ t^c_{\alpha\beta}S_F^{\beta\sigma}(v)S_A^{cd}(x_2) -S_F^{\alpha\beta}(\omega)t^d_{\beta\sigma}\right]
 A^{\bar\eta}(v-x_2)\bigg\}
\nonumber\\
&&
\hspace{5cm}
\times
\left| (\rmq)[p^+-k_1^+-k_2^+,x_3]^{\sigma}_{s''}, (\rmg)[k_2^+,x_2]^d_{\eta}, (\rmp) [k_1^+,x_1]^{\lambda}\right\rangle_D\ .
\nonumber
\eeq

Finally, to obtain the partonic cross section Eq. \eqref{Partonic_q_chan}, we need to calculate the expectation value of the number operator in the outgoing wave function Eq. \eqref{out_full}. The number operator will pick up the Fock state containing a photon, a gluon, and a quark, and therefore only the $O(g_e g_s)$ component of Eq. \eqref{out_full} is relevant for our purposes. The resulting cross section must then be averaged over the configurations of the target field $\alpha^-$, which the Wilson lines depend on. We denote such CGC averages by $\langle\ \cdot\ \rangle_{x_A}$ with $x_A$ referring to the small longitudinal momentum fraction of the gluons in the target wave function. The outcome can be organized as
\beq
\label{Q_initiated_full}
(2\pi)^9\frac{d\sigma^{qA\to \gamma gq+X}}{d^3\uq_1d^3\uq_2d^3\uq_3}&=&\frac{1}{2N_c}g_s^2g_e^2(2\pi)\delta(p^+-q_1^+-q_2^+-q_3^+) \frac{1}{2q_1^+}\frac{1}{2q_2^+}\nonumber\\
&&
\times\; \Big\langle {\rm I}_{\rm bef-bef}+{\rm I}_{\rm aft-aft}+{\rm I}_{\rm bef-aft}+{\rm I}_{\rm aft-bef}\Big\rangle_{x_A}\;,
\eeq 
where the subscript ``${\rm bef}$" stands for the contribution in which the photon is emitted before the gluon (the first $O(g_eg_s)$ term in the outgoing wave function Eq. \eqref{out_full}) and  the subscript ``${\rm aft}$" stands for the case where the gluon is emitted first. The first subscript denotes the configuration in the amplitude, the second one corresponds to the conjugate amplitude.

Let us present the explicit expressions of these contributions, starting with ${\rm I}_{\rm bef-bef}$ (for the remaining of this subsection, $\xi_i$ denotes $q_i^+/p^+$):
\beq
\label{bef_bef}
&&
{\rm I}_{\rm bef-bef}=\sum_{s's''}\sum_{\lambda\eta}\int_{wv,w'v',y_1z_1,y_2z_2,y_3z_3}
\!\!\!\!\!\!e^{iq_1\cdot(y_1-z_1)+iq_2\cdot(y_2-z_2)+iq_3\cdot(y_3-z_3)} \delta^{(2)}[w'-(\bar\xi_1v'+\xi_1y_1)]
\nonumber\\
&&
\times\; 
\delta^{(2)}[w-(\bar\xi_1v+\xi_1z_1)]
\delta^{(2)}\bigg[v'-\Big\{\Big(1-\frac{\xi_2}{\bar\xi_1}\Big)y_3+\frac{\xi_2}{\bar\xi_1}y_2\Big\}\bigg]
\delta^{(2)}\bigg[v-\Big\{\Big(1-\frac{\xi_2}{\bar\xi_1}\Big)z_3+\frac{\xi_2}{\bar\xi_1}z_2\Big\}\bigg]
\nonumber\\
&&
\times\; 
\bigg\{ 
\phi^{\lambda\bar\lambda}_{ss'}(\xi_1) \phi^{*\lambda\bar\lambda'}_{s\bar s'}(\xi_1)
\phi^{\eta\bar\eta}_{s's''}\Big(\frac{\xi_2}{\bar\xi_1}\Big)
\phi^{*\eta\bar\eta'}_{\bar s's''}\Big(\frac{\xi_2}{\bar\xi_1}\Big)\bigg\}
A^{\bar\eta}(z_3-z_2)A^{\bar\eta'}(y_3-y_2)\nonumber\\
&&
\times\;  \tr\bigg[
{\cal G}^{\dagger d \bar{\lambda}'}_{\rm bef}\bigg(\xi_1,\frac{\xi_2}{\bar\xi_1}; w',v',y_1,y_2,y_3\bigg)\; 
{\cal G}^{d\bar{\lambda}}_{\rm bef}\bigg(\xi_1,\frac{\xi_2}{\bar\xi_1};w,v,z_1,z_2,z_3\bigg) \bigg]\;,
\eeq
where $\tr\{\cdots\}$ indicates the trace over the fundamental color indices. We have introduced a compact notation, ${\cal G}^{d \bar{\lambda}}_{\rm  bef}$, to encode the Wilson line structure in the amplitude or complex conjugate amplitude 
\beq
\label{G_bef}
&&
{\cal G}^{d\bar{\lambda}}_{\rm bef}\bigg(\xi_1,\frac{\xi_2}{\bar\xi_1};w,v,z_1,z_2,z_3\bigg)_{\alpha\sigma}\nonumber\\
&&
=\bigg\{ \Big[t^cS_F(z_3)S_A^{cd}(z_2)-S_F(w)t^d\Big]{\cal A}^{\bar\lambda}\Big(\xi_1,v-z_1;\frac{\xi_2}{\bar\xi_1},z_3-z_2\Big)\nonumber\\
&&
-\Big[ \big\{ S_F(v)-S_F(w)\big\}t^d\Big]A^{\bar\lambda}(v-z_1)\bigg\}_{\alpha\sigma}\;,
\eeq
with the standard and modified Weizs\"acker-Williams fields defined in Eqs. \eqref{Standard_WW} and \eqref{Modified_WW}, respectively. Similarly, the ${\rm I}_{\rm aft-aft}$ contribution can be written
\beq
\label{aft-aft}
&&
{\rm I}_{\rm aft-aft}=\sum_{s's''}\sum_{\lambda\eta}\int_{wv,w'v',y_1z_1,y_2z_2,y_3z_3}
\!\!\!\!\!\!e^{iq_1\cdot(y_1-z_1)+iq_2\cdot(y_2-z_2)+iq_3\cdot(y_3-z_3)} \delta^{(2)}[w'-(\bar\xi_2v'+\xi_2y_2)]
\nonumber\\
&&
\times\; 
\delta^{(2)}[w-(\bar\xi_2v+\xi_2z_2)]
\delta^{(2)}\bigg[v'-\Big\{\Big(1-\frac{\xi_1}{\bar\xi_2}\Big)y_3+\frac{\xi_1}{\bar\xi_2}y_1\Big\}\bigg]
\delta^{(2)}\bigg[v-\Big\{\Big(1-\frac{\xi_1}{\bar\xi_2}\Big)z_3+\frac{\xi_1}{\bar\xi_2}z_1\Big\}\bigg]
\nonumber\\
&&
\times\; 
\bigg\{ 
\phi^{\eta\bar\eta}_{ss'}(\xi_2) \phi^{*\eta\bar\eta'}_{s\bar s'}(\xi_2)
\phi^{\lambda\bar\lambda}_{s's''}\Big(\frac{\xi_1}{\bar\xi_2}\Big)
\phi^{*\lambda\bar\lambda'}_{\bar s's''}\Big(\frac{\xi_1}{\bar\xi_2}\Big)\bigg\}
A^{\bar\lambda}(z_3-z_1)A^{\bar\lambda'}(y_3-y_1)\nonumber\\
&&
\times\;  \tr\bigg[
{\cal G}^{\dagger d \bar{\eta}'}_{\rm aft}\bigg(\xi_2,\frac{\xi_1}{\bar\xi_2};w',v',y_1,y_2,y_3\bigg)\; 
{\cal G}^{d\bar{\eta}}_{\rm aft}\bigg(\xi_2,\frac{\xi_1}{\bar\xi_2};w,v,z_1,z_2,z_3\bigg) \bigg]\;,
\eeq 
with 
\beq
\label{G_aft}
&&
{\cal G}^{d\bar{\eta}}_{\rm aft}\bigg(\xi_2,\frac{\xi_1}{\bar\xi_2};w,v,z_1,z_2,z_3\bigg)_{\sigma\alpha}=
\bigg\{ \Big[t^cS_F(z_3)S^{cd}_A(z_2)-S_F(w)t^d\Big]{\cal A}^{\bar \eta}\Big(\xi_2,v-z_2;\frac{\xi_1}{\bar\xi_2},z_3-z_1 \Big)
\nonumber\\
&&
\hspace{6cm}
-
\Big[t^cS_F(v)S^{cd}_A(z_2)-S_F(w)t^d\Big]A^{\bar\eta}(v-z_2)\bigg\}_{\sigma\alpha}\;.
\eeq
The remaining two contributions ${\rm I}_{\rm bef-aft}$ and ${\rm I}_{\rm aft-bef}$ can also be written in a compact form by using Eq. \eqref{G_bef} and Eq. \eqref{G_aft} as
\beq
\label{bef-aft}
&&
{\rm I}_{\rm  bef-aft}=\sum_{s's''}\sum_{\lambda\eta}\int_{wv,w'v',y_1z_1,y_2z_2,y_3z_3}
\!\!\!\!\!\!e^{iq_1\cdot(y_1-z_1)+iq_2\cdot(y_2-z_2)+iq_3\cdot(y_3-z_3)} \delta^{(2)}[w'-(\bar\xi_1v'+\xi_1y_1)]
\nonumber\\
&&
\times\; 
\delta^{(2)}[w-(\bar\xi_2v+\xi_2z_2)]
\delta^{(2)}\bigg[v'-\Big\{\Big(1-\frac{\xi_2}{\bar\xi_1}\Big)y_3+\frac{\xi_2}{\bar\xi_1}y_2\Big\}\bigg]
\delta^{(2)}\bigg[v-\Big\{\Big(1-\frac{\xi_1}{\bar\xi_2}\Big)z_3+\frac{\xi_1}{\bar\xi_2}z_1\Big\}\bigg]
\nonumber\\
&&
\times\; 
\bigg\{ 
\phi^{\eta\bar\eta}_{ss'}(\xi_2) \phi^{*\eta\bar\eta'}_{\bar s' s''}\Big(\frac{\xi_2}{\bar\xi_1}\Big)
\phi^{\lambda\bar\lambda}_{s's''}\Big(\frac{\xi_1}{\bar\xi_2}\Big)
\phi^{*\lambda\bar\lambda'}_{s\bar s'}(\xi_1)\bigg\}
A^{\bar\lambda}(z_3-z_1)A^{\bar\eta'}(y_3-y_2)\nonumber\\
&&
\times\;  \tr\bigg[
{\cal G}^{\dagger d\bar{\lambda}'}_{\rm bef}\bigg(\xi_1,\frac{\xi_2}{\bar\xi_1};w',v',y_1,y_2,y_3\bigg)\; 
{\cal G}^{d\bar{\eta}}_{\rm aft}\bigg(\xi_2,\frac{\xi_1}{\bar\xi_2};w,v,z_1,z_2,z_3\bigg)\bigg]
\eeq
and 
\beq
\label{aft-bef}
&&
{\rm I}_{\rm aft-bef}=\sum_{s's''}\sum_{\lambda\eta}\int_{wv,w'v',y_1z_1,y_2z_2,y_3z_3}
\!\!\!\!\!\!e^{iq_1\cdot(y_1-z_1)+iq_2\cdot(y_2-z_2)+iq_3\cdot(y_3-z_3)} \delta^{(2)}[w'-(\bar\xi_2v'+\xi_2y_2)]
\nonumber\\
&&
\times\; 
\delta^{(2)}[w-(\bar\xi_1v+\xi_1z_1)]
\delta^{(2)}\bigg[v'-\Big\{\Big(1-\frac{\xi_1}{\bar\xi_2}\Big)y_3+\frac{\xi_1}{\bar\xi_2}y_1\Big\}\bigg]
\delta^{(2)}\bigg[v-\Big\{\Big(1-\frac{\xi_2}{\bar\xi_1}\Big)z_3+\frac{\xi_2}{\bar\xi_1}z_2\Big\}\bigg]
\nonumber\\
&&
\times\; 
\bigg\{ 
\phi^{\lambda\bar\lambda}_{ss'}(\xi_1) \phi^{*\lambda\bar\lambda'}_{\bar s' s''}\Big(\frac{\xi_1}{\bar\xi_2}\Big)
\phi^{\eta\bar\eta}_{s's''}\Big(\frac{\xi_2}{\bar\xi_1}\Big)
\phi^{*\eta\bar\eta'}_{s\bar s'}(\xi_2)\bigg\}
A^{\bar\eta}(z_3-z_2)A^{\bar\lambda'}(y_3-y_1)\nonumber\\
&&
\times\;  \tr\bigg[
{\cal G}^{\dagger d\bar{\eta}'}_{\rm aft}\bigg(\xi_2,\frac{\xi_1}{\bar\xi_2};w',v',y_1,y_2,y_3\bigg)\; 
{\cal G}^{d\bar{\lambda}}_{\rm bef}\bigg(\xi_1,\frac{\xi_2}{\bar\xi_1};w,v,z_1,z_2,z_3\bigg)\bigg]\, . 
\eeq
Note that, in order to keep the expressions compact, we have not performed the trivial integrations over $w,w',v$ and $v'$. 

Each contribution to the production cross section of the quark initiated channel can be written as a function of Wilson lines in the fundamental representation, in terms of the standard dipole and quadrupole amplitudes. To this end, some color algebra needs to be performed, with the help of the Fierz identity  
\beq
\label{Fierz}
t^a_{\alpha\beta}t^a_{\sigma\lambda}=\frac{1}{2}\left[ \delta_{\alpha\lambda}\delta_{\beta\sigma}-\frac{1}{N_c}\delta_{\alpha\beta}\delta_{\sigma\lambda}\right]\,,
\eeq
and the identity that relates the adjoint and fundamental representations of a unitary matrix 
\beq
\label{Adj_to_Fundamental}
S^{ab}_A(x)=2\, \tr\left[t^aS_F(x)t^bS_F^{\dagger}(x)\right]\,.
\eeq
Then, the before-before contribution can be written as
\beq
\label{bef-bef-final}
&&
{\rm I}_{\rm  bef-bef}=
\int_{wv,w'v',y_1z_1,y_2z_2,y_3z_3}
\!\!\!\!\!\!e^{iq_1\cdot(y_1-z_1)+iq_2\cdot(y_2-z_2)+iq_3\cdot(y_3-z_3)} \delta^{(2)}[w'-(\bar\xi_1v'+\xi_1y_1)]
\nonumber\\
&&
\times\; 
\delta^{(2)}[w-(\bar\xi_1v+\xi_1z_1)]
\delta^{(2)}\bigg[v'-\Big\{\Big(1-\frac{\xi_2}{\bar\xi_1}\Big)y_3+\frac{\xi_2}{\bar\xi_1}y_2\Big\}\bigg]
\delta^{(2)}\bigg[v-\Big\{\Big(1-\frac{\xi_2}{\bar\xi_1}\Big)z_3+\frac{\xi_2}{\bar\xi_1}z_2\Big\}\bigg]
\nonumber\\
&&
\times\;  
8\, {\cal M}_q^{\bar\lambda\bar\lambda'; \bar\eta\bar\eta'}\bigg(\xi_1,\frac{\xi_2}{\bar\xi_1}\bigg)
A^{\bar\eta}(z_3-z_2)A^{\bar\eta'}(y_3-y_2) 
\bigg[ A^{\bar\lambda'}(v'-y_1)A^{\bar\lambda}(v-z_1)\, {\rm W}_{\rm bef-bef}^{(A A)}
\nonumber\\
&&
\hspace{0.5cm}
+\; 
{\cal A}^{\bar\lambda'}\bigg( \xi_1, v'-y_1;\frac{\xi_2}{\bar\xi_1}, y_3-y_2\bigg)
{\cal A}^{\bar\lambda}\bigg(\xi_1,v-z_1; \frac{\xi_2}{\bar\xi_1}, z_3-z_2\bigg) {\rm W}_{\rm bef-bef}^{(\cal A \cal A)}
\nonumber\\
&&
\hspace{0.5cm}
- \; 
{\cal A}^{\bar\lambda'}\bigg( \xi_1, v'-y_1;\frac{\xi_2}{\bar\xi_1}, y_3-y_2\bigg)
A^{\bar\lambda}(v-z_1)\, {\rm W}_{\rm bef-bef}^{({\cal A} A)}
\nonumber\\
&&
\hspace{0.5cm}
- \; 
A^{\bar\lambda'}(v'-y_1)
{\cal A}^{\bar\lambda}\bigg(\xi_1,v-z_1; \frac{\xi_2}{\bar\xi_1}, z_3-z_2\bigg)
{\rm W}_{\rm bef-bef}^{( A \cal A)} \bigg]\;.
\eeq
Let us explain the compact notation that we have introduced in Eq. \eqref{bef-bef-final}. The functions ${\rm W}_{\rm bef-bef}$ define the dipole and quadrupole structures of the before-before contribution, accompanied by the standard or modified Weizs\"acker-Williams fields as indicated in the superscripts. These functions are calculated using Eqs. \eqref{Fierz} and \eqref{Adj_to_Fundamental}, and their explicit expressions read 
\beq
&&
\label{W_bb_AA}
{\rm W}_{\rm bef-bef}^{(AA)}=\frac{N_c^2-1}{2}\Big[ s(v,v')+s(w,w')-s(v,w')-s(w,v')\Big]\;,\\
&&
\label{W_bb_A'A'}
{\rm W}_{\rm bef-bef}^{(\cal A A)}= \frac{N_c^2}{2}\Big[ Q(y_2,z_2,z_3,y_3)s(z_2,y_2)-s(y_2,y_3)s(w,y_2)-s(z_2,w')s(z_3,z_2)+s(w,w')\Big]\nonumber\\
&&
\hspace{1.8cm}
-\, 
\frac{1}{2}\Big[ s(z_3,y_3)-s(w,y_3)-s(z_3,w')+s(w,w')\Big]\;,
\\
&&
\label{W_bb_A'A}
{\rm W}_{\rm bef-bef}^{({\cal A} A)}=\frac{N_c^2}{2}\Big[ s(y_2,y_3)s(v,y_2)-s(y_2,y_3)s(w,y_2)-s(v,w')+s(w,w')\Big]
\nonumber\\
&&
\hspace{1.8cm}
- \, 
\frac{1}{2}\Big[ s(v,y_3)-s(w,y_3)+s(w,w')-s(v,w')\Big]\;,
\nonumber\\
&&
\hspace{1.6cm}
={\rm W}_{\rm bef-bef}^{(\cal A A)}(z_3\to v, z_2\to v)\;,
\\
&&
\label{W_bb_AA'}
{\rm W}_{\rm bef-bef}^{(A{\cal A} )}= \frac{N_c^2}{2}\Big[ s(z_3,z_2)s(z_2,v')-s(z_3,z_2)s(z_2,w')-s(w,v')+s(w,w')\Big]
\nonumber\\
&&
\hspace{1.8cm}
-\, 
\frac{1}{2}\Big[ s(z_3,v')-s(z_3,w')-s(w,v')+s(w,w')\Big] \;,
\nonumber\\
&&
\hspace{1.6cm}
={\rm W}_{\rm bef-bef}^{(\cal A A)}(y_3\to v, y_2\to v)\;,
\eeq 
were we have used the standard definitions of dipole and quadrupole amplitudes: 
\beq
\label{dip_amp}
s(x,y)&=&\frac{1}{N_c}\tr\big[S_F(x)S_F^\dagger(y)\big]\;,
 \\
 \label{quad_amp}
 Q(x,y,v,w)&=&\frac{1}{N_c}\tr \big[S_F(x)S^\dagger_F(y)S_F(v)S^\dagger_F(w)\big]\;.
\eeq
Moreover, we have calculated explicitly the product of the splitting amplitudes, by using their definition given in Eq. \eqref{splt_Amp} and summing over the spin and polarization indices. 
This product is encoded in the function ${\cal M}_q^{\bar\lambda\bar\lambda'; \bar\eta\bar\eta'}\bigg(\xi_1,\frac{\xi_2}{\bar\xi_1}\bigg)$ (see Appendix \ref{App:splittings} for the details of the calculation), defined as 
\beq
\sum_{s's''}\sum_{\lambda\eta}\bigg\{ \phi^{\lambda\bar\lambda}_{ss'}(\xi_1)\phi^{*\lambda\bar\lambda'}_{\bar s' s}(\xi_1)\bigg\} \bigg\{ \phi^{\eta\bar\eta}_{s's''}\bigg(\frac{\xi_2}{\bar\xi_1}\bigg)\phi^{*\eta\bar\eta'}_{s''\bar s'}\bigg(\frac{\xi_2}{\bar\xi_1}\bigg)\bigg\}= 8\, {\cal M}_q^{\bar\lambda\bar\lambda'; \bar\eta\bar\eta'}\bigg(\xi_1,\frac{\xi_2}{\bar\xi_1}\bigg)
\eeq
with 
\beq
\label{F_bef-bef}
\hspace{-0.5cm}
{\cal M}_q^{\bar\lambda\bar\lambda'; \bar\eta\bar\eta'}\bigg(\xi_1,\frac{\xi_2}{\bar\xi_1}\bigg)=(1+\bar\xi_1^2)\bigg[1+\bigg(1-\frac{\xi_2}{\bar\xi_1}\bigg)^2\bigg]\delta^{\bar\lambda\bar\lambda'}\delta^{\bar\eta\bar\eta'}\nonumber\\
-(2-\xi_1)\xi_1\bigg(2-\frac{\xi_2}{\bar\xi_1}\bigg)\frac{\xi_2}{\bar\xi_1}\epsilon^{\bar\lambda\bar\lambda'}\epsilon^{\bar\eta\bar\eta'}\;.
\eeq
The remaining after-after and the crossed contributions (after-before and before-after) are computed in the same way. The explicit expression for the after-after contribution reads
\beq
\label{aft-aft-final}
&&
{\rm I}_{\rm aft-aft}=
\int_{wv,w'v',y_1z_1,y_2z_2,y_3z_3}
\!\!\!\!\!\!e^{iq_1\cdot(y_1-z_1)+iq_2\cdot(y_2-z_2)+iq_3\cdot(y_3-z_3)} \delta^{(2)}[w'-(\bar\xi_2v'+\xi_2y_2)]
\nonumber\\
&&
\times\; 
\delta^{(2)}[w-(\bar\xi_2v+\xi_2z_2)]
\delta^{(2)}\bigg[v'-\Big\{\Big(1-\frac{\xi_1}{\bar\xi_2}\Big)y_3+\frac{\xi_1}{\bar\xi_2}y_1\Big\}\bigg]
\delta^{(2)}\bigg[v-\Big\{\Big(1-\frac{\xi_1}{\bar\xi_2}\Big)z_3+\frac{\xi_1}{\bar\xi_2}z_1\Big\}\bigg]
\nonumber\\
&&
\times\;  
8\, {\cal M}_q^{\bar\lambda\bar\lambda'; \bar\eta\bar\eta'}\bigg(\xi_2,\frac{\xi_1}{\bar \xi_2}\bigg)
A^{\bar\lambda}(z_3-z_1)A^{\bar\lambda'}(y_3-y_1) 
\bigg[ A^{\bar\eta'}(v'-y_2)A^{\bar\eta}(v-z_2)\, {\rm W}_{\rm aft-aft}^{(A A)}
\nonumber\\
&&
\hspace{0.5cm}
+\; 
{\cal A}^{\bar\eta'}\bigg( \xi_2, v'-y_2;\frac{\xi_1}{\bar\xi_2}, y_3-y_1\bigg)
{\cal A}^{\bar\eta}\bigg(\xi_2,v-z_2; \frac{\xi_1}{\bar\xi_2}, z_3-z_1\bigg) {\rm W}_{\rm aft-aft}^{(\cal A \cal A)}
\nonumber\\
&&
\hspace{0.5cm}
- \; 
{\cal A}^{\bar\eta'}\bigg( \xi_2, v'-y_2;\frac{\xi_1}{\bar\xi_2}, y_3-y_1\bigg)
A^{\bar\eta}(v-z_2)\, {\rm W}_{\rm aft-aft}^{({\cal A} A)}
\nonumber\\
&&
\hspace{0.5cm}
- \; 
A^{\bar\eta'}(v'-y_2)
{\cal A}^{\bar\eta}\bigg(\xi_2,v-z_2; \frac{\xi_1}{\bar\xi_2}, z_3-z_1\bigg)
{\rm W}_{\rm aft-aft}^{( A \cal A)} \bigg]\;.
\eeq
Here, the function ${\cal M}_q^{\bar\lambda\bar\lambda'; \bar\eta\bar\eta'}\bigg(\xi_2,\frac{\xi_1}{\bar\xi_2}\bigg)$, which accounts for the product of splitting amplitudes is given by Eq. \eqref{F_bef-bef} with $\xi_1$ and $\xi_2$ interchanged. We have introduced functions ${\rm W}_{\rm aft-aft}$, which encode the dipole and quadrupole structures that accompany the standard and modified Weizs\"acker-Williams fields in the after-after contribution. These functions are very similar to their counterparts ${\rm W}_{\rm bef-bef}$ calculated for the before-before contribution, Eqs. \eqref{W_bb_AA} - \eqref{W_bb_AA'}, as can be easily observed by comparing the Wilson line structures given at the amplitude level as ${\cal G}^{b\bar\lambda}_{\rm  bef}$ and ${\cal G}^{b\bar\eta}_{\rm aft}$ in Eqs. \eqref{G_bef} and \eqref{G_aft}. We obtain:
\beq
&&
{\rm W}_{\rm aft-aft}^{(\cal AA)}={\rm W}_{\rm bef-bef}^{(\cal AA)}\;,
\\
&&
{\rm W}_{\rm aft-aft}^{( A A)}={\rm W}_{\rm aft-aft}^{(\cal A \cal A)}(z_3\to v, y_3\to v')={\rm W}_{\rm bef-bef}^{(\cal AA)}(z_3\to v, y_3\to v')\;,\\
&&
{\rm W}_{\rm aft-aft}^{( {\cal A} A)}={\rm W}_{\rm aft-aft}^{(\cal A \cal A)}(z_3\to v)={\rm W}_{\rm bef-bef}^{(\cal AA)}(z_3\to v)\;,\\
&&
{\rm W}_{\rm aft-aft}^{( A \cal A)}={\rm W}_{\rm aft-aft}^{(\cal A \cal A)}( y_3\to v')={\rm W}_{\rm bef-bef}^{(\cal AA)}( y_3\to v')\;.
\eeq

In a similar way, the crossed contributions (after-before and before-after) can be calculated, reading:
\beq
\label{aft-bef-final}
&&
{\rm I}_{\rm aft-bef}=\int_{wv,w'v',y_1z_1,y_2z_2,y_3z_3}
\!\!\!\!\!\!e^{iq_1\cdot(y_1-z_1)+iq_2\cdot(y_2-z_2)+iq_3\cdot(y_3-z_3)}
\delta^{(2)}[w'-(\bar\xi_2 v'+\xi_2 y_2)]
\nonumber\\
&&
\times \; 
\delta^{(2)}[w-(\bar\xi_1 v+\xi_1z_1)] 
\delta^{(2)}\bigg[ v'-\Big\{ \Big( 1-\frac{\xi_1}{\bar\xi_2}\Big)y_3+\frac{\xi_1}{\bar\xi_2}y_1\Big\}\bigg]
\delta^{(2)}\bigg[ v-\Big\{ \Big(1-\frac{\xi_2}{\bar\xi_1}\Big)z_3+\frac{\xi_2}{\bar\xi_1}z_2\Big\}\bigg]
\nonumber\\
&&
\times \; 
8\, \widetilde{\cal M}_q^{\bar\lambda\bar\lambda';\bar\eta\bar\eta'}(\xi_1,\xi_2)\, A^{\bar\eta}(z_3-z_2)A^{\bar\lambda'}(y_3-y_1) \bigg[ 
A^{\bar\eta'}(v'-y_2) A^{\bar\lambda}(v-z_1){\rm W}_{\rm aft-bef}^{(AA)}
\nonumber\\
&&
\hspace{0.5cm}
+\, {\cal A}^{\bar\eta'}\Big( \xi_2, v'-y_2; \frac{\xi_1}{\bar\xi_2},y_3-y_1\Big) 
{\cal A}^{\bar\lambda}\Big( \xi_1,v-z_1; \frac{\xi_2}{\bar\xi_1}, z_3-z_2\Big)
{\rm W}_{\rm aft-bef}^{(\cal AA)}
\nonumber\\
&&
\hspace{0.5cm}
-\, 
{\cal A}^{\bar\eta'}\Big( \xi_2, v'-y_2; \frac{\xi_1}{\bar\xi_2},y_3-y_1\Big) 
A^{\bar\lambda}(v-z_1)
{\rm W}_{\rm aft-bef}^{({\cal A}A)}
\nonumber\\
&&
\hspace{0.5cm}
-\, 
A^{\bar\eta'}(v'-y_2)
{\cal A}^{\bar\lambda}\Big( \xi_1,v-z_1; \frac{\xi_2}{\bar\xi_1}, z_3-z_2\Big)
{\rm W}_{\rm aft-bef}^{(A{\cal A})}   \bigg]\;,
\eeq 
where the function $\widetilde{\cal M}_q^{\bar\lambda\bar\lambda';\bar\eta\bar\eta'}(\xi_1,\xi_2)$ is calculated explicitly using Eq. \eqref{splt_Amp} (see Appendix \ref{App:splittings} for the details of the calculation) and reads 
\beq
\label{F_cross}
\widetilde{\cal M}_q^{\bar\lambda\bar\lambda'; \bar\eta\bar\eta'}(\xi_1,\xi_2)=\bigg[1+\bar\xi_1\bigg(1-\frac{\xi_1}{\bar\xi_2}\bigg)\bigg]\bigg[1+\bar\xi_2\bigg(1-\frac{\xi_2}{\bar\xi_1}\bigg)\bigg]\delta^{\bar\lambda\bar\lambda'}\delta^{\bar\eta\bar\eta'}\nonumber\\
-\frac{\xi_2}{\bar\xi_2}\frac{\xi_1}{\bar\xi_1}\big( \bar\xi_1+\bar\xi_2\big)^2\epsilon^{\bar\lambda\bar\lambda'}\epsilon^{\bar\eta\bar\eta'}\;.
\eeq
We would like to note that, as expected, this function is symmetric under the exchange of $\xi_1$ and $\xi_2$, since it appears at the cross-section level and corresponds to the product of the amplitude in which the photon is emitted \textit{before} the gluon, and the complex amplitude in which the emission of the photon takes place \textit{after} the radiation of the gluon (interchanging the gluon and the photon also involves changing $\lambda \leftrightarrow \eta$, under which $\cal{M}$ is also symmetric).

Once more, we define functions ${\rm W}_{\rm aft-bef}$ which encode the dipole and quadrupole structure of the after-before contribution, and which again can be written in terms of the functions ${\rm W}_{\rm bef-bef}$ by exchanging the coordinates of the intermediate and final state quarks:
\beq
&&
{\rm W}_{\rm aft-bef}^{(\cal AA)}={\rm W}_{\rm bef-bef}^{(\cal AA)}\;,
\\
&&
{\rm W}_{\rm aft-bef}^{(A\cal A)}={\rm W}_{\rm aft-bef}^{(\cal AA)}(y_3\to v')={\rm W}_{\rm bef-bef}^{(\cal A\cal A)}(y_3\to v')\;,
\\
&&
{\rm W}_{\rm aft-bef}^{({\cal A}A)}={\rm W}_{\rm bef-bef}^{({\cal A}A)}\;,
\\
&&
{\rm W}_{\rm aft-bef}^{(AA)}={\rm W}_{\rm aft-bef}^{({\cal A}A)}(y_3\to v')={\rm W}_{\rm bef-bef}^{({\cal A} A)}(y_3 \to v')\;,
\eeq
where ${\rm W}_{\rm bef-bef}^{(\cal AA)}$ and ${\rm W}_{\rm bef-bef}^{({\cal A}A)}$ are given in Eqs. \eqref{W_bb_A'A'} and \eqref{W_bb_A'A} respectively. 

Finally, we obtain the following expression for the before-after contribution:
\beq
\label{bef-aft-final}
&&
{\rm I}_{\rm bef-aft}=\int_{wv,w'v',y_1z_1,y_2z_2,y_3z_3}
\!\!\!\!\!\!e^{iq_1\cdot(y_1-z_1)+iq_2\cdot(y_2-z_2)+iq_3\cdot(y_3-z_3)}
\delta^{(2)}[w'-(\bar\xi_1v'+\xi_1y_1)]
\nonumber\\
&&
\times \, 
\delta^{(2)}[w-(\bar\xi_2v+\xi_2z_2)] 
\delta^{(2)}\bigg[ v'-\Big\{ \Big( 1-\frac{\xi_2}{\bar\xi_1}\Big)y_3+\frac{\xi_2}{\bar\xi_1}y_2\Big\}\bigg]
\delta^{(2)}\bigg[ v-\Big\{ \Big( 1-\frac{\xi_1}{\bar\xi_2}\Big)z_3+\frac{\xi_1}{\bar\xi_2}z_1\Big\}\bigg]
\nonumber\\
&&
\times \, 
8\, \widetilde{\cal M}_q^{\bar\lambda\bar\lambda';\bar\eta\bar\eta'}(\xi_1,\xi_2) \, 
A^{\bar\lambda}(z_3-z_1)A^{\bar\eta'}(y_3-y_2)\bigg[
A^{\bar\lambda'}(v'-y_1)A^{\bar\eta}(v-z_2)
{\rm W}_{\rm bef-aft}^{(AA)}
\nonumber\\
&&
\hspace{0.5cm}
+ \, 
{\cal A}^{\bar\lambda'}\Big( \xi_1, v'-y_1; \frac{\xi_2}{\bar\xi_1}, y_3-y_2\Big) 
{\cal A}^{\bar\eta}\Big( \xi_2, v-z_2; \frac{\xi_1}{\bar\xi_2}, z_3-z_1\Big)
{\rm W}_{\rm bef-aft}^{(\cal AA)}
\nonumber\\
&&
\hspace{0.5cm}
-\, 
{\cal A}^{\bar\lambda'}\Big( \xi_1, v'-y_1; \frac{\xi_2}{\bar\xi_1}, y_3-y_2\Big) 
A^{\bar\eta}(v-z_2)
{\rm W}_{\rm bef-aft}^{({\cal A}A)}
\nonumber\\
&&
\hspace{0.5cm}
-\, 
A^{\bar\lambda'}(v'-y_1)
{\cal A}^{\bar\eta}\Big( \xi_2, v-z_2; \frac{\xi_1}{\bar\xi_2}, z_3-z_1\Big)
{\rm W}_{\rm bef-aft}^{(A\cal A)} \bigg]\;.
\eeq
As mentioned previously, the product of splitting functions in the amplitude and conjugate amplitude is symmetric under the exchange of the gluon and the photon, which corresponds to $\xi_1\leftrightarrow \xi_2$, $\lambda \leftrightarrow \eta$ and $\bar{\lambda} \leftrightarrow \bar{\eta}$, hence for this contribution the function $\widetilde{\cal M}_q^{\bar\lambda\bar\lambda';\bar\eta\bar\eta'}(\xi_1,\xi_2)$ is given by Eq. \eqref{F_cross} as well. Once more, the functions ${\rm W}_{\rm bef-aft}$ are written using the ones calculated in the before-before contribution and performing the appropriate change of coordinates, which gives 
\beq
&&
{\rm W}_{\rm bef-aft}^{(\cal A A)}={\rm W}_{\rm bef-bef}^{(\cal A A)}\;,
\\
&&
{\rm W}_{\rm bef-aft}^{({\cal A}A)}={\rm W}_{\rm bef-aft}^{({\cal A A})}(z_3\to v)={\rm W}_{\rm bef-bef}^{(\cal AA)}(z_3\to v)\;,
\\
&&
{\rm W}_{\rm bef-aft}^{(A\cal A)}={\rm W}_{\rm bef-bef}^{(A\cal A)}\;,
\\
&&
{\rm W}_{\rm bef-aft}^{(AA)}={\rm W}_{\rm bef-aft}^{(A\cal A)}( z_3\to v)={\rm W}_{\rm bef-bef}^{(A\cal A)}( z_3\to v)\;,
\eeq 
with ${\rm W}_{\rm bef-bef}^{(\cal AA)}$ and ${\rm W}_{\rm bef-bef}^{(A\cal A)}$ defined in Eqs. \eqref{W_bb_A'A'} and \eqref{W_bb_AA'}.

This concludes the calculation of the cross section for the production of a hard photon and two hard jets in the quark initiated channel. The final result at the partonic level is given by Eq. \eqref{Q_initiated_full}, with the terms ${\rm I}_{\rm  bef-bef}$, ${\rm I}_{\rm aft-aft}$, ${\rm I}_{\rm  aft-bef}$ and ${\rm I}_{\rm  bef-aft}$ presented in Eqs. \eqref{bef-bef-final}, \eqref{aft-aft-final}, \eqref{aft-bef-final} and \eqref{bef-aft-final}. In order to obtain the full cross section, the partonic cross section should be convolved with the quark distribution function $f_{q}(x_p, \mu^2)$ as stated in Eq. \eqref{Q_in_full_X_section}.


\subsection{Gluon initiated channel: $gA\to q\bq\gamma+X$}
Let us now consider the production in the gluon initiated channel. Just like in the quark initiated case, the partonic level cross section is defined as the expectation value of the number operator in the outgoing gluon state: 
\begin{figure}
\begin{centering}
\includegraphics[trim=4.8cm 21cm 1.6cm 1.5cm, clip=true,scale=1]{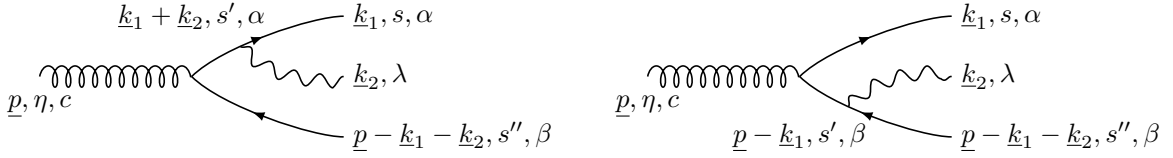}\par\end{centering}
\caption{The dressed gluon state to order $O(g_eg_s)$, with the photon being emitted either from the quark (left) or from the antiquark (right).}
\label{fig:gluon_channel_notations}  
\end{figure}
\beq
\label{X_sec_gluon_in}
&&
(2\pi)^9\frac{d\sigma^{gA\to q\gamma\bq +X}}{d^3\uq_1d^3\uq_2d^3\uq_3}(2\pi)\delta(p^+-q_1^+-q_2^+-q_3^+)
\nonumber\\
&&
\hspace{2.5cm}
=\frac{1}{2(N_c^2-1)}\sum_{c,\eta}{}_{\rm out}\Big\langle(\rmg)[p^+,0]^c_{\eta}\Big|O(\uq_1,\uq_2,\uq_3)\Big|(\rmg)[p^+,0]^c_{\eta}\Big\rangle_{\rm out}\;,
\eeq
where the normalization factor $1/2(N_c^2-1)$ comes from the averaging over the color and polarization indices of the outgoing gluon state. In order to calculate the production of a photon together with a quark-antiquark pair, we define the number operator in Eq. \eqref{X_sec_gluon_in} as follows:
\beq
\label{Num_Op_G_in}
O(\uq_1,\uq_2,\uq_3)=b^{\dagger \sigma}_t(\uq_1)b^{\sigma}_t(\uq_1)\gamma^\dagger_\delta(\uq_2)\gamma_\delta(\uq_2)d^{\dagger\kappa}_r(\uq_3) d^\kappa_r(\uq_3)\;,
\eeq
where $b^\sigma_t(\uq_1)$ is the annihilation operator of a dressed quark with color $\sigma$, spin $t$ and three-momentum $\uq_1$, and $\gamma_\delta(\uq_2)$ is the annihilation operator of the dressed photon state with polarization $\delta$ and three-momentum $\uq_2$. Finally, $d^\kappa_r(\uq_3)$ is the annihilation operator of the dressed antiquark with color $\kappa$, spin $r$ and three-momentum $\uq_3$.  The action of the creation and annihilation operators on the dressed states and on the Fock vacuum are defined in the standard way (see Eqs. \eqref{action_ann} and \eqref{action_cre}). As in the case of the quark initiated channel, it is more convenient to write the expectation value of the number operator in the mixed (longitudinal momentum and transverse coordinate) space:
\beq
\label{Num_Op_G_mixed}
&&
{}_{\rm out}\Big\langle(\rmg)[p^+,0]^c_{\eta}\Big|O(\uq_1,\uq_2,\uq_3)\Big|(\rmg)[p^+,0]^c_{\eta}\Big\rangle_{\rm out}=
\int_{y_1z_1y_2z_2y_3z_3} e^{iq_1\cdot(y_1-z_1)+ iq_2\cdot(y_2-z_2)+iq_3\cdot(y_3-z_3)}
\nonumber\\
&&
\hspace{3.3cm}
\, 
{}_{\rm out}\Big\langle(\rmg)[p^+,0]^c_{\eta}\Big| 
b^{\dagger \sigma}_t(q^+_1,y_1)b^{\sigma}_t(q^+_1,z_1)\gamma^\dagger_\delta(q^+_2,y_2)\gamma_\delta(q^+_2,z_2)
\nonumber\\
&&
\hspace{7cm}
\times\; 
 d^{\dagger\kappa}_r(q^+_3,y_3) d^\kappa_r(q^+_3,z_3)
\Big|(\rmg)[p^+,0]^c_{\eta}\Big\rangle_{\rm out}\;.
\eeq
We will adopt the same road map as introduced in the quark initiated channel. Namely, in order to evaluate Eq. \eqref{Num_Op_G_mixed}, we first need to calculate the outgoing wave function in the gluon initiated channel. This will be achieved in the following way: starting from the perturbative expression of the dressed gluon state written in terms of the bare states in full momentum space, we perform a Fourier transform to get the mixed space expression. This expression is then used to calculate the interaction with the target, which provides us with the outgoing gluon wave function in terms of the bare states. Finally, we rewrite the outgoing gluon wave function in terms of the dressed components.  In this subsection, we will only give the results of various steps of the calculation, and relegate the details to the appendix \ref{App:Derivation}.

In full momentum space, the perturbative expression of the dressed gluon state with longitudinal momentum $p^+$, vanishing transverse momentum, color $c$ and polarization $\eta$  at $O(g_eg_s)$ can be written as 
\beq
\label{dressed_gluon_mom}
&&
\big|(\rmg)[p^+,0]^c_\eta\big\rangle_D=Z^g \big| (\rmg) [p^+,0]^c_\eta\big\rangle_0
\nonumber\\
&&
\hspace{1cm}
+\, 
Z^{q\bar q}\, g_s\sum_{ss',\alpha\beta} \int \frac{dk^+_1}{2\pi} \frac{d^2k_1}{(2\pi)^2} \,  t^c_{\alpha\beta}\, 
\, F^{(1)}_{(\bf q\bar q)} \Big[ ({\bf q}) [k_1^+,k_1], ({\bf \bar q})[p^+-k^+_1, -k_1]\Big]^{\eta}_{s's} 
\nonumber\\
&&
\hspace{6cm}
\times \; 
\big| ({\bf q}) [k_1^+,k_1]_s^{\alpha}; ({\bf \bar q})[p^+-k_1^+,-k_1^+]_{s'}^\beta\big\rangle_0
\nonumber\\
&&
\hspace{1cm}
+\,
Z^{q\bar q \rmp} \, g_sg_e\sum_{ss''}\sum_{\alpha\beta\lambda}
\int   \frac{dk^+_1}{2\pi} \frac{d^2k_1}{(2\pi)^2}  \frac{dk^+_2}{2\pi} \frac{d^2k_2}{(2\pi)^2} \, t^c_{\alpha\beta}
\nonumber\\
&&
\hspace{1.2cm}
\times\, 
\bigg\{ 
F^{(2)}_{({\bf q\bar q-\bar q\rmp})}\Big[ ({\bf q}) [k_1^+,k_1], (\gamma) [k_2^+,k_2], ({\bf \bar q}) [p^+-k^+_1-k_2^+, -k_1-k_2]\Big]^{\eta,\lambda}_{ss''} 
\nonumber\\
&&
\hspace{1.5cm}
+\; 
F^{(2)}_{({\bf q\bar q-q\rmp})}\Big[ ({\bf q}) [k_1^+,k_1], (\rmp) [k_2^+,k_2], ({\bf \bar q}) [p^+-k^+_1-k_2^+, -k_1-k_2]\Big]^{\eta,\lambda}_{ss''} \bigg\}
\nonumber\\
&&
\hspace{1.2cm}
\times \,
\big| ({\bf q})[k_1^+,k_1]^{\alpha}_s, (\rmp)[k_2^+,k_2]^\lambda, ({\bf \bar q})[p^+-k_1^+-k_2^+, -k_1-k_2]^\beta_{s''}\big\rangle_0\;.
\eeq
Here $Z^g$, $Z^{q\bar q}$ and $Z^{q\bar q\gamma}$ are again the normalization functions which provide the virtual contributions to the production process. As in the case of the quark initiated channel, we neglect these contributions and set them equal to one, since here we are interested in the tree-level production of a quark, antiquark and photon, initiated by a gluon. Moreover, the instantaneous gluon contribution is neglected since it is suppressed by two powers of the final momenta of the jets, which are hard.

The function $F^{(1)}_{\bf q\bar q}$ defines the momentum structure of the gluon to quark-antiquark splitting. It is well known in the literature (see for example \cite{Elastic_vs_Inelastic}), but for the sake of completeness, the derivation of this function is presented in Appendix \ref{App:Derivation}. Its explicit expression reads
\beq
\label{F1}
F^{(1)}_{(\bf q\bar q)} \Big[ ({\bf q}) [k_1^+,k_1], ({\bf \bar q})[p^+-k^+_1,p -k_1]\Big]^{\eta}_{s's} = \Bigg[ \frac{-1}{\sqrt{2p^+}}\Bigg] \Psi^{\eta\bar\eta}_{ss'}(\xi_1)\frac{(\xi_1p-k_1)^{\bar\eta}}{(\xi_1p-k_1)^2}
\eeq
with 
\beq
\label{split_qbarq}
\Psi^{\eta\bar\eta}_{ss'}(\xi_1)=\Big[(1-2\xi_1)\delta^{\eta\bar\eta}\delta_{s,-s'}-i\epsilon^{\eta \bar\eta}\sigma^3_{s,-s'}\Big]
\eeq
where the longitudinal momentum ratio is $\xi_1\equiv k_1^+/p^+$. The functions $F^{(2)}_{(\bf q\bar q-\bar q\rmp)}$ and $F^{(2)}_{(\bf q\bar q- q\rmp)}$, which appear at $O(g_eg_s)$, define the momentum structure of two successive splittings: the splitting of the gluon in a quark-antiquark pair, followed by photon emission from the antiquark, or the gluon splitting followed by photon emission from the quark, respectively (see Fig. \ref{fig:gluon_channel_notations}). The derivation of these two functions is presented in appendix \ref{App:Derivation}, and the result reads
\beq
\label{F2_antiq-photon}
F^{(2)}_{(\bf q\bar q-\bar q\rmp)}\Big[ ({\bf q}) [k_1^+,k_1], (\rmp) [k_2^+,k_2], ({\bf \bar q}) [p^+-k^+_1-k_2^+, p-k_1-k_2]\Big]^{\eta,\lambda}_{ss''} \nonumber\\
= \bigg[\frac{-1}{\sqrt{2p^+}}\Psi^{\eta\bar\eta}_{ss'}(\xi_1)\bigg]\bigg[\frac{1}{\sqrt{2\xi_2p^+}}\phi^{\lambda\bar \lambda}_{s's''}\Big(\frac{\xi_2}{\bar\xi_1}\Big)\bigg]\hspace{3.3cm}\nonumber\\
\times\, 
 \frac{(\xi_1p-k_1)^{\bar\eta}}{(\xi_1p-k_1)^2}
 \frac{\xi_1[\xi_2(p-k_1)^{\bar\lambda}-\bar\xi_1k_2^{\bar\lambda}]}{\xi_2(\xi_1p-k_1)^2+\xi_1(\xi_2p-k_2)^2-(\xi_2k_1-\xi_1k_2)^2}
\eeq
and
\beq
\label{F2_q-photon}
F^{(2)}_{(\bf q\bar q- q\rmp)}\Big[ ({\bf q}) [k_1^+,k_1], (\rmp) [k_2^+,k_2], ({\bf \bar q}) [p^+-k^+_1-k_2^+, p-k_1-k_2]\Big]^{\eta,\lambda}_{ss''}
\hspace{1cm}\nonumber\\
=\bigg[ \frac{-1}{\sqrt{2p^+}}\Psi^{\eta\bar\eta}_{s's''}(\xi_1+\xi_2)\bigg]
\bigg[ \frac{1}{\sqrt{2\xi_2p^+}}\phi^{\lambda\bar\lambda}_{ss'}\Big(\frac{\xi_2}{\xi_1+\xi_2}\Big)\bigg]\hspace{3.9cm}\\
\times\; \frac{\big[(\xi_1+\xi_2)p-(k_1+k_2)\big]^{\bar\eta}}{\big[(\xi_1+\xi_2)p-(k_1+k_2)\big]^2}
\frac{(1-\xi_1-\xi_2)(\xi_2k_1-\xi_1k_2)^{\bar\lambda}}{\xi_2(\xi_1p-k_1)^2+\xi_1(\xi_2p-k_2)^2-(\xi_2k_1-\xi_1k_2)^2}\nonumber
\eeq
with the longitudinal momentum fractions defined as in the first part of the previous subsection: $\xi_2\equiv k_2^+/p^+$, $\bar\xi_1\equiv 1-\xi_1$ and $\bar\xi_2\equiv 1-\xi_2$. The splitting amplitudes for gluon to quark-antiquark ($\Psi^{\eta\bar\eta}_{ss'}(\xi_1)$) and for quark to quark-photon ($\phi^{\lambda\bar\lambda}_{ss'}(\xi_1)$) are defined in Eqs. \eqref{split_qbarq} and \eqref{splt_Amp}, respectively. 

The next step is to perform the two dimensional Fourier transform on the dressed gluon state given in Eq. \eqref{dressed_gluon_mom} in order to write it in the mixed space. We again refer to the appendix \ref{App:Dressed_in_mixed_space} for the details of the calculation, and write the final result as 
\beq
\label{dressed_gluon_mixed_sp}
&&
\big| ({\bf g})[p^+,0]^c_\eta\big\rangle_D=\int_w\big|({\bf g})[p^+,w]^c_{\eta}\big\rangle_0
\\
&&
+\, 
g_s\int\frac{dk_1^+}{2\pi} \, t^c_{\alpha\beta} \int_{wvx_1} 
\bigg[\frac{(-i)}{\sqrt{2p^+}}\Psi^{\eta\bar\eta}_{ss'}(\xi_1)\bigg] A^{\bar\eta}(v-x_1) \, 
\delta^{(2)}[w-(\xi_1x_1+\bar\xi_1v)] 
\nonumber\\
&&
\hspace{7cm}
\times\, 
\big| ({\bf q})[k_1^+,x_1]^\alpha_s; ({\bf \bar q})[p^+-k_1^+,v]^\beta_{s'}\big\rangle_0
\nonumber\\
&&
+\, 
g_sg_e \int \frac{dk_1^+}{2\pi}\frac{dk_2^+}{2\pi} \, t^c_{\alpha\beta} \int_{wvx_1x_2x_3}
\Bigg\{ \delta^{(2)}[w-(\xi_1x_1+\bar\xi_1v)] \delta^{(2)}\bigg[v-\bigg\{\bigg(1-\frac{\xi_2}{\bar\xi_1}\bigg)x_3+\frac{\xi_2}{\bar\xi_1}x_2\bigg\}\bigg]
\nonumber\\
&&
\times \, 
\bigg[\frac{(-i)}{\sqrt{2p^+}}\Psi^{\eta\bar\eta}_{ss'}(\xi_1)\bigg]
\bigg[\frac{(+i)}{\sqrt{2\xi_2p^+}}\phi^{\lambda\bar\lambda}_{s's''}\Big(\frac{\xi_2}{\bar\xi_1}\Big)\bigg]
A^{\bar\lambda}(x_3-x_2){\cal A}^{\bar\eta}\bigg(\xi_1, v-x_1; \frac{\xi_2}{\bar\xi_1},x_3-x_2\bigg)
\nonumber\\
&&
+\; 
\delta^{(2)}\Big[w-\big\{ (\xi_1+\xi_2)v+(1-\xi_1-\xi_2)x_3\big\}\Big]
\delta^{(2)}\bigg[ v-\bigg\{\bigg(1-\frac{\xi_2}{\xi_1+\xi_2}\bigg)x_1+\frac{\xi_2}{\xi_1+\xi_2}x_2\bigg\}\bigg] 
\nonumber\\
&&
\times\, 
\bigg[ \frac{(-i)}{\sqrt{2p^+}}\Psi^{\eta\bar\eta}_{s's''}(\xi_1+\xi_2)\bigg]\bigg[\frac{(+i)}{\sqrt{2\xi_2p^+}}\phi^{\lambda\bar\lambda}_{ss'}\bigg(\frac{\xi_2}{\xi_1+\xi_2}\bigg)\bigg]
A^{\bar\lambda}(x_2-x_1)
\nonumber\\
&&
\times\; 
{\cal A}^{\bar\eta}\bigg((1-\xi_1-\xi_2),v-x_3; \frac{\xi_2}{\xi_1+\xi_2}, x_2-x_1\bigg)\Bigg\}
\nonumber\\
&&
\times\; 
\big| ({\bf q})[k_1^+,x_1]^\alpha_s; (\rmp)[k_2^+,x_2]^\lambda; ({\bf \bar q})[p^+-k_1^+-k_2^+,x_3]^\beta_{s''}\big\rangle_0
\nonumber
\eeq
where $A^{\bar\eta}$ and ${\cal A}^{\bar\eta}$ are the standard and modified Weizs\"acker-Williams fields, given in Eqs. \eqref{Standard_WW} and \eqref{Modified_WW}, as in the case of the quark initiated channel. The standard one arises from the single splitting whereas the modified one is originated from the contribution with two successive splittings.

Just like in the quark initiated channel, from the above expression in mixed Fourier space, the eikonal interaction with the target can be incorporated in a straightforward way by inserting the appropriate Wilson lines, yielding the outgoing wave function in terms of the bare components
\beq
\label{out_gluon_bare}
&&
\big| ({\bf g})[p^+,0]^c_{\eta}\big\rangle_{\rm out}= \int_w S_A^{cd}(w)\big|({\bf g})[p^+,w]^d_{\eta}\big\rangle_0
\\
&&
+\;
g_s\int \frac{dk_1^+}{2\pi}\int_{wvx_1}
\Big[S^\dagger_F(v)t^cS_F(x_1)\Big]^{\beta\alpha} 
\delta^{(2)}\big[ w-(\xi_1x_1+\bar\xi_1v)\big]
\frac{(-i)}{\sqrt{2p^+}}\Psi^{\eta\bar\eta}_{ss'}(\xi_1)
\; A^{\bar\eta}(v-x_1)
\nonumber\\
&&
\hspace{9cm}
\times\; 
\big| ({\bf q})[k_1^+,x_1]^\alpha_s; ({\bf \bar q})[p^+-k_1^+,v]^\beta_{s'}\big\rangle_0
\nonumber
\\
&&
+\; 
g_sg_e\int \frac{dk_1^+}{2\pi}\frac{dk_2^+}{2\pi}\int_{wvx_1x_2x_3} \Big[ S^\dagger_F(x_3)t^cS_F(x_1)\Big]^{\beta\alpha}
\nonumber\\
&&
\times\; \Bigg\{
\delta^{(2)}[w-(\xi_1x_1+\bar\xi_1v)] 
\delta^{(2)}\bigg[ v-\bigg\{\bigg(1-\frac{\xi_2}{\bar\xi_1}\bigg)x_3+\frac{\xi_2}{\bar\xi_1}x_2\bigg\}\bigg]
\frac{(-i)}{\sqrt{2p^+}}\Psi^{\eta\bar\eta}_{ss'}(\xi_1)
\frac{(+i)}{\sqrt{2\xi_2p^+}}\phi^{\lambda\bar\lambda}_{s's''}\bigg(\frac{\xi_2}{\bar\xi_1}\bigg)
\nonumber\\
&&
\hspace{0.5cm}
\times\; 
A^{\bar\lambda}(x_3-x_2){\cal A}^{\bar\eta}\Big(\xi_1, v-x_1;\frac{\xi_2}{\bar\xi_1},x_3-x_2\Big) 
\nonumber\\
&&
\hspace{0.5cm}
+\; 
\delta^{(2)}\big[ w-\big\{(\xi_1+\xi_2)v+(1-\xi_1-\xi_2)x_3\big\}\big]
\delta^{(2)}\bigg[v-\bigg\{\bigg(1-\frac{\xi_2}{\xi_1+\xi_2}\bigg)x_1+\frac{\xi_2}{\xi_1+\xi_2}x_2\bigg\}\bigg]
\nonumber\\
&&
\hspace{0.5cm}
\times\; 
\frac{(-i)}{\sqrt{2p^+}}\Psi^{\eta\bar\eta}_{s's''}(\xi_1+\xi_2)\, 
\frac{(+i)}{\sqrt{2\xi_2p^+}}\phi^{\lambda\bar\lambda}_{ss'}\bigg(\frac{\xi_2}{\xi_1+\xi_2}\bigg)
\nonumber\\
&&
\hspace{0.5cm}
\times\; 
A^{\bar\lambda}(x_2-x_1){\cal A}^{\bar\eta}\bigg( (1-\xi_1-\xi_2), v-x_3; \frac{\xi_2}{\xi_1+\xi_2},x_2-x_1\bigg)
\Bigg\}
\nonumber\\
&&
\hspace{6cm}
\times\; 
\big| ({\bf q})[k_1^+,x_1]^\alpha_s; (\rmp)[k_2^+,x_2]^\lambda; ({\bf \bar q})[p^+-k_1^+-k_2^+,x_3]^\beta_{s''}\big\rangle_0
\nonumber
\eeq
The final step is to rewrite the outgoing gluon wave function in terms of the dressed states instead of the bare ones. A schematic derivation of this step is presented in appendix \ref{App:bare2dressed}, and the result is:
\beq
\label{gluon_out_final}
&&
\big| ({\bf g})[p^+,0]^c_\eta\big\rangle_{\rm out}=\int_w S_A^{cd}(w)\big|({\bf g})[p^+,w]^d_\eta\big\rangle_D
\\
&&
+\; 
g_s\int\frac{dk_1^+}{2\pi}\int_{wvx_1}
\Big[ S^\dagger_F(v)t^cS_F(x_1)-S_A^{cd}(w)t^d\Big]_{\beta\alpha}
\frac{(-i)}{\sqrt{2p^+}}\Psi^{\eta\bar\eta}_{ss'}(\xi_1) A^{\bar\eta}(v-x_1)
\nonumber\\
&&
\times\; 
\delta^{(2)}[w-(\xi_1x_1+\bar\xi_1v)]
\big| ({\bf q})[k_1^+,x_1]^\alpha_s; ({\bf \bar q})[p^+-k_1^+,v]^\beta_{s'}\big\rangle_D
\nonumber
\\
&&
+\; 
g_sg_e\int \frac{dk_1^+}{2\pi}\frac{dk_2^+}{2\pi}\int_{wvx_1x_2x_3} \Bigg\{
\delta^{(2)}[w-(\xi_1x_1+\bar\xi_1v)]
\delta^{(2)}\bigg[v-\bigg\{ \bigg(1-\frac{\xi_2}{\bar\xi_1}\bigg)x_3+\frac{\xi_2}{\bar\xi_1}x_2\bigg\}\bigg]
\nonumber\\
&&
\times\; 
\frac{(-i)}{\sqrt{2p^+}}\Psi^{\eta\bar\eta}_{ss'}(\xi_1) 
\frac{(+i)}{\sqrt{2\xi_2p^+}}\phi^{\lambda\bar\lambda}_{s's''}\bigg(\frac{\xi_2}{\bar\xi_1}\bigg)
A^{\bar\lambda}(x_3-x_2) 
\Big[{\cal G}^c_{{ \bar q\gamma}}(\xi_1,\xi_2; w,v,x_1,x_2,x_3)\Big]^{\bar\eta}_{\beta\alpha}
\nonumber\\
&&
+\; 
\delta^{(2)}\big[w-\big\{(\xi_1+\xi_2)v+(1-\xi_1-\xi_2)x_3\big\} \big]
\delta^{(2)}\bigg[ v- \bigg\{ \bigg(1-\frac{\xi_2}{\xi_1+\xi_2}\bigg)x_1+\frac{\xi_2}{\xi_1+\xi_2}x_2\bigg\}\bigg]
\nonumber\\
&&
\times\; 
\frac{(-i)}{\sqrt{2p^+}}\Psi^{\eta\bar\eta}_{s's''}(\xi_1+\xi_2) 
\frac{(+i)}{\sqrt{2\xi_2p^+}}\phi^{\lambda\bar\lambda}_{ss'}\bigg( \frac{\xi_2}{\xi_1+\xi_2}\bigg)
A^{\bar\lambda}(x_2-x_1)
\Big[{\cal G}^c_{{ q\gamma}}(\xi_1,\xi_2; w,v,x_1,x_2,x_3)\Big]^{\bar\eta}_{\beta\alpha}
\Bigg\}
\nonumber\\
&&
\times\; 
\big| ({\bf q})[k_1^+,x_1]^\alpha_s; (\rmp)[k_2^+,x_2]^\lambda; ({\bf \bar q})[p^+-k_1^+-k_2^+,x_3]^\beta_{s''}\big\rangle_D
\nonumber
\eeq 
where we have introduced a compact notation when writing the quark-antiquark-photon component of the outgoing gluon wave function. The two contributions to this component  originate either from photon emission by the antiquark or by the quark, as is indicated by the subscript of the newly introduced function ${\cal G}$:


\beq
\label{antiquark_photon}
\Big[{\cal G}_{\bar q\gamma}^c(\xi_1,\xi_2; w,v,x_1,x_2,x_3)\Big]^{\bar\eta}_{\beta\alpha}\hspace{5.4cm}\nonumber\\
=\Big[S^\dagger_F(x_3)t^cS_F(x_1)-S_A^{cd}(w)t^d\Big]_{\beta\alpha}
{\cal A}^{\bar\eta}\bigg( \xi_1, v-x_1; \frac{\xi_2}{\bar\xi_1},x_3-x_2\bigg) 
\nonumber\\
-\Big[S^\dagger_F(v)t^cS_F(x_1)-S_A^{cd}(w)t^d\Big]_{\beta\alpha}A^{\bar\eta}(v-x_1)\;,\hspace{2.8cm}
\eeq
and 
\beq
\label{quark_photon}
\hspace{-1.8cm}
\Big[{\cal G}_{q\gamma}^c(\xi_1,\xi_2; w,v,x_1,x_2,x_3)\Big]^{\bar\eta}_{\beta\alpha}\hspace{7.6cm}\nonumber\\\
=\Big[S^\dagger_F(x_3)t^cS_F(x_1)-S_A^{cd}(w)t^d\Big]_{\beta\alpha}
{\cal A}^{\bar\eta}\bigg(1-\xi_1-\xi_2, v-x_3; \frac{\xi_2}{\xi_1+\xi_2},x_2-x_1\bigg) 
\nonumber\\
-\Big[S^\dagger_F(x_3)t^cS_F(v)-S_A^{cd}(w)t^d\Big]_{\beta\alpha}A^{\bar\eta}(v-x_3)\;.\hspace{5.1cm}
\eeq

The partonic cross section is once again obtained from the expectation value of the number operator Eq. \eqref{Num_Op_G_in} in the outgoing gluon wave function Eq. \eqref{gluon_out_final}. Note that, just like in the quark initiated channel, the number operator only will only extract the three-particle Fock state we are interested in, hence the dressed gluon and dressed quark-antiquark states in the outgoing gluon wave function can be neglected. The result is:
\beq
\label{gluon_channel_partonic}
(2\pi)^9\frac{d\sigma^{gA\to q\gamma\bar q+X}}{d^3\uq_1d^3\uq_2d^3\uq_3}
=\frac{1}{2(N_c^2-1)}g_s^2g_e^2(2\pi)\delta(p^+-q_1^+-q_2^+-q_3^+)\frac{1}{2p^+}\frac{1}{2q_2^+}
\nonumber\\
\times\; 
\Big\langle {\rm I}_{\bar q\gamma -\bar q\gamma}+{\rm I}_{q\gamma - q\gamma}+{\rm I}_{\bar q\gamma - q\gamma}+{\rm I}_{q\gamma -\bar q\gamma} \Big\rangle_{x_A}
\eeq  
where the subscript of each term $\rm{I}$ stands for the photon radiation from the quark ($q\gamma$) or from the antiquark ($\bar q\gamma$) - in the amplitude and in the complex conjugate amplitude.

Let us calculate each of the terms separately and discuss their properties. We start with the contribution that stems from the emission of the photon from the antiquark both in the amplitude and in the complex conjugate amplitude, ${\rm I}_{\bar q\gamma -\bar q\gamma}$ (from now on, $\xi_i=q^+_i/p^+$): 
\beq
\label{barq_gamma_barq_gamma_w_Tr}
&&
{\rm I}_{\bar q\gamma -\bar q\gamma}=\int_{ww'vv'y_1z_1y_2z_2y_3z_3}e^{iq_1\cdot(y_1-z_1)+iq_2\cdot(y_2-z_2)+iq_3\cdot(y_3-z_3)}
\delta^{(2)}[w'-(\xi_1y_1+\bar\xi_1v')]
\nonumber\\
&&
\times\; 
\delta^{(2)}[w-(\xi_1z_1+\bar\xi_1v)]
\delta^{(2)}\bigg[v'-\bigg\{\bigg(1-\frac{\xi_2}{\bar\xi_1}\bigg)y_3+\frac{\xi_2}{\bar\xi_1}y_2\bigg\}\bigg]
\delta^{(2)}\bigg[v-\bigg\{\bigg(1-\frac{\xi_2}{\bar\xi_1}\bigg)z_3+\frac{\xi_2}{\bar\xi_1}z_2\bigg\}\bigg]
\nonumber\\
&&
\times\;
A^{\bar\lambda'}(y_3-y_2)A^{\bar\lambda}(z_3-z_2)
\big[{\Psi^{\eta\bar\eta'}_{s\bar s'}(\xi_1)}\big]^*
\Psi^{\eta\bar\eta}_{ss'}(\xi_1)
\bigg[\phi^{\lambda\bar\lambda'}_{\bar s's''}\bigg(\frac{\xi_2}{\bar\xi_1}\bigg)\bigg]^*
\phi^{\lambda\bar\lambda}_{s's''}\bigg(\frac{\xi_2}{\bar\xi_1}\bigg)
\nonumber\\
&&
\times\; 
\tr\Big\{ \big[{{\cal G}^c_{\bar q\gamma}}(\xi_1,\xi_2; w',v',y_1,y_2,y_3)]^\dagger_{\bar\eta'} 
\, 
\big[{\cal G}^c_{\bar q\gamma}(\xi_1,\xi_2; w,v,z_1,z_2,z_3)]_{\bar\eta}\Big\}
\eeq
Similarly, the contribution that originates from the emission of the photon from the quark both in the amplitude and in the complex conjugate amplitude reads
\beq
\label{q_gamma_q_gamma_Cont}
&&
{\rm I}_{q\gamma-q\gamma}= 
\int_{ww'vv'y_1z_1y_2z_2y_3z_3}e^{iq_1\cdot(y_1-z_1)+iq_2\cdot(y_2-z_2)+iq_3\cdot(y_3-z_3)}
\nonumber\\
&&
\times\; 
\delta^{(2)}\big[ w'-\{(\xi_1+\xi_2)v'+(1-\xi_1-\xi_2)y_3\}\big]
\delta^{(2)}\big[ w-\{(\xi_1+\xi_2)v+(1-\xi_1-\xi_2)z_3\}\big]
\nonumber\\
&&
\times\; 
\delta^{(2)}\bigg[ v'-\bigg\{\bigg( 1-\frac{\xi_2}{\xi_1+\xi_2}\bigg)y_1+\frac{\xi_2}{\xi_1+\xi_2}y_2\bigg\}\bigg]
\delta^{(2)}\bigg[ v-\bigg\{\bigg( 1-\frac{\xi_2}{\xi_1+\xi_2}\bigg)z_1+\frac{\xi_2}{\xi_1+\xi_2}z_2\bigg\}\bigg]
\nonumber
\\
&&
\times\; 
A^{\bar\lambda'}(y_2-y_1)A^{\bar\lambda}(z_2-z_1)
\big[\Psi^{\eta\bar\eta'}_{\bar s's''}(\xi_1+\xi_2)\big]^*
\Psi^{\eta\bar\eta}_{s's''}(\xi_1+\xi_2)
\bigg[ \phi^{\lambda\bar\lambda'}_{s\bar s' }\bigg(\frac{\xi_2}{\xi_1+\xi_2}\bigg)\bigg]^*
\phi^{\lambda\bar\lambda}_{ss'}\bigg(\frac{\xi_2}{\xi_1+\xi_2}\bigg)
\nonumber\\
&&
\times\; 
\tr\Big\{ \big[{{\cal G}^c_{q\gamma}}(\xi_1,\xi_2; w',v',y_1,y_2,y_3)]^\dagger_{\bar\eta'} 
\, 
\big[{\cal G}^c_{q\gamma}(\xi_1,\xi_2; w,v,z_1,z_2,z_3)]_{\bar\eta}\Big\}
\eeq

Finally, the crossed contributions which originate from the emission of the photon from the quark in the amplitude and from the antiquark in the complex conjugate amplitude (or vice versa) can be written as 
\beq
\label{bar_q_gamma_q_gamma_Cont}
&&
{\rm I}_{\bar q\gamma- q\gamma}=
\int_{ww'vv'y_1z_1y_2z_2y_3z_3}e^{iq_1\cdot(y_1-z_1)+iq_2\cdot(y_2-z_2)+iq_3\cdot(y_3-z_3)}
\nonumber\\
&&
\times\; 
\delta^{(2)}[w'-(\xi_1y_1+\bar\xi_1v')] \delta^{(2)}\big[ w-\{(\xi_1+\xi_2)v+(1-\xi_1-\xi_2)z_3\}\big]
\nonumber\\
&&
\times\; 
\delta^{(2)}\bigg[v'-\bigg\{\bigg(1-\frac{\xi_2}{\bar\xi_1}\bigg)y_3+\frac{\xi_2}{\bar\xi_1}y_2\bigg\}\bigg]
\delta^{(2)}\bigg[ v-\bigg\{\bigg( 1-\frac{\xi_2}{\xi_1+\xi_2}\bigg)z_1+\frac{\xi_2}{\xi_1+\xi_2}z_2\bigg\}\bigg]
\nonumber\\
&&
\times\; 
A^{\bar\lambda'}(y_3-y_2) A^{\bar\lambda}(z_2-z_1) 
\big[{\Psi^{\eta\bar\eta'}_{\bar s's}(\xi_1)}\big]^*
\Psi^{\eta\bar\eta}_{s''s'}(\xi_1+\xi_2)
\bigg[\phi^{\lambda\bar\lambda'}_{s''\bar s'}\bigg(\frac{\xi_2}{\bar\xi_1}\bigg)\bigg]^*
\phi^{\lambda\bar\lambda}_{s's}\bigg(\frac{\xi_2}{\xi_1+\xi_2}\bigg)
\nonumber\\
&&
\times\; 
\tr\Big\{ \big[{{\cal G}^c_{\bar q\gamma}}(\xi_1,\xi_2; w',v',y_1,y_2,y_3)]^\dagger_{\bar\eta'} 
\, 
\big[{\cal G}^c_{q\gamma}(\xi_1,\xi_2; w,v,z_1,z_2,z_3)]_{\bar\eta}\Big\}
\eeq  
and
\beq
\label{q_gamma_bar_q_gamma_Cont}
&&
{\rm I}_{q\gamma- \bar q\gamma}=
\int_{ww'vv'y_1z_1y_2z_2y_3z_3}e^{iq_1\cdot(y_1-z_1)+iq_2\cdot(y_2-z_2)+iq_3\cdot(y_3-z_3)}
\nonumber\\
&&
\times\; 
\delta^{(2)}\big[ w'-\{(\xi_1+\xi_2)v'+(1-\xi_1-\xi_2)y_3\}\big]
\delta^{(2)}[w-(\xi_1z_1+\bar\xi_1v)]
\nonumber\\
&&
\times\; 
\delta^{(2)}\bigg[ v'-\bigg\{\bigg( 1-\frac{\xi_2}{\xi_1+\xi_2}\bigg)y_1+\frac{\xi_2}{\xi_1+\xi_2}y_2\bigg\}\bigg]
\delta^{(2)}\bigg[v-\bigg\{\bigg(1-\frac{\xi_2}{\bar\xi_1}\bigg)z_3+\frac{\xi_2}{\bar\xi_1}z_2\bigg\}\bigg]
\nonumber\\
&&
\times\;
A^{\bar\lambda'}(y_2-y_1) A^{\bar\lambda}(z_3-z_2)
\big[\Psi^{\eta\bar\eta'}_{\bar s's}(\xi_1+\xi_2)\big]^*
\Psi^{\eta\bar\eta}_{s''s'}(\xi_1)
\bigg[ \phi^{\lambda\bar\lambda'}_{s''\bar s'}\bigg(\frac{\xi_2}{\xi_1+\xi_2}\bigg)\bigg]^*
\phi^{\lambda\bar\lambda}_{s's}\bigg(\frac{\xi_2}{\bar\xi_1}\bigg)
\nonumber\\
&&
\times\; 
\tr\Big\{ \big[{{\cal G}^c_{q\gamma}}(\xi_1,\xi_2; w',v',y_1,y_2,y_3)]^\dagger_{\bar\eta'} 
\, 
\big[{\cal G}^c_{\bar q\gamma}(\xi_1,\xi_2; w,v,z_1,z_2,z_3)]_{\bar\eta}\Big\}
\eeq

Each of these four contributions to the production cross section, Eqs. \eqref{barq_gamma_barq_gamma_w_Tr}-\eqref{q_gamma_bar_q_gamma_Cont}, can be simplified by performing a similar color algebra as in the case of quark initiated channel, with the help of identities Eqs. \eqref{Fierz} and \eqref{Adj_to_Fundamental}. Then, using the standard definitions of the dipole and quadrupole amplitudes: Eqs. \eqref{dip_amp} and \eqref{quad_amp}, one can organize the contribution $I_{\bar q\gamma -\bar q\gamma}$ to the cross section as follows:
\beq
\label{barq_gamma_squared_final}
&&
I_{\bar q\gamma -\bar q\gamma}=\int_{ww'vv'y_1z_1y_2z_2y_3z_3}e^{iq_1\cdot(y_1-z_1)+iq_2\cdot(y_2-z_2)+iq_3\cdot(y_3-z_3)}
\nonumber\\
&&
\times\; 
\delta^{(2)}[w'-(\xi_1y_1+\bar\xi_1v')]\delta^{(2)}[w-(\xi_1z_1+\bar\xi_1v)]\nonumber\\
&&
\times
\delta^{(2)}\bigg[v'-\bigg\{\bigg(1-\frac{\xi_2}{\bar\xi_1}\bigg)y_3+\frac{\xi_2}{\bar\xi_1}y_2\bigg\}\bigg]
\delta^{(2)}\bigg[v-\bigg\{\bigg(1-\frac{\xi_2}{\bar\xi_1}\bigg)z_3+\frac{\xi_2}{\bar\xi_1}z_2\bigg\}\bigg]
\nonumber\\
&&
\times\;
A^{\bar\lambda'}(y_3-y_2)A^{\bar\lambda}(z_3-z_2)
\, 8\, {\cal M}^{\bar\lambda\bar\lambda'; \bar\eta\bar\eta'}_{\bar q\bar q}\bigg(\xi_1,\frac{\xi_2}{\bar\xi_1}\bigg)
\bigg[ 
A^{\bar\eta'}(v'-y_1)
A^{\bar\eta}(v-z_1)
{\rm W}_{\bar q\gamma-\bar q \gamma}^{(AA)}
\nonumber\\
&&
+\, 
{\cal A}^{\bar\eta'}\bigg( \xi_1,v'-y_1; \frac{\xi_2}{\bar\xi_1}, y_3-y_2\bigg)
{\cal A}^{\bar\eta}\bigg( \xi_1,v-z_1; \frac{\xi_2}{\bar\xi_1}, z_3-z_2\bigg)
{\rm W}_{\bar q\gamma-\bar q \gamma}^{(\A\A)}
\nonumber\\
&&
-\, 
{\cal A}^{\bar\eta'}\bigg( \xi_1,v'-y_1; \frac{\xi_2}{\bar\xi_1}, y_3-y_2\bigg)
A^{\bar\eta}(v-z_1)
{\rm W}_{\bar q\gamma-\bar q \gamma}^{(\A A)}
\nonumber\\
&&
-\, 
A^{\bar\eta'}(v'-y_1)
{\cal A}^{\bar\eta}\bigg( \xi_1,v-z_1; \frac{\xi_2}{\bar\xi_1}, z_3-z_2\bigg)
{\rm W}_{\bar q\gamma-\bar q \gamma}^{(A \A)} \bigg]
\eeq
where the functions ${\rm W}_{\bar q\gamma - \bar q\gamma}$ encode the dipole and quadrupole structures that accompany the pairs of the standard and modified Weizs\"acker-Williams fields, indicated in the superscripts. As is obvious from the definition of the function ${\cal G}_{\bar q\gamma}$ in Eq. \eqref{antiquark_photon}, it is enough to compute the explicit expression of one of these functions, from which the remaining ones can be obtained by the exchange of coordinates. We thus present the explicit expression of the function ${\rm W}_{\bar q\gamma -\bar q\gamma}^{(AA)}$, which reads
\beq
{\rm W}_{\bar q\gamma -\bar q\gamma}^{(AA)}&=&\frac{N_c^2}{2}\bigg[
s(z_1,y_1)s(v',v)+s(w,w')s(w',w)-s(w,y_1)s(v',w)-s(z_1,w')s(w,v)\bigg]
\nonumber\\
&&
-\frac{1}{2}\bigg[ 1+Q(z_1,y_1,v',v)-s(v',y_1)-s(z_1,v)\bigg]\;,
\eeq
and the remaining ones can be written as
\beq
{\rm W}^{(\A\A)}_{\bar q\gamma -\bar q\gamma}&=&{\rm W}^{(AA)}_{\bar q\gamma -\bar q\gamma}(v'\to y_3, v\to z_3)\;,
\\
{\rm W}^{(\A A)}_{\bar q\gamma -\bar q\gamma}&=&{\rm W}^{(AA)}_{\bar q\gamma -\bar q\gamma}(v'\to y_3)\;,
\\
{\rm W}^{(A\A)}_{\bar q\gamma -\bar q\gamma}&=&{\rm W}^{(AA)}_{\bar q\gamma -\bar q\gamma}(v\to z_3)\;.
\eeq
Moreover, in Eq. \eqref{barq_gamma_squared_final}, we have performed the summation over the spin and polarization indices of the product of splitting amplitudes, which can be computed in a straightforward manner (see Appendix \ref{App:splittings} for the details of the calculation):
\beq
\sum_{s'\bar s'}\sum_{\lambda\eta} \bigg\{\Psi^{\eta\bar\eta}_{ss'}(\xi_1) \Psi^{*\eta\bar\eta'}_{s\bar s'}(\xi_1)\bigg\}
\bigg\{ \phi^{\lambda\bar\lambda}_{s's''}\bigg(\frac{\xi_2}{\bar\xi_1}\bigg)
\phi^{*\lambda\bar\lambda'}_{\bar s' s''}\bigg(\frac{\xi_2}{\bar\xi_1}\bigg)\bigg\}
=8\, {\cal M}^{\bar\lambda\bar\lambda'; \bar\eta\bar\eta'}_{\bar q\bar q}\bigg(\xi_1,\frac{\xi_2}{\bar\xi_1}\bigg)\;,
\eeq
where 
\beq
\label{product_splitt_squared}
{\cal M}^{\bar\lambda\bar\lambda'; \bar\eta\bar\eta'}_{\bar q\bar q}\bigg(\xi_1,\frac{\xi_2}{\bar\xi_1}\bigg)=&&
\Big[\xi_1^2+(1-\xi_1)^2\Big] 
\bigg[1+\bigg(1-\frac{\xi_2}{\bar\xi_1}\bigg)^2\bigg]
\delta^{\bar\lambda\bar\lambda'}
\delta^{\bar\eta\bar\eta'}
\nonumber\\
&&
+\; 
(1-2\xi_1)\frac{\xi_2}{\bar\xi_1}\bigg(2-\frac{\xi_2}{\bar\xi_1}\bigg)\epsilon^{\bar\lambda\bar\lambda'}\epsilon^{\bar\eta\bar\eta'}\;.
\eeq

The second contribution to the cross section originates from the emission of the photon from the quark both in the amplitude and in the complex conjugate amplitude, and can be computed in the same way:
\beq
\label{q_gamma_squared_final}
&&
I_{q\gamma-q\gamma}= 
\int_{ww'vv'y_1z_1y_2z_2y_3z_3}e^{iq_1\cdot(y_1-z_1)+iq_2\cdot(y_2-z_2)+iq_3\cdot(y_3-z_3)}
\nonumber\\
&&
\times\; 
\delta^{(2)}\big[ w'-\{(\xi_1+\xi_2)v'+(1-\xi_1-\xi_2)y_3\}\big]
\delta^{(2)}\big[ w-\{(\xi_1+\xi_2)v+(1-\xi_1-\xi_2)z_3\}\big]
\nonumber\\
&&
\times\; 
\delta^{(2)}\bigg[ v'-\bigg\{\bigg( 1-\frac{\xi_2}{\xi_1+\xi_2}\bigg)y_1+\frac{\xi_2}{\xi_1+\xi_2}y_2\bigg\}\bigg]
\delta^{(2)}\bigg[ v-\bigg\{\bigg( 1-\frac{\xi_2}{\xi_1+\xi_2}\bigg)z_1+\frac{\xi_2}{\xi_1+\xi_2}z_2\bigg\}\bigg]
\nonumber
\\
&&
\times\; 
A^{\bar\lambda'}(y_2-y_1)A^{\bar\lambda}(z_2-z_1) \, 
8\, {\cal M}_{qq}^{\bar\lambda\bar\lambda'; \bar\eta\bar\eta'}\bigg(\xi_1+\xi_2, \frac{\xi_2}{\xi_1+\xi_2}\bigg)
\bigg[
 A^{\bar\eta'}(v'-y_3)
 A^{\bar\eta}(v-z_3)\, 
 {\rm W}^{(AA)}_{q\gamma- q\gamma}
\nonumber\\
&&
+\; 
{\cal A}^{\bar\eta'}\bigg( 1-\xi_1-\xi_2, v'-y_3; \frac{\xi_2}{\xi_1+\xi_2}, y_2-y_1\bigg)
{\cal A}^{\bar\eta'}\bigg( 1-\xi_1-\xi_2, v-z_3; \frac{\xi_2}{\xi_1+\xi_2}, z_2-z_1\bigg)
 {\rm W}^{(\A\A)}_{q\gamma- q\gamma}
\nonumber\\
&&
-\, 
{\cal A}^{\bar\eta'}\bigg( 1-\xi_1-\xi_2, v'-y_3; \frac{\xi_2}{\xi_1+\xi_2}, y_2-y_1\bigg)
A^{\bar\eta}(v-z_3)\, 
{\rm W}^{(\A A)}_{q\gamma- q\gamma}
\nonumber\\
&&
-\, 
A^{\bar\eta'}(v'-y_3)
{\cal A}^{\bar\eta'}\bigg( 1-\xi_1-\xi_2, v-z_3; \frac{\xi_2}{\xi_1+\xi_2}, z_2-z_1\bigg)
{\rm W}^{(A \A)}_{q\gamma- q\gamma}\bigg]\;.
\eeq
The functions ${\rm W}_{q\gamma-q\gamma}$, which encode the dipole and quadrupole structure of the contribution that originates from the emission of the photon from the quark both in the amplitude and in the complex conjugate amplitude, have very similar structure as ${\rm W}_{\bar q\gamma-\bar q\gamma}$, as can be easily seen from the comparison of the functions ${\cal G}_{q\gamma}$ and ${\cal G}_{\bar q\gamma}$. Therefore, in order to compute these functions, it is again enough to perform the color algebra in one of the terms, while the rest can be read off from the explicitly calculated one. Keeping this in mind, we have
\beq
{\rm W}_{q\gamma-q\gamma}^{(AA)}&=& \frac{N_c^2}{2}\Big[ s(v,v')s(y_3,z_3)+s(w,w')s(w',w)-s(w,v')s(y_3,w')-s(v,w')s(w,z_3)\Big]\nonumber
\\
&&
-\, 
\frac{1}{2}\Big[1+Q(v,v',y_3,z_3)-s(y_3,v')-s(v,z_3)\Big]\;,
\eeq
from which the remaining ones can be obtained as follows
\beq
{\rm W}_{q\gamma-q\gamma}^{(\A\A)}&=&{\rm W}_{q\gamma-q\gamma}^{(AA)}(v'\to y_1, v\to z_1)\;,\\
{\rm W}_{q\gamma-q\gamma}^{(\A A)}&=&{\rm W}_{q\gamma-q\gamma}^{(AA)}(v'\to y_1)\;,\\
{\rm W}_{q\gamma-q\gamma}^{(A\A)}&=&{\rm W}_{q\gamma-q\gamma}^{(AA)}(v\to z_1)\;.
\eeq
Moreover, in Eq. \eqref{q_gamma_squared_final}, we have performed the summation over the spin and polarization indices (see Appendix \ref{App:splittings} for the details of the calculation):
\beq
&&
\hspace{-2cm}
\sum_{s'\bar s'}\sum_{\lambda\eta} 
\bigg\{
\Psi^{\eta\bar\eta}_{s's''}(\xi_1+\xi_2) \Psi^{*\eta\bar\eta'}_{\bar s's''}(\xi_1+\xi_2)
\bigg\}
\nonumber\\
&&
\times\, 
\bigg\{
\phi^{\lambda\bar\lambda}_{ss'}\bigg(\frac{\xi_2}{\xi_1+\xi_2}\bigg)
 \phi^{*\lambda\bar\lambda'}_{s\bar s' }\bigg(\frac{\xi_2}{\xi_1+\xi_2}\bigg)
\bigg\}
=8\, {\cal M}_{qq}^{\bar\lambda\bar\lambda'; \bar\eta\bar\eta'}\bigg(\xi_1+\xi_2, \frac{\xi_2}{\xi_1+\xi_2}\bigg)\;,
\eeq
where 
\beq
\label{M_qq}
{\cal M}_{qq}^{\bar\lambda\bar\lambda'; \bar\eta\bar\eta'}\bigg(\xi_1+\xi_2, \frac{\xi_2}{\xi_1+\xi_2}\bigg)&=&
\Big[ (\xi_1+\xi_2)^2+(1-\xi_1-\xi_2)^2\Big]\bigg[ 1+\bigg( 1-\frac{\xi_2}{\xi_2+\xi_1}\bigg)^2\bigg] \delta^{\bar\lambda\bar\lambda'}\delta^{\bar\eta\bar\eta'}
\nonumber\\
&&
-\Big[ 1-2(\xi_1+\xi_2)\Big]\frac{\xi_2}{\xi_1+\xi_2}\bigg( 2-\frac{\xi_2}{\xi_1+\xi_2}\bigg)\epsilon^{\bar\lambda\bar\lambda'}\epsilon^{\bar\eta\bar\eta'}
\eeq

Finally, the crossed contributions, i.e. photon emission from the antiquark in the amplitude and from the quark in the complex conjugate amplitude (or vice versa), can be computed in a very similar way. The first contribution reads
\beq
\label{crossed_1_fin}
&&
{\rm I}_{\bar q\gamma- q\gamma}=
\int_{ww'vv'y_1z_1y_2z_2y_3z_3}e^{iq_1\cdot(y_1-z_1)+iq_2\cdot(y_2-z_2)+iq_3\cdot(y_3-z_3)}
\nonumber\\
&&
\times\; 
\delta^{(2)}[w'-(\xi_1y_1+\bar\xi_1v')] \delta^{(2)}\big[ w-\{(\xi_1+\xi_2)v+(1-\xi_1-\xi_2)z_3\}\big]
\nonumber\\
&&
\times\; 
\delta^{(2)}\bigg[v'-\bigg\{\bigg(1-\frac{\xi_2}{\bar\xi_1}\bigg)y_3+\frac{\xi_2}{\bar\xi_1}y_2\bigg\}\bigg]
\delta^{(2)}\bigg[ v-\bigg\{\bigg( 1-\frac{\xi_2}{\xi_1+\xi_2}\bigg)z_1+\frac{\xi_2}{\xi_1+\xi_2}z_2\bigg\}\bigg]
\nonumber\\
&&
\times\; 
A^{\bar\lambda'}(y_3-y_2) A^{\bar\lambda}(z_2-z_1) \, (-8)\, \widetilde{\cal M}^{\bar\lambda\bar\lambda'; \bar\eta\bar\eta'}_{g}(\xi_1,\xi_2)\bigg[
A^{\bar\eta'}(v'-y_1)
A^{\bar\eta}(v-z_3)
{\rm W}^{(AA)}_{\bar q\gamma-q\gamma}
\nonumber\\
&&
+\,
{\cal A}^{\bar\eta'}\bigg( \xi_1, v'-y_1; \frac{\xi_2}{\bar\xi_1}, y_3-y_2\bigg)
{\cal A}^{\bar\eta}\bigg( 1-\xi_1-\xi_2, v-z_3; \frac{\xi_2}{\xi_1+\xi_2}, z_2-z_1\bigg)
{\rm W}^{(\A\A)}_{\bar q\gamma-q\gamma}
\nonumber\\
&&
-\, 
{\cal A}^{\bar\eta'}\bigg( \xi_1, v'-y_1; \frac{\xi_2}{\bar\xi_1}, y_3-y_2\bigg)
A^{\bar\eta}(v-z_3)
{\rm W}^{(\A A)}_{\bar q\gamma-q\gamma}
\nonumber\\
&&
-\, 
A^{\bar\eta'}(v'-y_1)
{\cal A}^{\bar\eta}\bigg( 1-\xi_1-\xi_2, v-z_3; \frac{\xi_2}{\xi_1+\xi_2}, z_2-z_1\bigg)
{\rm W}^{(A \A)}_{\bar q\gamma-q\gamma}\bigg]
\eeq  
Obviously, the functions ${\rm W}_{\bar q\gamma-q\gamma}$ have similar properties as those of the squared contributions, since the trace over the fundamental indices for this contribution mixes the functions ${\cal G}_{q\gamma}$ and ${\cal G}_{\bar q\gamma}$. 
We again write explicitly the function ${\rm W}^{(A A)}_{\bar q\gamma-q\gamma}$, and obtain the others by exchanging the coordinates: 
\beq
{\rm W}^{(A A)}_{\bar q\gamma-q\gamma}&=&\frac{N_c^2}{2}\Big[ s(v,y_1)s(v',z_3)+s(w,w')s(w',w)-s(w,y_1)s(v',w')-s(v,w')s(w,z_3)\Big]\
\nonumber\\
&&
-\, 
\frac{1}{2}\Big[ 1+ Q(v,y_1,v',z_3)-s(v',y_1)-s(v,z_3)\Big]\;,
\eeq
and:
\beq
{\rm W}^{(\A\A)}_{\bar q\gamma-q\gamma}&=&{\rm W}^{(A A)}_{\bar q\gamma-q\gamma}(v'\to y_3, v\to z_1)\;,\\
{\rm W}^{(\A A)}_{\bar q\gamma-q\gamma}&=&{\rm W}^{(A A)}_{\bar q\gamma-q\gamma}(v'\to y_3)\;,\\
{\rm W}^{(A \A)}_{\bar q\gamma-q\gamma}&=&{\rm W}^{(A A)}_{\bar q\gamma-q\gamma}(v\to z_1)\;.
\eeq
The product of the splitting amplitudes for the crossed contributions is different from the one corresponding to the squared contributions,  due to the fact that here the longitudinal momentum ratios are mixed. However, the calculation can be performed following the same lines, and yields (see Appendix \ref{App:splittings} for the details of the calculation):
\beq
\sum_{s'\bar s'}\sum_{\lambda\eta}
\bigg\{
\Psi^{\eta\bar\eta}_{s''s'}(\xi_1+\xi_2)
{\Psi^{*\eta\bar\eta'}_{\bar s's}(\xi_1)}
\bigg\}
\bigg\{
\phi^{\lambda\bar\lambda}_{s's}\bigg(\frac{\xi_2}{\xi_1+\xi_2}\bigg)
\phi^{*\lambda\bar\lambda'}_{s''\bar s'}\bigg(\frac{\xi_2}{\bar\xi_1}\bigg)
\bigg\}= -8\, \widetilde{\cal M}^{\bar\lambda\bar\lambda'; \bar\eta\bar\eta'}_{g}(\xi_1,\xi_2)
\nonumber\\
\eeq

with 
\beq
\label{splitting_mixed}
\widetilde{\cal M}^{\bar\lambda\bar\lambda'; \bar\eta\bar\eta'}_{g}(\xi_1,\xi_2)&=&
\Big[ \bar\xi_1+(2\xi_1-1)(\xi_1+\xi_2)\Big]
\bigg(2-\frac{\xi_2}{\bar\xi_1}-\frac{\xi_2}{\xi_1+\xi_2}\bigg)\delta^{\bar\lambda\bar\lambda'}\delta^{\bar\eta\bar\eta'}
\nonumber\\
&-&
\frac{\xi_2\big[\bar\xi_1-(\xi_1+\xi_2)\big]^2}{\bar\xi_1(\xi_1+\xi_2)}\; 
\epsilon^{\bar\lambda\bar\lambda'}\epsilon^{\bar\eta\bar\eta'}.
\eeq

Finally, the last contribution to the production cross section in the gluon initiated channel is the remaining crossed term, which can be written as 
\beq
\label{crossed_2_fin}
&&
{\rm I}_{q\gamma- \bar q\gamma}=
\int_{ww'vv'y_1z_1y_2z_2y_3z_3}e^{iq_1\cdot(y_1-z_1)+iq_2\cdot(y_2-z_2)+iq_3\cdot(y_3-z_3)}
\nonumber\\
&&
\times\; 
\delta^{(2)}\big[ w'-\{(\xi_1+\xi_2)v'+(1-\xi_1-\xi_2)y_3\}\big]
\delta^{(2)}[w-(\xi_1z_1+\bar\xi_1v)]
\nonumber\\
&&
\times\; 
\delta^{(2)}\bigg[ v'-\bigg\{\bigg( 1-\frac{\xi_2}{\xi_1+\xi_2}\bigg)y_1+\frac{\xi_2}{\xi_1+\xi_2}y_2\bigg\}\bigg]
\delta^{(2)}\bigg[v-\bigg\{\bigg(1-\frac{\xi_2}{\bar\xi_1}\bigg)z_3+\frac{\xi_2}{\bar\xi_1}z_2\bigg\}\bigg]
\nonumber\\
&&
\times\;
A^{\bar\lambda'}(y_2-y_1) A^{\bar\lambda}(z_3-z_2)
\, (-8)\, \widetilde{\cal M}^{\bar\lambda\bar\lambda'; \bar\eta\bar\eta'}_{g}(\xi_1,\xi_2)
\bigg[ 
A^{\bar\eta'}(v'-y_3)
A^{\bar\eta}(v-z_3) 
{\rm W}_{q\gamma-\bar q\gamma}^{(AA)} 
\nonumber\\
&&
 +\; 
{\cal A}^{\bar\eta'}\bigg(1-\xi_1-\xi_2),v'-y_3; \frac{\xi_2}{\xi_1+\xi_2}, y_2-y_1\bigg)
{\cal A}^{\bar\eta}\bigg( \xi_1,v-z_1; \frac{\xi_2}{\bar\xi_1}, z_3-z_2\bigg) 
{\rm W}_{q\gamma-\bar q\gamma}^{(\A\A)} 
\nonumber\\
&&
-\; 
{\cal A}^{\bar\eta'}\bigg(1-\xi_1-\xi_2),v'-y_3; \frac{\xi_2}{\xi_1+\xi_2}, y_2-y_1\bigg)
A^{\bar\eta}(v-z_3) 
{\rm W}_{q\gamma-\bar q\gamma}^{(\A A)}
\nonumber\\
&&
-\; 
A^{\bar\eta'}(v'-y_3)
{\cal A}^{\bar\eta}\bigg( \xi_1,v-z_1; \frac{\xi_2}{\bar\xi_1}, z_3-z_2\bigg) 
{\rm W}_{q\gamma-\bar q\gamma}^{(A \A)}\bigg]
\eeq
where the functions that encode the dipole and quadrupole structures are given by: 
\beq
{\rm W}_{q\gamma-\bar q\gamma}^{(A A)}&=&\frac{N_c^2}{2}\Big[ s(z_1,v')s(y_3,v)+s(w,w')s(w',w)-s(w,v')s(y_3,w')-s(z_1,w')s(w,v)\Big]
\nonumber\\
&&
-\, \frac{1}{2}\Big[ 1+Q(z_1,v',y_3,v)-s(y_3,v')-s(z_1,v)\Big]\;,
\eeq
with 
\beq
{\rm W}_{q\gamma-\bar q\gamma}^{(\A \A)}&=&{\rm W}_{q\gamma-\bar q\gamma}^{(A A)}(v'\to y_1; v\to z_3)\;,\\
{\rm W}_{q\gamma-\bar q\gamma}^{(\A A)}&=&{\rm W}_{q\gamma-\bar q\gamma}^{(A A)}(v'\to y_1)\;,\\
{\rm W}_{q\gamma-\bar q\gamma}^{(A \A)}&=&{\rm W}_{q\gamma-\bar q\gamma}^{(A A)}(v\to z_3)\;.
\eeq
The product of the splitting amplitudes is the same as the previous crossed contribution ($\bar q\gamma-q\gamma$ contribution), and is given in Eq. \eqref{splitting_mixed}. 

This concludes the calculation of the production cross section in the gluon initiated channel. At the partonic level, the final result is given by Eq. \eqref{gluon_channel_partonic}, where the final results of each contribution ${\rm I}_{\bar q\gamma-\bar q\gamma}$, ${\rm I}_{q\gamma-q\gamma}$, ${\rm I}_{\bar q\gamma- q\gamma}$ and ${\rm I}_{q\gamma-\bar q\gamma}$ are presented in Eqs. \eqref{barq_gamma_squared_final}, \eqref{q_gamma_squared_final}, \eqref{crossed_1_fin} and \eqref{crossed_2_fin} respectively. 

\section{Correlation limit and gluon TMDs}

In the large$-N_c$ limit \cite{firstlowxTMDs}, and later keeping $N_c$ finite \cite{Marquet:2016cgx}, it was shown that in the so-called correlation limit $|q_1+q_2|\ll|q_1|,|q_2|$ (corresponding to nearly back-to-back jets), the dilute-dense CGC expression for forward dijets production could be written as:
\begin{eqnarray}
\frac{d\sigma^{qA\rightarrow {\rm dijets}+X}}{d^3\uq_1d^3\uq_2}&=&
2\pi\delta(p^+-q_1^+-q_2^+) \sum_i H_{qg\to qg}^{(i)}\ \mathcal{F}_{qg}^{(i)}(x_A,q_1+q_2)\label{eq:tmd-dijet1}\\
\frac{d\sigma^{gA\rightarrow {\rm dijets}+X}}{d^3\uq_1d^3\uq_2}&=&
2\pi\delta(p^+-q_1^+-q_2^+)\sum_i \Big(H_{gg\to q\bar{q}}^{(i)}+\frac12 H_{gg\to gg}^{(i)}\Big)\mathcal{F}_{gg}^{(i)}(x_A,q_1+q_2)
\label{eq:tmd-dijet2}
\end{eqnarray}
coinciding with the small-x limit of the TMD factorization formula for forward dijets. In Eqs. \eqref{eq:tmd-dijet1}-\eqref{eq:tmd-dijet2}, $\mathcal{F}_{ag}^{(i)}$ denotes several distinct TMD gluon distributions, with different operator definitions. They are accompanied by on-shell hard factors (i.e. evaluated with $q_1=-q_2$) denoted $H_{ag\to cd}^{(i)}$. To extract those formulae from the CGC expressions, one performs a small dipole-size expansion of the Wilson line content of the cross section (which involves functions similar to our W's), except for those dipole sizes that are Fourier conjugate to the small transverse momentum $q_1+q_2$, and those very Fourier transformations of the resulting CGC correlators are then identified with the gluon TMDs. 

It is natural to ask the question whether this equivalence between the CGC and TMD frameworks, in the overlapping domain
of availability, can be extended to three-parton final states. The goal of the present work is precisely to answer that question, starting with -- from the color structure point-of-view -- a simple process, before tackling the case of three jets (the quark initiated cross section was recently obtained in \cite{Iancu:2018aa}) in a future publication. 
In the following, we show that one can indeed extract a TMD factorization formula from our CGC expressions for dijet+photon production obtained in the previous section. That is, we check that one can define a TMD regime for the 3-particle final-state, in which Eq. \eqref{Q_initiated_full} in the quark initiated channel, and Eq. \eqref{gluon_channel_partonic} in the gluon initiated channel, factorize in a similar way as Eqs. \eqref{eq:tmd-dijet1} and \eqref{eq:tmd-dijet2}. Not surprisingly, this regime is characterized by $|{\rm P}|\ll|q_1|,|q_2|,|q_3|$, with the small transverse momentum defined as:
\begin{equation}
\label{P_def}
{\rm P}=q_1+q_2+q_3\;.
\end{equation}

\subsection{Correlation limit: quark channel}

In order to study the correlation limit in the quark initiated channel, we organize the terms in a similar way as in the previous section, i.e. we study the correlation limit of the {\it bef-bef}, {\it aft-aft} contributions and the crossed ones {\it bef-aft} and {\it aft-bef} separately.

\subsubsection{bef-bef contribution}
In the quark initiated channel, we begin the analysis by considering the {\it bef-bef} contribution, and start from Eq. \eqref{bef_bef}. In the correlation limit, we should expand our result in powers of the small dipole sizes, as is discussed at beginning of this section. The dipole sizes in this contribution are 
\beq
&&
r_g=z_3-z_2 \;, \hspace{1cm} r'_g=y_3-y_2\;,\\
&&
r_\gamma=v-z_1 \; , \hspace{1.1cm} r'_\gamma=v'-y_1\;.
\eeq
After performing the above change of variables, the {\it bef-bef} contribution reads
\beq
&&
{\rm I_{\rm bef-bef}}= \int_{z_3y_3, r_gr'_g, r_\gamma r'_\gamma} e^{i{\rm P}\cdot(y_3-z_3)-q_1\cdot(r'_\gamma-r_\gamma)-i{\rm Q}\cdot(r'_g-r_g)} \, 8 \, {\cal M}_q^{\bar\lambda\bar\lambda'; \bar\eta\bar\eta'}\bigg(\xi_1, \frac{\xi_2}{\bar\xi_1}\bigg)
\nonumber\\
&&
\times\; 
A^{\bar\eta}(r_g)A^{\bar\eta'}(r'_g) \; 
\tr\bigg[ 
{\cal \bar{G}}^{\dagger d\bar\lambda'}_{\rm bef}\bigg(\xi_1,\frac{\xi_2}{\bar\xi_1}; y_3,r'_g,r'_\gamma\bigg)
{\cal \bar{G}}^{d\bar\lambda}_{\rm bef}\bigg(\xi_1,\frac{\xi_2}{\bar\xi_1}; z_3,r_g,r_\gamma\bigg)
\bigg]\;, \label{eq:I_bef_bef}
\eeq
where we have introduced the auxiliary transverse momentum
\beq
\label{Q_def}
{\rm Q}=q_2+\frac{\xi_2}{\bar\xi_1}q_1
\eeq
and performed the trivial integrations over $w$, $w'$, $v$ and $v'$. The function ${\cal M}_q^{\bar\lambda\bar\lambda'; \bar\eta\bar\eta'}$ accounts for the product of the splitting amplitudes and is defined in Eq.\eqref{F_bef-bef}.

The function ${\cal \bar{G}}^{d\bar\lambda}_{\rm bef}$ is easily obtained from ${\cal G}^{d\bar\lambda}_{\rm bef}$ after the aforementioned change of variable, and proper replacement of $w$ and $v$ (or $w'$ and $v'$). It encodes the Wilson line structure of the amplitude for the case in which the photon is emitted before the gluon. This function should be expanded in powers of the small dipole sizes ($r_g$ or $r_\gamma$) up to the first nontrivial order, which in this case is the first order. The result of the expansion reads:
\beq
\label{expanded_G_bef}
&&
{\cal \bar{G}}^{d\bar\lambda}_{\rm bef}\bigg(\xi_1, \frac{\xi_2}{\bar\xi_1}; z_3,r_g,r_\gamma\bigg)_{\alpha\beta}= \bigg\{ 
\xi_1r_\gamma^i\bigg[ \A^{\bar\lambda}\bigg(\xi_1,r_\gamma; \frac{\xi_2}{\bar\xi_1},r_g\bigg)-A^{\bar\lambda}(r_\gamma)\bigg](\partial^iS_{z_3})t^d
\nonumber\\
&&
-\; 
r_g^i \A^{\bar\lambda}\bigg(\xi_1,r_\gamma; \frac{\xi_2}{\bar\xi_1},r_g\bigg) 
\bigg[ \bigg(1-\frac{\xi_2}{\bar\xi_1}\bigg)\partial^iS_{z_3}t^d+S_{z_3}t^d(\partial^iS^\dagger_{z_3})S_{z_3}\bigg]\bigg\}_{\alpha\beta}\;.
\eeq
where from now on $S_{x}$ stands for the fundamental Wilson line $S_F(x)$. Pluggin the above result into Eq. \eqref{eq:I_bef_bef} and computing the trace over the fundamental color indices, we can write after some color algebra the {\it bef-bef} contribution as: 
\beq
\label{bef-bef-expanded-Before-integration}
&&
{\rm I_{bef-bef}}= \int_{z_3y_3, r_gr'_g, r_\gamma r'_\gamma} e^{i{\rm P}\cdot(y_3-z_3)-iq_1\cdot(r'_\gamma-r_\gamma)-i{\rm Q}\cdot(r'_g-r_g)} \, 8 \, {\cal M}^{\lambda\bar\lambda; \eta\bar\eta}_q\bigg(\xi_1, \frac{\xi_2}{\bar\xi_1}\bigg)
A^{\bar\eta}(r_g)A^{\bar\eta'}(r'_g)
\nonumber\\
&&
\times\; 
\frac{N_c^2}{2}\bigg\{
\bigg\lgroup
\xi_1r'^j_\gamma\bigg[ \A^{\bar\lambda'}\bigg(\xi_1, r'_\gamma; \frac{\xi_2}{\bar\xi_1}, r'_g\bigg)-A^{\bar\lambda'}(r'_\gamma)\bigg]
-\bigg(1-\frac{\xi_2}{\bar\xi_1}\bigg)r'^j_g\A^{\bar\lambda'}\bigg(\xi_1, r'_\gamma; \frac{\xi_2}{\bar\xi_1}, r'_g\bigg)\bigg\rgroup
\nonumber\\
&&
\hspace{1cm}
\times\; 
\bigg\lgroup
\xi_1r^i_\gamma\bigg[ \A^{\bar\lambda}\bigg(\xi_1, r_\gamma; \frac{\xi_2}{\bar\xi_1}, r_g\bigg)-A^{\bar\lambda}(r_\gamma)\bigg]
-\bigg(1-\frac{\xi_2}{\bar\xi_1}\bigg)r^i_g\A^{\bar\lambda}\bigg(\xi_1, r_\gamma; \frac{\xi_2}{\bar\xi_1}, r_g\bigg)\bigg\rgroup
\nonumber\\
&&
\hspace{1cm}
\times\; 
\frac{1}{N_c}\tr\big[(\partial^iS_{z_3})(\partial^jS^\dagger_{y_3})\big]
\nonumber\\
&&
\hspace{1cm}
-\; 
\bigg\lgroup 
\xi_1r'^j_\gamma\bigg[ \A^{\bar\lambda'}\bigg(\xi_1, r'_\gamma; \frac{\xi_2}{\bar\xi_1}, r'_g\bigg)-A^{\bar\lambda'}(r'_\gamma)\bigg] 
+\frac{\xi_2}{\bar\xi_1}r'^j_g  \A^{\bar\lambda'}\bigg(\xi_1, r'_\gamma; \frac{\xi_2}{\bar\xi_1}, r'_g\bigg)\bigg\rgroup
\nonumber\\
&&
\hspace{1cm}
\times\; 
\bigg\lgroup 
\xi_1r^i_\gamma\bigg[ \A^{\bar\lambda}\bigg(\xi_1, r_\gamma; \frac{\xi_2}{\bar\xi_1}, r_g\bigg)-A^{\bar\lambda}(r_\gamma)\bigg] 
+\frac{\xi_2}{\bar\xi_1}r^i_g  \A^{\bar\lambda}\bigg(\xi_1, r_\gamma; \frac{\xi_2}{\bar\xi_1}, r_g\bigg)\bigg\rgroup
\nonumber\\
&&
\hspace{1cm}
\times\; 
\frac{1}{N_c^2}\frac{1}{N_c}\tr\big[(\partial^iS_{z_3})(\partial^jS^\dagger_{y_3})\big]
\nonumber\\
&&
\hspace{1cm}
-\; 
\bigg\lgroup r'^j_g  \A^{\bar\lambda'}\bigg(\xi_1, r'_\gamma; \frac{\xi_2}{\bar\xi_1}, r'_g\bigg) \bigg\rgroup
\bigg\lgroup r^i_g  \A^{\bar\lambda}\bigg(\xi_1, r_\gamma; \frac{\xi_2}{\bar\xi_1}, r_g\bigg) \bigg\rgroup
\nonumber\\
&&
\hspace{1cm}
\times\; 
\frac{1}{N_c}\tr\big[ S^\dagger_{z_3}(\partial^iS_{z_3})S^\dagger_{y_3}(\partial^jS_{y_3})\big]\frac{1}{N_c}\tr\big[S_{z_3}S^\dagger_{y_3}\big]
\bigg\}\;.
\eeq
Note that the $z_3$ and $y_3$ dependences only enter this expression through the Wilson lines, and as a result, one can already recognize the definitions of two different gluon TMDs (in the low-$x$ limit) \cite{firstlowxTMDs, Marquet:2017xwy}, which will emerge after the target averaging of ${\rm I_{bef-bef}}$:
\beq
\label{Fq_1}
\int_{y_3z_3}e^{i{\rm P}\cdot(y_3-z_3)}\left\langle\tr\big[(\partial^iS_{z_3})(\partial^jS^\dagger_{y_3})\big]\right\rangle_{x_A}&=&
{\rm P}^i{\rm P}^j \int_{y_3z_3}e^{i{\rm P}\cdot(y_3-z_3)} N_c \left\langle s(z_3,y_3)\big]\right\rangle_{x_A}
\nonumber\\
&=&g^2_s\,(2\pi)^3\, \frac{{\rm P}^i{\rm P}^j}{4\rm P^2}{\cal F}^{(1)}_{qg}(x_A, {\rm P})\;,
\eeq 
and
\beq
\label{Fq_2}
\int_{z_3 y_3}e^{i{\rm P}\cdot(y_3-z_3)}
\left\langle\tr\big[ S^\dagger_{z_3}(\partial^iS_{z_3})S^\dagger_{y_3}(\partial^jS_{y_3})\big]\tr\big[S_{z_3}S^\dagger_{y_3}\big]\right\rangle_{x_A}&&\nonumber \\
=-g^2_s  (2\pi)^3\, N_c\, \frac{1}{4}
\bigg[
\frac{1}{2}\delta^{ij}{\cal F}^{(2)}_{qg}(x_A, {\rm P})-\frac{1}{2}\bigg(\delta^{ij}-2\frac{{\rm P}^i{\rm P}^j}{\rm P^2}\bigg){\cal H}^{(2)}_{qg}(x_A,{\rm P})\bigg]\;.
\eeq
The gluon TMD defined in Eq. \eqref{Fq_2} consists of two parts, corresponding to unpolarized (${\cal F}^{(2)}_{qg}$) and linearly-polarized (${\cal H}^{(2)}_{qg}$) distributions inside the unpolarized target. For the so-called (fundamental) dipole gluon TMD defined in Eq. \eqref{Fq_1}, the simpler Wilson line structure implies that ${\cal H}^{(1)}_{qg}={\cal F}^{(1)}_{qg}$.

In order to get the final expression for the {\it bef-bef} contribution to the correlation limit of the production cross section in the quark initiated channel, the integrals over $r_\gamma$, $r_g$, $r'_\gamma$ and $r'_g$ should be performed, which can be done using the following generic integrals:
\beq
\label{integral_1}
\int_{r_g r_\gamma}e^{i{\rm K}\cdot r_\gamma +i{\rm Q}\cdot r_g} \,  A^{\bar\eta}(r_g)\,  r_g^i \,  {\cal A}^{\bar\lambda}\bigg(\xi_1,r_\gamma; \frac{\xi_2}{\bar\xi_1},r_g\bigg)=i \frac{{\rm K}^\lambda}{{\rm K}^2}\frac{1}{{\rm Q}^2+c_0{\rm K}^2}\bigg[ \delta^{\bar\eta i}-2\frac{{\rm Q}^{\bar\eta}{\rm Q}^i}{{\rm Q}^2+c_0{\rm K}^2}\bigg]\hspace{1cm}
\eeq
\beq
\label{integral_2}
\int_{r_g r_\gamma}e^{i{\rm K}\cdot r_\gamma +i{\rm Q}\cdot r_g} \,  A^{\bar\eta}(r_g)\,  r_{\gamma}^i \,  \bigg[{\cal A}^{\bar\lambda}\bigg(\xi_1,r_\gamma; \frac{\xi_2}{\bar\xi_1},r_g\bigg) -A^{\bar\lambda}(r_\gamma)\bigg]
&=&-i\frac{{\rm Q}^{\bar\eta}}{{\rm Q}^2}
\frac{1}{{\rm K}^2+c_0^{-1}{\rm Q}^2}
\nonumber\\
&&
\hspace{-1cm}
\times\; 
\bigg[\delta^{i\bar\lambda} -2\frac{{\rm K}^i{\rm K}^{\bar\lambda}}{{\rm K}^2+c_0^{-1}{\rm Q}^2}\bigg]
\eeq
with 
\beq
c_0=\frac{1}{\xi_1}\frac{\xi_2}{\bar\xi_1}
\bigg(1-\frac{\xi_2}{\bar\xi_1}\bigg)\;.
\eeq
The detailed calculation of these integrals is presented in Appendix \ref{App:Integrals}. Putting the pieces together, we can write the {\it bef-bef} contribution as 
\beq
\label{bef-bef_cor_final}
\hspace{-0.5cm}
\Big\langle{\rm I_{bef-bef}}\Big\rangle_{x_A}&=&{\cal M}_q^{\bar \lambda\bar\lambda'; \bar \eta\bar\eta'}\bigg(\xi_1, \frac{\xi_2}{\bar\xi_1}\bigg)\, g^2_s\, (2\pi)^3\, N_c
\bigg\{
{\left[{\rm H}^{(1)}_{qg}\right]}_{\it bef-bef}^{\bar\lambda\bar\lambda';\bar\eta\bar\eta'; ij} \frac{{\rm P}^i{\rm P}^j}{\rm P^2}
{\cal F}^{(1)}_{qg}(x_A, {\rm P})
\nonumber\\
&&
\hspace{-1cm}
+\; 
{\left[{\rm H}^{(2)}_{qg}\right]}_{\it bef-bef}^{\bar\lambda\bar\lambda';\bar\eta\bar\eta'; ij}
\bigg[\frac{1}{2}\delta^{ij}{\cal F}^{(2)}_{qg}(x_A, {\rm P})
-\; 
\frac{1}{2}\bigg(\delta^{ij}-2\frac{{\rm P}^i{\rm P}^j}{\rm P^2}\bigg){\cal H}^{(2)}_{qg}(x_A, {\rm P})\bigg]\bigg\}\;,
\eeq
where the hard parts that accompany the TMDs are defined as 
\beq
\label{hard_part_1_bef_bef}
{\left[{\rm H}^{(1)}_{qg}\right]}_{\it bef-bef}^{\bar\lambda\bar\lambda';\bar\eta\bar\eta'; ij}&=&
\bigg\lgroup \xi_1\Pi^{\bar\eta' ; \bar\lambda' j}[{\rm Q}; c_0^{-1}, q_1]+\bigg(1-\frac{\xi_2}{\bar\xi_1}\bigg)\Pi^{\bar\lambda'; \bar\eta' j}[ q_1; c_0, {\rm Q}]\bigg\rgroup
\nonumber\\
&&
\hspace{-0.4cm}
\times
\; 
\bigg\lgroup \xi_1\Pi^{\bar\eta; \bar\lambda i}[ {\rm Q}; c_0^{-1}, q_1]+\bigg(1-\frac{\xi_2}{\bar\xi_1}\bigg)\Pi^{\bar\lambda ;\bar\eta i}[ q_1; c_0, {\rm Q}]\bigg\rgroup
\nonumber\\
&&
\hspace{-1cm}
-\, 
\frac{1}{N_c^2}
\bigg\lgroup \xi_1\Pi^{\bar\eta' ; \bar\lambda' j}[{\rm Q}; c_0^{-1}, q_1]- \frac{\xi_2}{\bar\xi_1} \Pi^{\bar\lambda' ;  \bar\eta' j}[ q_1; c_0, {\rm Q}]\bigg\rgroup
\nonumber\\
&&
\hspace{-0.4cm}
\times\, 
\bigg\lgroup \xi_1 \Pi^{\bar\eta ; \bar\lambda i}[{\rm Q}; c_0^{-1}, q_1]-\frac{\xi_2}{\bar\xi_1}\Pi^{\bar\lambda ; \bar\eta i}[ q_1; c_0, {\rm Q}]\bigg\rgroup\;,
\eeq
and 
\beq
\label{hard_part_2_bef_bef}
{\left[{\rm H}^{(2)}_{qg}\right]}_{\it bef-bef}^{\bar\lambda\bar\lambda';\bar\eta\bar\eta'; ij}=\Big\lgroup\Pi^{\bar\lambda' ;  \bar\eta' j}[ q_1; c_0, {\rm Q}] \Big\rgroup\Big\lgroup\Pi^{\bar\lambda ; \bar\eta i}[q_1; c_0, {\rm Q}]\Big\rgroup\;,
\eeq
where we have introduced the compact notation $\Pi^{i; jk}[p; c_0, q]$ given by: 
\beq
\label{def_pi}
\Pi^{i; jk}[p; c_0, q]\equiv \bigg(\frac{p^i}{p^2}\bigg)\bigg\{\frac{1}{q^2+c_0p^2}\bigg[\delta^{jk}-2\frac{q^jq^k}{q^2+c_0p^2}\bigg]\bigg\}\;.
\eeq
Eq.~\eqref{bef-bef_cor_final} is the final result for the {\it bef-bef} contribution to the production cross section in the quark initiated channel given in terms of gluon TMDs.

\subsubsection{aft-aft contribution}
The same procedure can be performed to take the correlation limit of the {\it aft-aft} contribution, i.e., we first identify the dipole sizes in which we will expand to linear order, after which we integrate them out and identify the gluon TMDs that appear. The small dipole sizes for this contribution are:
\beq
r_g&=&v-z_2 \; , \hspace{1.1cm} r'_g=v'-y_2\;, \\
r_\gamma&=& z_3-z_1\; , \hspace{1cm} r'_\gamma=y_3-y_1\;.
\eeq
After performing this change of variables, the {\it aft-aft} contribution Eq. \eqref{aft-aft} can be written as 
\beq
&&
{\rm I}_{\rm aft-aft}=\int_{z_3y_3, r_gr'_g, r_\gamma r'_\gamma}e^{i{\rm P}\cdot(y_3-z_3)-i{\rm K}\cdot(r'_\gamma-r_\gamma)-iq_2\cdot(r'_g-r_g)} \, 8\, {\cal M}^{\bar\lambda\bar\lambda'; \bar\eta\bar\eta'}_q \bigg(\xi_2, \frac{\xi_1}{\bar \xi_2}\bigg)
\nonumber\\
&&
\times\; A^{\bar\lambda}(r_\gamma)A^{\bar\lambda'}(r'_\gamma) \; 
\tr\bigg[ {\cal \bar{G}}^{\dagger d\bar\eta'}_{\rm aft}\bigg(\xi_2, \frac{\xi_1}{\bar\xi_2}; y_3,r'_g, r'_\gamma\bigg){\cal \bar{G}}^{d\bar\eta}_{\rm aft}\bigg( \xi_2, \frac{\xi_1}{\bar\xi_2}; z_3, r_g, r_\gamma\bigg)\bigg]
\eeq
Here, we have performed the integrations over $w$, $w'$, $v$ and $v'$. As before, the function ${\cal M}_q^{\bar\lambda\bar\lambda';\bar\eta\bar\eta'}$ is the product of the splitting amplitudes which is defined in Eq.~\eqref{F_bef-bef}, the function ${\cal \bar{G}}^{d\bar\eta}_{\rm aft}$ is obtained from ${\cal G}^{d\bar\eta}_{\rm aft}$ (Eq.~\eqref{G_aft}) after the aforementioned change of variable and replacement of $w$ and $v$, and we have introduced the new auxiliary momentum $\rm K$, defined as:
\beq
\label{K_def}
{\rm K}&=&q_1+\frac{\xi_1}{\bar\xi_2}q_2\;.
\eeq 

The next step is to Taylor expand the function ${\cal \bar{G}}^{d\bar\eta}_{\rm aft}$ that encodes the Wilson line structure of the amplitude in the case of the photon emission after the gluon emission. To first nontrivial order in the dipole sizes, this function reads
\beq
\label{expanded_G_aft}
&&
{\cal \bar{G}}^{d\bar\eta}_{\rm aft}\bigg( \xi_2, \frac{\xi_1}{\bar\xi_2}; z_3, r_g, r_\gamma\bigg)_{\alpha\beta}=-\bigg[ \bar \xi_2 r_g^i(\partial^iS_{z_3})t^d+\bigg(\frac{\xi_1}{\bar\xi_2}r_\gamma^i+r_g^i\bigg)S_{z_3}t^d(\partial^iS^\dagger_{z_3})S_{z_3}\bigg]_{\alpha\beta}
\nonumber\\
&&
\times\; 
\bigg[ {\cal A}^{\bar\eta}\bigg(\xi_2, r_g; \frac{\xi_1}{\bar\xi_2}, r_\gamma\bigg)-A^{\bar\eta}(r_g)\bigg]
+\frac{\xi_1}{\bar\xi_2}r_\gamma^i\Big[S_{z_3}t^dS^\dagger_{z_3}(\partial^i S_{z_3})\Big]_{\alpha\beta}A^{\bar\eta}(r_g)\;.
\eeq
Using the above expansion, one can compute the trace over the fundamental color indices. After some color algebra and using the Fierz identity given in Eq.~\eqref{Fierz}, the {\it aft-aft} contribution can be organized in the most convenient way as 
\beq
&&
{\rm I}_{\rm aft-aft}=\int_{z_3y_3, r_gr'_g, r_\gamma r'_\gamma}e^{i{\rm P}\cdot(y_3-z_3)-i{\rm K}\cdot(r'_\gamma-r_\gamma)-iq_2\cdot(r'_g-r_g)} \, 8\, {\cal M}^{\bar\lambda\bar\lambda'; \bar\eta\bar\eta'}_q \bigg(\xi_2, \frac{\xi_1}{\bar \xi_2}\bigg)
 A^{\bar\lambda}(r_\gamma)A^{\bar\lambda'}(r'_\gamma)
\nonumber\\
&&
\times\; \; 
\frac{N_c^2}{2}\bigg\{
\bigg\lgroup \bar\xi_2 r'^j_g \bigg[ {\cal A}^{\bar\eta'}\bigg( \xi_2, r'_g; \frac{\xi_1}{\bar\xi_2}, r'_\gamma\bigg)-A^{\bar\eta'}(r'_g)\bigg] \bigg\rgroup
\bigg\lgroup \bar\xi_2 r^i_g \bigg[ {\cal A}^{\bar\eta}\bigg( \xi_2, r_g; \frac{\xi_1}{\bar\xi_2}, r_\gamma\bigg)-A^{\bar\eta}(r_g)\bigg] \bigg\rgroup
\nonumber\\
&&
\times\; \frac{1}{N_c}\tr\big[ (\partial^iS_{z_3})(\partial^jS^\dagger_{y_3})\big]\nonumber\\
&&
-\; 
\frac{1}{N_c^2}
\bigg\lgroup \frac{\xi_1}{\bar\xi_2} r'^j_\gamma{\cal A}^{\bar\eta'}\bigg( \xi_2, r'_g; \frac{\xi_1}{\bar\xi_2}, r'_\gamma\bigg)
+\xi_2r'^j_g\bigg[ {\cal A}^{\bar\eta'}\bigg( \xi_2, r'_g; \frac{\xi_1}{\bar\xi_2}, r'_\gamma\bigg)-A^{\bar\eta'}(r'_g)\bigg] \bigg\rgroup
\nonumber\\
&&
\times\; 
\bigg\lgroup \frac{\xi_1}{\bar\xi_2} r^i_\gamma{\cal A}^{\bar\eta}\bigg( \xi_2, r_g; \frac{\xi_1}{\bar\xi_2}, r_\gamma\bigg)
+\xi_2r^i_g\bigg[ {\cal A}^{\bar\eta}\bigg( \xi_2, r_g; \frac{\xi_1}{\bar\xi_2}, r_\gamma\bigg)-A^{\bar\eta}(r_g)\bigg] \bigg\rgroup
\frac{1}{N_c}\tr\big[ (\partial^iS_{z_3})(\partial^jS^\dagger_{y_3})\big]
\nonumber\\
&&
-\; 
\bigg\lgroup \frac{\xi_1}{\bar\xi_2} r'^j_\gamma{\cal A}^{\bar\eta'}\bigg( \xi_2, r'_g; \frac{\xi_1}{\bar\xi_2}, r'_\gamma\bigg)
+r'^j_g\bigg[ {\cal A}^{\bar\eta'}\bigg( \xi_2, r'_g; \frac{\xi_1}{\bar\xi_2}, r'_\gamma\bigg)-A^{\bar\eta'}(r'_g)\bigg] \bigg\rgroup
\nonumber\\
&&
\times\; 
\bigg\lgroup \frac{\xi_1}{\bar\xi_2} r^i_\gamma{\cal A}^{\bar\eta}\bigg( \xi_2, r_g; \frac{\xi_1}{\bar\xi_2}, r_\gamma\bigg)
+r^i_g\bigg[ {\cal A}^{\bar\eta}\bigg( \xi_2, r_g; \frac{\xi_1}{\bar\xi_2}, r_\gamma\bigg)-A^{\bar\eta}(r_g)\bigg] \bigg\rgroup
\nonumber\\
&&
\times\; 
\frac{1}{N_c}\tr\big[S_{z_3}S^\dagger_{y_3}\big] \frac{1}{N_c}\tr\big[ S^\dagger_{z_3}(\partial^iS_{z_3})S^\dagger_{y_3}(\partial^jS_{y_3})\big]\bigg\}\;.
\eeq
As in the case of {\it bef-bef} contribution, the integrals over the dipole sizes  $r_g$, $r'_g$, $r_\gamma$ and $r'_\gamma$ are factorized from the rest of the expression. They can be performed with the help of the generic integrals given in Eqs. \eqref{integral_1} and \eqref{integral_2}, and we use the same compact notation introduced in Eq. \eqref{def_pi} to organize the resulting expression. Moreover, the remaining integrals over $z_3$ and $y_3$ have the same structure as the definition of the TMDs introduced in Eqs. \eqref{Fq_1} and \eqref{Fq_2}. Hence, we can write the correlation limit of the  
{\it aft-aft} contribution to the cross section in the quark initiated channel as:
\beq
\label{aft-aft_cor_final}
\Big\langle {\rm I_{aft-aft}}\Big\rangle_{x_A}&=& {\cal M}_q^{\bar \lambda\bar\lambda'; \bar \eta\bar\eta'}\bigg(\xi_2, \frac{\xi_1}{\bar\xi_2}\bigg)\, g_s^2 \, (2\pi)^3 \, N_c\bigg\{
{\left[{\rm H}^{(1)}_{qg}\right]}_{\it aft-aft}^{\bar\lambda\bar\lambda';\bar\eta\bar\eta'; ij} \frac{{\rm P}^i{\rm P}^j}{\rm P^2} {\cal F}^{(1)}_{qg}(x_A, {\rm P})
\nonumber\\
&+&
{\left[{\rm H}^{(2)}_{qg}\right]}_{\it aft-aft}^{\bar\lambda\bar\lambda';\bar\eta\bar\eta'; ij}
\bigg[\frac{1}{2}\delta^{ij}{\cal F}^{(2)}_{qg}(x_A, {\rm P})
-\; 
\frac{1}{2}\bigg(\delta^{ij}-2\frac{{\rm P}^i{\rm P}^j}{\rm P^2}\bigg){\cal H}^{(2)}_{qg}(x_A, {\rm P})\bigg]\bigg\}\;,\hspace{1cm}
\eeq
where the hard parts are given by 
\beq
\label{hard_part_1_aft_aft}
{\left[{\rm H}^{(1)}_{qg}\right]}_{\it aft-aft}^{\bar\lambda\bar\lambda';\bar\eta\bar\eta'; ij}&=&\bigg\lgroup \bar\xi_2 \Pi^{\bar\eta', \bar\lambda' j}\big[{\rm K}; c_1^{-1},q_2\big]\bigg\rgroup
\bigg\lgroup \bar\xi_2\Pi^{\bar\eta, \bar\lambda i}\big[ {\rm K}; c_1^{-1}, q_2\Big]\bigg\rgroup\nonumber\\
&&
\hspace{-1cm}
-\,\frac{1}{N_c^2}
\bigg\lgroup 
\xi_2\Pi^{\bar\lambda', \bar\eta'j}\big[{\rm K}; c_1^{-1}, q_2\big]
-
\frac{\xi_1}{\bar\xi_2}\Pi^{\bar\eta', \bar\lambda'j}\big[q_2; c_1,{\rm K}\big]
\bigg\rgroup
\nonumber\\
&&
\hspace{-0.4cm}
\times\; 
\bigg\lgroup
 \xi_2\Pi^{\bar\lambda, \bar\eta i}\big[{\rm K}; c_1^{-1}, q_2\big]
  -
\frac{\xi_1}{\bar\xi_2}\Pi^{\bar\eta, \bar\lambda i}\big[q_2; c_1,{\rm K}\big]
 \bigg\rgroup\;,
\eeq
and 
\beq
\label{hard_part_2_aft_aft}
{\left[{\rm H}^{(2)}_{qg}\right]}_{\it aft-aft}^{\bar\lambda\bar\lambda';\bar\eta\bar\eta'; ij}&=&
\bigg\lgroup 
\Pi^{\bar\lambda', \bar\eta'j}\big[{\rm K}; c_1^{-1}, q_2\big]
-
\frac{\xi_1}{\bar\xi_2}\Pi^{\bar\eta', \bar\lambda'j}\big[q_2; c_1,{\rm K}\big]
\bigg\rgroup
\nonumber\\
&&
\hspace{-0.4cm}
\times\; 
\bigg\lgroup 
\Pi^{\bar\lambda, \bar\eta i}\big[{\rm K}; c_1^{-1}, q_2\big]
-
\frac{\xi_1}{\bar\xi_2}\Pi^{\bar\eta, \bar\lambda i}\big[q_2; c_1,{\rm K}\big]
\bigg\rgroup\;.
\eeq
In these expressions, the variable $c_1$ is defined as the following function of the longitudinal momentum fractions $\xi_1$ and $\xi_2$:
\beq
c_1=\frac{1}{\xi_2}\frac{\xi_1}{\bar\xi_2}\bigg(1-\frac{\xi_1}{\bar\xi_2}\bigg)\;.
\eeq
Eq.~\eqref{aft-aft_cor_final} is the final result for the {\it aft-aft} contribution to the correlation limit of the production cross section in the quark initiated channel, given in terms of the TMDs. 

\subsubsection{bef-aft contribution}
The remaining two contributions are the crossed ones. We first consider the {\it bef-aft} case. The small dipole sizes for this contribution are 
\beq
r_g&=&v-z_2 \; , \hspace{1.1cm} r'_g=y_3-y_2\;, \\
r_\gamma&=& z_3-z_1\; , \hspace{1cm} r'_\gamma=v'-y_1\;.
\eeq
After this change of variables, the {\it bef-aft} contribution given in Eq. \eqref{bef-aft} yields:  
\beq
&&
{\rm I}_{\rm bef-aft}=\int_{z_3y_3, r_gr'_g, r_\gamma r'_\gamma}e^{i{\rm P}\cdot(y_3-z_3)-i\big[q_1\cdot r'_\gamma-{\rm K}\cdot r_\gamma\big]-i\big[{\rm Q}\cdot r'_g-q_2\cdot r_g\big]} 
8\; \widetilde{\cal M}^{\bar\lambda\bar\lambda'; \bar\eta\bar\eta'}_q(\xi_1,\xi_2)\nonumber\\
&&
\times\; 
A^{\bar\lambda}(r_\gamma)A^{\bar\eta'}(r'_g)\tr\bigg[ {\cal \bar{G}}^{\dagger d\bar\lambda'}_{\rm bef}\bigg(\xi_1, \frac{\xi_2}{\bar\xi_1}; y_3,r'_g,r'_\gamma\bigg) 
{\cal \bar{G}}^{d\bar\eta}_{\rm aft}\bigg(\xi_2, \frac{\xi_1}{\bar\xi_2}; z_3,r_g,r_\gamma\bigg)\bigg]\;,
\eeq
where $\widetilde{\cal M}^{\bar\lambda\bar\lambda';\bar\eta\bar\eta'}_q(\xi_1,\xi_2)$ is the function that accounts for the product of the splitting amplitudes for the crossed contributions, defined in Eq. \eqref{F_cross}. The transverse momenta ${\rm P}, {\rm Q}$ and ${\rm K}$ are defined in Eqs. \eqref{P_def}, \eqref{Q_def} and \eqref{K_def}. The small dipole expansion of the functions ${\cal \bar{G}}^{\dagger d\bar\lambda'}_{\rm bef}$ and ${\cal \bar{G}}^{d\bar\eta}_{\rm aft}$ was already performed above, hence we can simply plug in the results of this expansion (Eqs. \eqref{expanded_G_bef} and \eqref{expanded_G_aft}) to cast the {\it bef-aft} contribution into the following form:
\beq
&&
{\rm I}_{\rm bef-aft}=-
\int_{z_3y_3, r_gr'_g, r_\gamma r'_\gamma}e^{i{\rm P}\cdot(y_3-z_3)-i\big[q_1\cdot r'_\gamma-{\rm K}\cdot r_\gamma\big]-i\big[{\rm Q}\cdot r'_g-q_2\cdot r_g\big]} \nonumber\\
&&\times\;8\widetilde{\cal M}^{\bar\lambda\bar\lambda'; \bar\eta\bar\eta'}_q(\xi_1,\xi_2)
A^{\bar\lambda}(r_\gamma)A^{\bar\eta'}(r'_g)
\nonumber\\
&&
\times\; 
\frac{N_c^2}{2}\bigg\{
\bigg\lgroup
\xi_1r'^j_\gamma\bigg[ \A^{\bar\lambda'}\bigg(\xi_1, r'_\gamma; \frac{\xi_2}{\bar\xi_1}, r'_g\bigg)-A^{\bar\lambda'}(r'_\gamma)\bigg]
-\bigg(1-\frac{\xi_2}{\bar\xi_1}\bigg)r'^j_g\A^{\bar\lambda'}\bigg(\xi_1, r'_\gamma; \frac{\xi_2}{\bar\xi_1}, r'_g\bigg)\bigg\rgroup
\nonumber\\
&&
\times\; 
\bigg\lgroup \bar\xi_2 r^i_g \bigg[ {\cal A}^{\bar\eta}\bigg( \xi_2, r_g; \frac{\xi_1}{\bar\xi_2}, r_\gamma\bigg)-A^{\bar\eta}(r_g)\bigg] \bigg\rgroup
\frac{1}{N_c}\tr\big[(\partial^iS_{z_3})(\partial^jS^\dagger_{y_3})\big]
\nonumber\\
&&
+\; \frac{1}{N_c^2}
\bigg\lgroup 
\xi_1r'^j_\gamma\bigg[ \A^{\bar\lambda'}\bigg(\xi_1, r'_\gamma; \frac{\xi_2}{\bar\xi_1}, r'_g\bigg)-A^{\bar\lambda'}(r'_\gamma)\bigg] 
+\frac{\xi_2}{\bar\xi_1}r'^j_g  \A^{\bar\lambda'}\bigg(\xi_1, r'_\gamma; \frac{\xi_2}{\bar\xi_1}, r'_g\bigg)\bigg\rgroup
\nonumber\\
&&
\times\; 
\bigg\lgroup \frac{\xi_1}{\bar\xi_2} r^i_\gamma{\cal A}^{\bar\eta}\bigg( \xi_2, r_g; \frac{\xi_1}{\bar\xi_2}, r_\gamma\bigg)
+\xi_2r^i_g\bigg[ {\cal A}^{\bar\eta}\bigg( \xi_2, r_g; \frac{\xi_1}{\bar\xi_2}, r_\gamma\bigg)-A^{\bar\eta}(r_g)\bigg] \bigg\rgroup
\frac{1}{N_c}\tr\big[ (\partial^iS_{z_3})(\partial^jS^\dagger_{y_3})\big]
\nonumber\\
&&
+\; 
\bigg\lgroup r'^j_g  \A^{\bar\lambda'}\bigg(\xi_1, r'_\gamma; \frac{\xi_2}{\bar\xi_1}, r'_g\bigg) \bigg\rgroup
\bigg\lgroup \frac{\xi_1}{\bar\xi_2} r^i_\gamma{\cal A}^{\bar\eta}\bigg( \xi_2, r_g; \frac{\xi_1}{\bar\xi_2}, r_\gamma\bigg)
+r^i_g\bigg[ {\cal A}^{\bar\eta}\bigg( \xi_2, r_g; \frac{\xi_1}{\bar\xi_2}, r_\gamma\bigg)-A^{\bar\eta}(r_g)\bigg] \bigg\rgroup
\nonumber\\
&&
\times\; 
\frac{1}{N_c}\tr\big[S_{z_3}S^\dagger_{y_3}\big] \frac{1}{N_c}\tr\big[ S^\dagger_{z_3}(\partial^iS_{z_3})S^\dagger_{y_3}(\partial^jS_{y_3})\big]\bigg\}\;.
\eeq
Again, the integrations over $z_3$ and $y_3$ are factorized from the rest of the expression, and the structure of the parts that depend on the dipole sizes in the amplitude and in the complex conjugate amplitude is very similar to the {\it bef-bef} and {\it aft-aft} contributions. Thus, the integrals over $r_g,r'_g, r_\gamma$ and $r'_\gamma$ can be performed easily using the generic integrals given in Eqs. \eqref{integral_1} and \eqref{integral_2}. Once again, one recognizes the definition of the TMDs presented in Eqs. \eqref{Fq_1}  and \eqref{Fq_2}, such that the overall result of {\it bef-aft} contribution reads
\beq
\label{bef-aft_cor_final}
\Big\langle {\rm I_{bef-aft}}\Big\rangle_{x_A} &=& -\widetilde{\cal M}_q^{\bar \lambda\bar\lambda'; \bar \eta\bar\eta'}(\xi_1,\xi_2)\, g_s^2 (2\pi)^3 N_c
\bigg\{
{\left[{\rm H}^{(1)}_{qg}\right]}_{\it bef-aft}^{\bar\lambda\bar\lambda';\bar\eta\bar\eta'; ij} \frac{{\rm P}^i{\rm P}^j}{\rm P^2}{\cal F}^{(1)}_{qg}(x_A, {\rm P})
\nonumber\\
&+&
{\left[{\rm H}^{(2)}_{qg}\right]}_{\it bef-aft}^{\bar\lambda\bar\lambda';\bar\eta\bar\eta'; ij}
\bigg[\frac{1}{2}\delta^{ij}{\cal F}^{(2)}_{qg}(x_A, {\rm P})
-\; 
\frac{1}{2}\bigg(\delta^{ij}-2\frac{{\rm P}^i{\rm P}^j}{\rm P^2}\bigg){\cal H}^{(2)}_{qg}(x_A, {\rm P})\bigg]\bigg\}\;,\hspace{1cm}
\eeq
with the hard parts (written in terms of the compact notation that is introduced in Eq.~\eqref{def_pi}) being
\beq
\label{hard_part_1_bef_aft}
{\left[{\rm H}^{(1)}_{qg}\right]}_{\it bef-aft}^{\bar\lambda\bar\lambda';\bar\eta\bar\eta'; ij}&=&
\bigg\lgroup \xi_1\Pi^{\bar\eta', \bar\lambda'j}\big[ Q; c^{-1}_0,q_1\big]+\bigg(1-\frac{\xi_2}{\bar\xi_1}\bigg)\Pi^{\bar\lambda', \bar\eta'j}\big[q_1; c_0, Q\big]\bigg\rgroup
\nonumber\\
&&
\times\; 
\bigg\lgroup \bar\xi_2\Pi^{\bar\lambda,\bar\eta i}\big[K; c_1^{-1},q_2\big]\bigg\rgroup
\nonumber\\
&&
\hspace{-0.6cm}
+\; \frac{1}{N_c^2} 
\bigg\lgroup \xi_1\Pi^{\bar\eta', \bar\lambda'j}\big[Q;c_0^{-1},q_1\big]-\frac{\xi_2}{\bar\xi_1}\Pi^{\bar\lambda',\bar\eta'j}\big[q_1;c_0,Q\big]\bigg\rgroup
\nonumber\\
&&
\times\; 
\bigg\lgroup 
\xi_2\Pi^{\bar\lambda,\bar\eta i}\big[K,c_1^{-1},q_2\big]
-
\frac{\xi_1}{\bar\xi_2}\Pi^{\bar\eta,\bar\lambda i}\big[q_2;c_1,K\big]
\bigg\rgroup\;,
\eeq
and
\beq
\label{hard_part_2_bef_aft}
{\left[{\rm H}^{(2)}_{qg}\right]}_{\it bef-aft}^{\bar\lambda\bar\lambda';\bar\eta\bar\eta'; ij}=
\bigg\lgroup \Pi^{\bar\lambda',\bar\eta' j}\big[q_1;c_0,Q\big]\bigg\rgroup
\bigg\lgroup 
\Pi^{\bar\lambda,\bar\eta i}\big[K;c_1^{-1},q_2\big]
-
\frac{\xi_1}{\bar\xi_2}\Pi^{\bar\eta,\bar\lambda i}\big[q_2;c_1,K\big]
\bigg\rgroup.
\eeq

\subsubsection{aft-bef contribution}
The last contribution to the correlation limit of the production cross section in the quark initiated channel is the {\it aft-bef} case. Its calculation is exactly the same as the {\it bef-aft} contribution, with the amplitude and complex conjugate amplitude interchanged. The dipole sizes are 
\beq
r_g&=&z_3-z_2 \; , \hspace{1cm} r'_g=v'-y_2\;, \\
r_\gamma&=& v-z_1\; , \hspace{1.1cm} r'_\gamma=y_3-y_1\;,
\eeq
and after performing this change of variables, the {\it aft-bef} contribution can be written as:
\beq
&&
{\rm I}_{\rm aft-bef}=
\int_{z_3y_3, r_gr'_g, r_\gamma r'_\gamma}e^{i{\rm P}\cdot(y_3-z_3)-i\big[{\rm K}\cdot r'_\gamma-q_1\cdot r_\gamma\big]-i\big[q_2\cdot r'_g-{\rm Q}\cdot r_g\big]} 
8\; \widetilde{\cal M}_q^{\bar\lambda\bar\lambda'; \bar\eta\bar\eta'}(\xi_1,\xi_2)\nonumber\\
&&
\times\; 
A^{\bar\eta}(r_g)A^{\bar\lambda'}(r'_\gamma)
\tr\bigg[ {\cal \bar{G}}^{\dagger d\bar\eta'}_{\rm aft}\bigg(\xi_2, \frac{\xi_1}{\bar\xi_2}; y_3, r'_g, r'_\gamma\bigg){\cal \bar{G}}^{d\bar\lambda}_{\rm bef}\bigg(\xi_1, \frac{\xi_2}{\bar\xi_1}; z_3, r_g, r_\gamma\bigg)\bigg]\;.
\eeq
Using Eqs. \eqref{expanded_G_bef} and \eqref{expanded_G_aft} for the expanded expressions of the functions ${\cal \bar{G}}_{\rm bef}^{d\bar\lambda}$ and ${\cal \bar{G}}_{\rm aft}^{\dagger d\bar\eta'}$, we obtain:
\beq
&&
{\rm I}_{\rm aft-bef}=-
\int_{z_3y_3, r_gr'_g, r_\gamma r'_\gamma}e^{i{\rm P}\cdot(y_3-z_3)-i\big[{\rm K}\cdot r'_\gamma-q_1\cdot r_\gamma\big]-i\big[q_2\cdot r'_g-{\rm Q}\cdot r_g\big]} 
8\; \widetilde{\cal M}^{\bar\lambda\bar\lambda'; \bar\eta\bar\eta'}(\xi_1,\xi_2)\nonumber\\
&&
\times\;  A^{\bar\eta}(r_g)A^{\bar\lambda'}(r'_\gamma)
\frac{N_c^2}{2}\bigg\{ 
\bigg\lgroup \bar\xi_2 r'^j_g \bigg[ {\cal A}^{\bar\eta'}\bigg( \xi_2, r'_g; \frac{\xi_1}{\bar\xi_2}, r'_\gamma\bigg)-A^{\bar\eta'}(r'_g)\bigg] \bigg\rgroup
\nonumber\\
&&
\times\; 
\bigg\lgroup
\xi_1r^i_\gamma\bigg[ \A^{\bar\lambda}\bigg(\xi_1, r_\gamma; \frac{\xi_2}{\bar\xi_1}, r_g\bigg)-A^{\bar\lambda}(r_\gamma)\bigg]
-\bigg(1-\frac{\xi_2}{\bar\xi_1}\bigg)r^i_g\A^{\bar\lambda}\bigg(\xi_1, r_\gamma; \frac{\xi_2}{\bar\xi_1}, r_g\bigg)\bigg\rgroup
\nonumber\\
&&
\times\; 
\frac{1}{N_c}\tr\big[(\partial^iS_{z_3})(\partial^jS^\dagger_{y_3})\big]
\nonumber\\
&&
+\; 
\frac{1}{N_c^2}
\bigg\lgroup \frac{\xi_1}{\bar\xi_2} r'^j_\gamma{\cal A}^{\bar\eta'}\bigg( \xi_2, r'_g; \frac{\xi_1}{\bar\xi_2}, r'_\gamma\bigg)
+\xi_2r'^j_g\bigg[ {\cal A}^{\bar\eta'}\bigg( \xi_2, r'_g; \frac{\xi_1}{\bar\xi_2}, r'_\gamma\bigg)-A^{\bar\eta'}(r'_g)\bigg] \bigg\rgroup
\nonumber\\
&&
\times\; 
\bigg\lgroup 
\xi_1r^i_\gamma\bigg[ \A^{\bar\lambda}\bigg(\xi_1, r_\gamma; \frac{\xi_2}{\bar\xi_1}, r_g\bigg)-A^{\bar\lambda}(r_\gamma)\bigg] 
+\frac{\xi_2}{\bar\xi_1}r^i_g  \A^{\bar\lambda}\bigg(\xi_1, r_\gamma; \frac{\xi_2}{\bar\xi_1}, r_g\bigg)\bigg\rgroup
\frac{1}{N_c}\tr\big[(\partial^iS_{z_3})(\partial^jS^\dagger_{y_3})\big]
\nonumber\\
&&
+\; 
\bigg\lgroup \frac{\xi_1}{\bar\xi_2} r'^j_\gamma{\cal A}^{\bar\eta'}\bigg( \xi_2, r'_g; \frac{\xi_1}{\bar\xi_2}, r'_\gamma\bigg)
+r'^j_g\bigg[ {\cal A}^{\bar\eta'}\bigg( \xi_2, r'_g; \frac{\xi_1}{\bar\xi_2}, r'_\gamma\bigg)-A^{\bar\eta'}(r'_g)\bigg] \bigg\rgroup
\nonumber\\
&&
\times\; 
\bigg\lgroup r^i_g  \A^{\bar\lambda}\bigg(\xi_1, r_\gamma; \frac{\xi_2}{\bar\xi_1}, r_g\bigg) \bigg\rgroup
\frac{1}{N_c}\tr\big[ S^\dagger_{z_3}(\partial^iS_{z_3})S^\dagger_{y_3}(\partial^jS_{y_3})\big]\frac{1}{N_c}\tr\big[S_{z_3}S^\dagger_{y_3}\big]
\bigg\}\;.
\eeq
Performing the integrals over de dipole sizes using Eqs. \eqref{integral_1} and \eqref{integral_2}, and writing the remaining integrals over $z_3$ and $y_3$ in terms of the gluon TMDs, we find the following final result for the {\it aft-bef} contribution:
\beq
\label{aft-bef_cor_final}
\Big\langle{\rm I_{aft-bef}}\Big\rangle_{x_A}&=& -\widetilde{\cal M}_q^{\bar \lambda\bar\lambda'; \bar \eta\bar\eta'}(\xi_1,\xi_2)\, g_s^2\, (2\pi)^3\, N_c
\bigg\{
{\left[{\rm H}^{(1)}_{qg}\right]}_{\it aft-bef}^{\bar\lambda\bar\lambda';\bar\eta\bar\eta'; ij} \frac{{\rm P}^i{\rm P}^j}{\rm P^2}
{\cal F}^{(1)}_{qg}(x_A, {\rm P})
\nonumber\\
&+& 
{\left[{\rm H}^{(2)}_{qg}\right]}_{\it aft-bef}^{\bar\lambda\bar\lambda';\bar\eta\bar\eta'; ij}
\bigg[\frac{1}{2}\delta^{ij}{\cal F}^{(2)}_{qg}(x_A, {\rm P})
-\; 
\frac{1}{2}\bigg(\delta^{ij}-2\frac{{\rm P}^i{\rm P}^j}{\rm P^2}\bigg){\cal H}^{(2)}_{qg}(x_A, {\rm P})\bigg]\bigg\}\;,\hspace{1cm}
\eeq
with the hard parts 
\beq
\label{hard_part_1_aft_bef}
{\left[{\rm H}^{(1)}_{qg}\right]}_{\it aft-bef}^{\bar\lambda\bar\lambda';\bar\eta\bar\eta'; ij}&=&
\!\! \bigg\lgroup \bar\xi_2\Pi^{\bar\lambda',\bar\eta' j}\big[ {\rm K}; c_1^{-1},q_2\big]\bigg\rgroup
\bigg\lgroup \xi_1\Pi^{\bar\eta,\bar\lambda i}\big[{\rm Q}; c_0^{-1},q_1\big]+\bigg(1-\frac{\xi_2}{\bar\xi_1}\bigg)\Pi^{\bar\lambda, \bar\eta i}\big[q_1;c_0,{\rm Q}\big]\bigg\rgroup 
\nonumber\\
&&
\hspace{-1.1cm}
+\, \frac{1}{N_c^2}
\bigg\lgroup 
\xi_2\Pi^{\bar\lambda',\bar\eta' j}\big[{\rm K}; c_1^{-1},q_2\big]
-
\frac{\xi_1}{\bar\xi_2}\Pi^{\bar\eta',\bar\lambda' j}\big[q_2; c_1, {\rm K}\big]
\bigg\rgroup
\nonumber\\
&&
\hspace{-0.5cm}
\times\; 
\bigg\lgroup \xi_1\Pi^{\bar\eta,\bar\lambda i}\big[{\rm Q};c_0^{-1},q_1\big]-\frac{\xi_2}{\bar\xi_1}\Pi^{\bar\lambda,\bar\eta i}\big[q_1; c_0,{\rm Q}\big]\bigg\rgroup\;,
\eeq
and
\beq
\label{hard_part_2_aft_bef}
{\left[{\rm H}^{(2)}_{qg}\right]}_{\it aft-bef}^{\bar\lambda\bar\lambda';\bar\eta\bar\eta'; ij}=
\bigg\lgroup
\Pi^{\bar\lambda',\bar\eta' j}\big[{\rm K}; c_1^{-1},q_2\big]
-
\frac{\xi_1}{\bar\xi_2}\Pi^{\bar\lambda',\bar\eta' j}\big[q_2; c_1,{\rm K}\big]
\bigg\rgroup
\bigg\lgroup \Pi^{\bar\lambda, \bar\eta i}\big[q_1;c_0,{\rm Q}\big] \bigg\rgroup.
\eeq

To conclude this subsection, we calculated the correlation limit of the partonic cross section $ qA \to qg\gamma$, and the result is given in Eq. \eqref{Q_initiated_full}, with the separate  contributions ${\rm I}_{\rm bef-bef}$, ${\rm I}_{\rm aft-aft}$, ${\rm I}_{\rm bef-aft}$ and ${\rm I}_{\rm aft-bef}$ presented in Eqs. \eqref{bef-bef_cor_final}, \eqref{aft-aft_cor_final}, \eqref{bef-aft_cor_final} and \eqref{aft-bef_cor_final}, respectively.

\subsection{Correlation limit: gluon channel}
Let us now study the correlation limit of the production cross section in the gluon initiated channel. Again, we distinguish four separate contributions: photon emission from the antiquark both in the amplitude and in the complex conjugate amplitude ($\bar q\gamma-\bar q\gamma$ contribution), photon emission from the quark both in the amplitude and in the complex conjugate amplitude ($q\gamma-q\gamma$ contribution) and the crossed contributions where the photon is emitted from the quark in the amplitude and from the antiquark in the complex conjugate amplitude, and vice versa ($\bar q\gamma-q\gamma$ and $q\gamma-\bar q\gamma$ contributions). 

\subsubsection{$\bar q\gamma-\bar q\gamma$ contribution} 
We will expand in the small dipole sizes $r_q$ and $\bar r_{\bar q}$ (in the amplitude) and $r'_q$ and $\bar r'_{\bar q}$ (in the complex conjugate amplitude), defined as:
\beq
&&
r_q=v-z_1 \; , \hspace{1.1cm} r'_q=v'-y_1\;,\\
&&
\bar r_{\bar q}=z_3-z_2\; ,  \hspace{1cm} \bar r'_{\bar q}=y_3-y_2\;.
\eeq
Here, $r_q$ corresponds to the transverse size of the dipole formed by the intermediate antiquark (before the photon emission) and the final state quark in the amplitude. Similarly, $\bar r_{\bar q}$ is the transverse size of the dipole formed by the final antiquark and final photon in the amplitude. The primed ones ($r'_q$ and $\bar r'_{\bar q}$) correspond to the their counterparts in the complex conjugate amplitude.

After the above change of variables, the $\bar q\gamma-\bar q\gamma$ contribution given in Eq. \eqref{barq_gamma_barq_gamma_w_Tr} can be written as 
\beq
\label{bar_q_gamma_bar_q_gamma_Before_expansion}
&&
{\rm I}_{ \bar q\gamma-\bar q\gamma}=\int_{z_3y_3, r_qr'_q, \bar r_{\bar q}\bar r'_{\bar q}} 
e^{i{\rm P}\cdot (y_3-z_3)-i{\rm Q}\cdot (\bar r'_{\bar q}-\bar r_{\bar q})-iq_1\cdot(r'_q-r_q)}
\; 8\, {\cal M}_{\bar q\bar q}^{\bar\lambda\bar\lambda'; \bar\eta\bar\eta'}\bigg(\xi_1,\frac{\xi_2}{\bar\xi_1}\bigg)
\nonumber\\
&&
\times\; 
A^{\bar\lambda'}(\bar r'_{\bar q}) \, A^{\bar\lambda}(\bar r_{\bar q}) \; 
\tr\bigg[ 
{\cal \bar{G}}^{\dagger c\bar\eta'}_{\bar q\gamma}\bigg(\xi_1,\xi_2; y_3, r'_q,\bar r'_{\bar q}\bigg)
{\cal \bar{G}}^{c\bar\eta}_{\bar q\gamma}\bigg(\xi_1,\xi_2;z_3,r_q,\bar r_{\bar q}\bigg)\bigg]\;,
\eeq
where we have performed the trivial integrations over $w$, $w'$, $v$ and $v'$, and where ${\cal M}_g^{\bar\lambda\bar\lambda'; \bar\eta\bar\eta'}$ is the function that accounts for the product of splitting amplitudes, defined in Eq. \eqref{product_splitt_squared}. Moreover, the auxiliary hard transverse momentum $Q$ is
\beq
\label{def:Q}
{\rm Q}=q_2+\frac{\xi_2}{\bar\xi_1}q_1\;,
\eeq
which is the same formal expression as in the previous subsection, although now particle 1 refers to the final-state quark and particle 2 to the final-state photon. The function ${\cal \bar{G}}_{\bar q\gamma}^{c\bar\eta}$ encodes the Wilson line structure corresponding to the photon emission from the antiquark in the amplitude, and is obtained from ${\cal G}_{\bar q\gamma}^{c\bar\eta}$ (whose explicit expression is presented in Eq.~\eqref{antiquark_photon}) after the replacement of $w$ and $v$ and the change of variable. Expanding to first nontrivial order in powers of small dipole sizes, one obtains
\beq
\label{expanded_G_bar_q_gamma}
&&
{\cal \bar{G}}^{c\bar\eta}_{\bar q\gamma}\bigg(\xi_1,\xi_2;z_3,r_q,\bar r_{\bar q}\bigg)_{\alpha\beta}
= -\bar\xi_1r^i_q \Big[\A^{\bar\eta}\bigg(\xi_1,r_q;\frac{\xi_2}{\bar\xi_1},\bar r_{\bar q}\bigg)-A^{\bar \eta}(r_q)\bigg]
\; \Big[S^\dagger_{z_3}t^c(\partial^iS_{z_3})\Big]_{\alpha\beta}
\\
&&
+\; 
\Bigg\{ \frac{\xi_2}{\bar\xi_1}\bar r^i_{\bar q} \, \A^{\bar \eta}\bigg(\xi_1,r_q;\frac{\xi_2}{\bar\xi_1},\bar r_{\bar q}\bigg)
+\xi_1r_q^i\bigg[ \A^{\bar \eta}\bigg(\xi_1,r_q;\frac{\xi_2}{\bar\xi_1},\bar r_{\bar q}\bigg)-A^{\bar\eta}(r_q)\bigg]
\bigg\}
\Big[(\partial^iS^\dagger_{z_3})t^cS_{z_3}\Big]_{\alpha\beta}\nonumber
\eeq
The next step is to plug the expanded expression given above into Eq. \eqref{bar_q_gamma_bar_q_gamma_Before_expansion}, and perform the trace over the fundamental color indices. After using the Fierz identity given in Eq. \eqref{Fierz}, the final result reads 
\beq
\label{bar_q_gamma_bar_q_gamma_expanded}
&&
{\rm I}_{ \bar q\gamma-\bar q\gamma}=\int_{z_3y_3, r_qr'_q, \bar r_{\bar q}\bar r'_{\bar q}} 
e^{i{\rm P}\cdot (y_3-z_3)-i{\rm Q}\cdot (\bar r'_{\bar q}-\bar r_{\bar q})-iq_1\cdot(r'_q-r_q)}
\; 8\, {\cal M}_{\bar q\bar q}^{\bar\lambda\bar\lambda'; \bar\eta\bar\eta'}\bigg(\xi_1,\frac{\xi_2}{\bar\xi_1}\bigg)A^{\bar\lambda'}(\bar r'_{\bar q}) \, A^{\bar\lambda}(\bar r_{\bar q})
\nonumber\\
&&
\times\; 
\bigg\{
\bigg\lgroup
\frac{\xi_2}{\bar\xi_1} \bar r'^j_{\bar q}\A^{\bar\eta'}\bigg(\xi_1,r'_q;\frac{\xi_2}{\bar\xi_1},\bar r'_{\bar q}\bigg)
+\xi_1r'^j_q\bigg[\A^{\bar \eta'}\bigg(\xi_1,r'_q;\frac{\xi_2}{\bar\xi_1},\bar r'_{\bar q}\bigg)-A^{\bar\eta'}(r'_q)\bigg]
\bigg\rgroup
\nonumber\\
&&
\hspace{0.3cm}
\times \;
\bigg\lgroup
\frac{\xi_2}{\bar\xi_1} \bar r^i_{\bar q}\A^{\bar\eta}\bigg(\xi_1,r_q;\frac{\xi_2}{\bar\xi_1},\bar r_{\bar q}\bigg)
+\xi_1r^i_q\bigg[\A^{\bar \eta}\bigg(\xi_1,r_q;\frac{\xi_2}{\bar\xi_1},\bar r_{\bar q}\bigg)-A^{\bar\eta}(r_q)\bigg]
\bigg\rgroup
\nonumber\\
&&
\hspace{0.3cm} 
\times\;
\frac{1}{2} \, \tr\big[S_{z_3}S^\dagger_{y_3}\big] \, \tr\big[(\partial^jS_{y_3})(\partial^iS^\dagger_{z_3})\big]
\nonumber\\
&&
\hspace{0.3cm} 
+\; 
\bigg\lgroup\bar \xi_1r'^j_q\bigg[ \A^{\bar \eta'}\bigg(\xi_1,r'_q;\frac{\xi_2}{\bar\xi_1},\bar r'_{\bar q}\bigg)-A^{\bar\eta'}(r'_q)\bigg]\bigg\rgroup
\bigg\lgroup
\bar \xi_1r^i_q \bigg[\A^{\bar \eta}\bigg(\xi_1,r_q;\frac{\xi_2}{\bar\xi_1},\bar r_{\bar q}\bigg)-A^{\bar\eta}(r_q)\bigg]
\bigg\rgroup
\nonumber\\
&&
\hspace{0.3cm} 
\times\; 
\frac{1}{2} \, \tr\big[S_{y_3}S^\dagger_{z_3}\big] \, \tr\big[(\partial^iS_{z_3})(\partial^jS^\dagger_{y_3})\Big]
\nonumber\\
&&
\hspace{0.3cm} 
-\; 
\bigg\lgroup
\frac{\xi_2}{\bar\xi_1} \bar r'^j_{\bar q}\A^{\bar\eta'}\bigg(\xi_1,r'_q;\frac{\xi_2}{\bar\xi_1},\bar r'_{\bar q}\bigg)
+\xi_1r'^j_q\bigg[\A^{\bar \eta'}\bigg(\xi_1,r'_q;\frac{\xi_2}{\bar\xi_1},\bar r'_{\bar q}\bigg)-A^{\bar\eta'}(r'_q)\bigg]
\bigg\rgroup
\nonumber\\
&&
\hspace{0.3cm} 
\times\; 
\bigg\lgroup
\bar \xi_1r^i_q \bigg[\A^{\bar \eta}\bigg(\xi_1,r_q;\frac{\xi_2}{\bar\xi_1},\bar r_{\bar q}\bigg)-A^{\bar\eta}(r_q)\bigg]
\bigg\rgroup
\frac{1}{2} \, \tr\Big[(\partial^jS_{y_3})S^\dagger_{z_3}\big] \, \tr\big[(\partial^iS_{z_3})S^\dagger_{y_3}\big]
\nonumber\\
&&
\hspace{0.3cm} 
-\; 
\bigg\lgroup\bar \xi_1r'^j_q\bigg[ \A^{\bar \eta'}\bigg(\xi_1,r'_q;\frac{\xi_2}{\bar\xi_1},\bar r'_{\bar q}\bigg)-A^{\bar\eta'}(r'_q)\bigg]\bigg\rgroup
\nonumber\\
&&
\hspace{0.3cm} 
\times\; 
\bigg\lgroup
\frac{\xi_2}{\bar\xi_1} \bar r^i_{\bar q}\A^{\bar\eta}\bigg(\xi_1,r_q;\frac{\xi_2}{\bar\xi_1},\bar r_{\bar q}\bigg)
+\xi_1r^i_q\bigg[\A^{\bar \eta}\bigg(\xi_1,r_q;\frac{\xi_2}{\bar\xi_1},\bar r_{\bar q}\bigg)-A^{\bar\eta}(r_q)\bigg]
\bigg\rgroup
\nonumber\\
&&
\hspace{0.3cm} 
\times\; 
\frac{1}{2} \, \tr\big[(\partial^jS^\dagger_{y_3})S_{z_3}\big] \, \tr\big[S_{y_3}(\partial^iS^\dagger_{z_3})\big]
\nonumber\\
&&
\hspace{0.3cm} 
+\; 
\bigg\lgroup
\frac{\xi_2}{\bar\xi_1} \bar r'^j_{\bar q}\A^{\bar\eta'}\bigg(\xi_1,r'_q;\frac{\xi_2}{\bar\xi_1},\bar r'_{\bar q}\bigg)
+r'^j_q\bigg[\A^{\bar \eta'}\bigg(\xi_1,r'_q;\frac{\xi_2}{\bar\xi_1},\bar r'_{\bar q}\bigg)-A^{\bar\eta'}(r'_q)\bigg]
\bigg\rgroup
\nonumber\\
&&
\hspace{0.3cm} 
\times\;
\bigg\lgroup
\frac{\xi_2}{\bar\xi_1} \bar r^i_{\bar q}\A^{\bar\eta}\bigg(\xi_1,r_q;\frac{\xi_2}{\bar\xi_1},\bar r_{\bar q}\bigg)
+r^i_q\bigg[\A^{\bar \eta}\bigg(\xi_1,r_q;\frac{\xi_2}{\bar\xi_1},\bar r_{\bar q}\bigg)-A^{\bar\eta}(r_q)\bigg]
\bigg\rgroup
\nonumber\\
&&
\hspace{0.3cm} 
\times\; 
\frac{1}{2N_c}\tr\big[(\partial^jS_{y_3})S^\dagger_{z_3}(\partial S_{z_3})S^\dagger_{y_3}\big] \Bigg\}\;.
\eeq
In the above equation, the dependence on the transverse dipole sizes $r_q$, $\bar r_{\bar q}$, $r'_q$ and $\bar r'_{\bar q}$ is factorized from the $z_3$ and $y_3$ dependence. We can therefore perform the integrations over those dipole sizes, for which we use the results Eqs. \eqref{integral_1} and Eq. \eqref{integral_2} (employing again the definition of the pseudo-projector $\Pi^{i, jk}[p;c_0,q]$ given in Eq. \eqref{def_pi}). The result is:
\beq
\label{barq_gamma_barq_gamma_before_TMDs}
&&
{\rm I}_{\bar q\gamma-\bar q\gamma}=8\, {\cal M}_{\bar q\bar q}^{\bar\lambda\bar\lambda';\bar\eta\bar\eta'}\bigg(\xi_1,\frac{\xi_2}{\bar\xi_1}\bigg)
\int_{z_3y_3}e^{i{\rm P}\cdot(y_3-z_3)}
\nonumber\\
&&
\times
\bigg\{ \!
\bigg\lgroup \frac{\xi_2}{\bar\xi_1}\Pi^{\bar\eta';\bar\lambda'j}\big[q_1;b_0,{\rm Q}\big]-\xi_1\Pi^{\bar\lambda';\bar\eta'j}\big[{\rm Q}; b_0^{-1},q_1\big]\bigg\rgroup  \!\!
\nonumber\\
&&
\hspace{0.5cm}
\times
\bigg\lgroup \frac{\xi_2}{\bar\xi_1}\Pi^{\bar\eta;\bar\lambda i}\big[q_1;b_0,{\rm Q}\big]-\xi_1\Pi^{\bar\lambda;\bar\eta i}\big[{\rm Q};b_0^{-1},q_1\big]\bigg\rgroup\frac{1}{2}\, \tr\big[S_{z_3}S^\dagger_{y_3}\big]\, \tr\big[(\partial^jS_{y_3})(\partial^iS^\dagger_{z_3})\big]
\nonumber\\
&&
\hspace{0.5cm}
+
\bigg\lgroup \bar\xi_1\Pi^{\bar\lambda';\bar\eta'j}\big[{\rm Q}; b_0^{-1},q_1\big]\bigg\rgroup
\bigg\lgroup \bar\xi_1\Pi^{\bar\lambda;\bar\eta i}\big[{\rm Q};b_0^{-1},q_1\big]\bigg\rgroup
\frac{1}{2}\, \tr\big[S_{y_3}S^\dagger_{z_3}\big] \, \tr\big[(\partial^iS_{z_3})(\partial^jS^\dagger_{y_3})\big]
\nonumber\\
&&
\hspace{0.5cm}
+
\bigg\lgroup \frac{\xi_2}{\bar\xi_1}\Pi^{\bar\eta';\bar\lambda'j}\big[q_1;b_0,{\rm Q}\big]-\xi_1\Pi^{\bar\lambda';\bar\eta'j}\big[{\rm Q}; b_0^{-1},q_1\big]\bigg\rgroup 
\bigg\lgroup \bar\xi_1\Pi^{\bar\lambda;\bar\eta i}\big[{\rm Q};b_0^{-1},q_1\big]\bigg\rgroup
\nonumber\\
&&
\hspace{0.5cm}
\times
\frac{1}{2}\, \tr\big[(\partial^jS_{y_3})S^\dagger_{z_3}\big]\, \tr\big[(\partial^iS_{z_3})S^\dagger_{y_3}\big]
\nonumber\\
&&
\hspace{0.5cm}
+
\bigg\lgroup \bar\xi_1\Pi^{\bar\lambda';\bar\eta'j}\big[{\rm Q}; b_0^{-1},q_1\big]\bigg\rgroup
\bigg\lgroup \frac{\xi_2}{\bar\xi_1}\Pi^{\bar\eta;\bar\lambda i}\big[q_1;b_0,{\rm Q}\big]-\xi_1\Pi^{\bar\lambda;\bar\eta i}\big[{\rm Q};b_0^{-1},q_1\big]\bigg\rgroup
\nonumber\\
&&
\hspace{0.5cm}
\times
\frac{1}{2}\, \tr\big[(\partial^jS^\dagger_{y_3})S_{z_3}\big]\, \tr\big[S_{y_3}(\partial^iS^\dagger_{z_3})\big]
\nonumber\\
&&
\hspace{0.5cm}
+
\bigg\lgroup \frac{\xi_2}{\bar\xi_1}\Pi^{\bar\eta';\bar\lambda'j}\big[q_1;b_0,{\rm Q}\big]-\Pi^{\bar\lambda';\bar\eta'j}\big[{\rm Q}; b_0^{-1},q_1\big]\bigg\rgroup  \!\!
\nonumber\\
&&
\hspace{0.5cm}
\times
\bigg\lgroup \frac{\xi_2}{\bar\xi_1}\Pi^{\bar\eta;\bar\lambda i}\big[q_1;b_0,{\rm Q}\big]-\Pi^{\bar\lambda;\bar\eta i}\big[{\rm Q};b_0^{-1},q_1\big]\bigg\rgroup\frac{1}{2N_c}\, \tr\big[(\partial^jS_{y_3})S^\dagger_{z_3}(\partial^iS_{z_3})S^\dagger_{y_3}\big]
\bigg\}\;,
\eeq
where 
\beq
b_0=\frac{1}{\xi_1}\frac{\xi_2}{\bar\xi_1}\bigg(1-\frac{\xi_2}{\bar\xi_1}\bigg)\;.
\eeq
Our next step is to perform the integrations over the remaining transverse coordinates $z_3$ and $y_3$. As in the case of quark initiated channel, these will be performed by using the definitions of gluon TMDs. However, let us first manipulate this formula to put it into a more elegant form. First, consider the first two terms in Eq. \eqref{barq_gamma_barq_gamma_before_TMDs}, which we will denote in the following with the superscript (1,2). It can easily be seen that the momentum structure of the parts that multiply the double trace operator, is symmetric under the exchange of $i\leftrightarrow j$, since the function ${\cal M}_g^{\bar\lambda\bar\lambda';\bar\eta\bar\eta'}$ is symmetric under the simultaneous exchanges $\bar\lambda\leftrightarrow\bar\lambda'$ and $\bar\eta\leftrightarrow\bar\eta'$.  Using this symmetry property, together with the exchange of the coordinates $z_3\leftrightarrow y_3$, these two terms can be written as 
\beq
\label{barq_gamma_barq_gamma_12_Inter}
&&
\hspace{-0.3cm}
{\rm I}_{\bar q\gamma-\bar q\gamma}^{(1,2)}= 8\, {\cal M}_{\bar q\bar q}^{\bar\lambda\bar\lambda';\bar\eta\bar\eta'}\bigg(\xi_1,\frac{\xi_2}{\bar\xi_1}\bigg)\frac{1}{2}
\bigg\{ \!
\bigg\lgroup \frac{\xi_2}{\bar\xi_1}\Pi^{\bar\eta';\bar\lambda'j}\big[q_1;b_0,{\rm Q}\big]-\xi_1\Pi^{\bar\lambda';\bar\eta'j}\big[{\rm Q}; b_0^{-1},q_1\big]\bigg\rgroup  \!\!
\\
&&
\hspace{-0.3cm}
\times
\bigg\lgroup \frac{\xi_2}{\bar\xi_1}\Pi^{\bar\eta;\bar\lambda i}\big[q_1;b_0,{\rm Q}\big]-\xi_1\Pi^{\bar\lambda;\bar\eta i}\big[{\rm Q};b_0^{-1},q_1\big]\bigg\rgroup
\nonumber\\
&&
\hspace{-0.3cm}
\times
\frac{1}{2}
\int_{z_3y_3} \! \!\bigg[
e^{i{\rm P}\cdot (y_3-z_3)}\tr\big[S_{z_3}S^\dagger_{y_3}\big]\tr\big[(\partial^jS_{y_3})(\partial^iS^\dagger_{z_3})\big]
+
e^{-i{\rm P}\cdot (y_3-z_3)} \tr\big[S_{y_3}S^\dagger_{z_3}\big]\tr\big[(\partial^iS_{z_3})(\partial^jS^\dagger_{y_3})\big] \!
\bigg]
\nonumber\\
&&
\hspace{-0.3cm}
+
\bigg\lgroup \bar\xi_1\Pi^{\bar\lambda';\bar\eta'j}\big[{\rm Q}; b_0^{-1},q_1\big]\bigg\rgroup
\bigg\lgroup \bar\xi_1\Pi^{\bar\lambda;\bar\eta i}\big[{\rm Q};b_0^{-1},q_1\big]\bigg\rgroup
\nonumber\\
&&
\hspace{-0.3cm}
\times
\frac{1}{2}
\int_{z_3y_3}\!\!
\bigg[ e^{i{\rm P}\cdot(y_3-z_3)}\tr\big[S_{y_3}S^\dagger_{z_3}\big]\tr\big[(\partial^iS_{z_3})(\partial^jS^\dagger_{y_3})\big]\nonumber\\
&&+e^{-i{\rm P}\cdot(y_3-z_3)}\tr\big[S_{z_3}S^\dagger_{y_3}\big]\tr\big[(\partial^jS_{y_3})(\partial^iS^\dagger_{z_3})\big]\!\bigg]\bigg\}\;.
\nonumber
\eeq
The definition of the first TMD we thus encounter in the gluon channel is:
\beq
\label{def:TMD_1}
&&
\int_{z_3y_3}e^{i{\rm P}\cdot(y_3-z_3)} \left\langle \tr\big[S_{z_3}S^\dagger_{y_3}\big]\, \tr\big[(\partial^{j}S_{y_3})(\partial^iS^\dagger_{z_3})\big]\right\rangle_{x_A} =g_s^2 (2\pi)^3N_c\frac{1}{4}
\nonumber\\
&&
\hspace{3cm}
\times
\bigg[\frac{1}{2}\delta^{ij}{\cal F}^{(1)}_{gg}(x_A,{\rm P})
-\frac{1}{2}
\bigg(\delta^{ij}-2\frac{\rm P^i\rm P^j}{\rm P^2}\bigg)
{\cal H}^{(1)}_{gg}(x_A,{\rm P})\bigg]\;.
\eeq
Using this definition, we rewrite (after target averaging) Eq. \eqref{barq_gamma_barq_gamma_12_Inter} as:
\beq
\label{barq_gamma_barq_gamma_12_Final}
&&
\hspace{-0.3cm}
\Big\langle{\rm I}_{\bar q\gamma-\bar q\gamma}^{(1,2)}\Big\rangle_{x_A}={\cal M}_{\bar q\bar q}^{\bar\lambda\bar\lambda';\bar\eta\bar\eta'}\bigg(\xi_1,\frac{\xi_2}{\bar\xi_1}\bigg)g_s^2(2\pi)^3N_c\frac{1}{2}
\\
&&
\hspace{-0.3cm}
\times
\bigg\{\!
\bigg\lgroup \frac{\xi_2}{\bar\xi_1}\Pi^{\bar\eta';\bar\lambda'j}\big[q_1;b_0,{\rm Q}\big]-\xi_1\Pi^{\bar\lambda';\bar\eta'j}\big[{\rm Q}; b_0^{-1},q_1\big]\bigg\rgroup  \!\!
\bigg\lgroup \frac{\xi_2}{\bar\xi_1}\Pi^{\bar\eta;\bar\lambda i}\big[q_1;b_0,{\rm Q}\big]-\xi_1\Pi^{\bar\lambda;\bar\eta i}\big[{\rm Q};b_0^{-1},q_1\big]\bigg\rgroup
\nonumber\\
&&
+
\bigg\lgroup \bar\xi_1\Pi^{\bar\lambda';\bar\eta'j}\big[{\rm Q}; b_0^{-1},q_1\big]\bigg\rgroup
\bigg\lgroup \bar\xi_1\Pi^{\bar\lambda;\bar\eta i}\big[{\rm Q};b_0^{-1},q_1\big]\bigg\rgroup
\bigg\}
\nonumber\\
&&
\hspace{-0.3cm}
\times
\bigg\{
\frac{1}{2}\delta^{ij}\bigg[{\cal F}^{(1)}_{gg}(x_A, {\rm P})+{\cal F}^{(1)}_{gg}(x_A, -{\rm P})\bigg]
-\frac{1}{2}\bigg(\delta^{ij}-2\frac{\rm P^i\rm P^j}{\rm P^2}\bigg)
\bigg[{\cal H}^{(1)}_{gg}(x_A,{\rm P})+{\cal H}^{(1)}_{gg}(x_A,-{\rm P})\bigg]\bigg\}\;.
\nonumber
\eeq

We can perform similar manipulations to rewrite the third and fourth terms of the Taylor expansion of ${\rm I}_{\bar q\gamma-\bar q\gamma}$, Eq. \eqref{barq_gamma_barq_gamma_before_TMDs}, which we denote with the superscript (3,4):
\beq
\label{barq_gamma_barq_gamma_34}
&&
{\rm I}_{\bar q\gamma-\bar q\gamma}^{(3,4)}=8\, {\cal M}_{\bar q\bar q}^{\bar\lambda\bar\lambda';\bar\eta\bar\eta'}\bigg(\xi_1,\frac{\xi_2}{\bar\xi_1}\bigg)\frac{1}{2}\int_{z_3y_3}e^{i{\rm P}\cdot(y_3-z_3)} 
\nonumber\\
&&
\times
\bigg\{
\bigg\lgroup \frac{\xi_2}{\bar\xi_1}\Pi^{\bar\eta';\bar\lambda'j}\big[q_1;b_0,{\rm Q}\big]-\xi_1\Pi^{\bar\lambda';\bar\eta'j}\big[{\rm Q}; b_0^{-1},q_1\big]\bigg\rgroup 
\bigg\lgroup \bar\xi_1\Pi^{\bar\lambda;\bar\eta i}\big[{\rm Q};b_0^{-1},q_1\big]\bigg\rgroup
\nonumber\\
&&
\hspace{0.4cm}
\times
\frac{1}{2}
\bigg[ \tr\big[(\partial^iS_{z_3})S^\dagger_{y_3}\big]\tr\big[(\partial^jS_{y_3})S^\dagger_{z_3}\big]+
\tr\big[(\partial^jS^\dagger_{z_3})S_{y_3}\big] \tr\big[ (\partial^iS^\dagger_{y_3})S_{z_3}\big]\bigg]
\nonumber\\
&&
\hspace{0.4cm}
+
\bigg\lgroup \bar\xi_1\Pi^{\bar\lambda';\bar\eta'j}\big[{\rm Q}; b_0^{-1},q_1\big]\bigg\rgroup
\bigg\lgroup \frac{\xi_2}{\bar\xi_1}\Pi^{\bar\eta;\bar\lambda i}\big[q_1;b_0,{\rm Q}\big]-\xi_1\Pi^{\bar\lambda;\bar\eta i}\big[{\rm Q};b_0^{-1},q_1\big]\bigg\rgroup
\nonumber\\
&&
\hspace{0.4cm}
\times\frac{1}{2}
\bigg[
\tr\big[ (\partial^jS_{z_3})S^\dagger_{y_3}\big]\tr\big[(\partial^iS_{y_3})S^\dagger_{z_3}\big]+
\tr\big[(\partial^iS^\dagger_{z_3})S_{y_3}\big] \tr\big[(\partial^jS^\dagger_{y_3})S_{z_3}\big] \bigg]\bigg\}\;.
\eeq 
Introducing the second gluon TMD which we encounter in the gluon channel: 
\beq
\frac{1}{2} \int_{z_3y_3}e^{i{\rm P}\cdot (y_3-z_3)}\, \left\langle \tr\big[(\partial^iS_{z_3})S^\dagger_{y_3}\big] \, \tr\big[(\partial^jS_{y_3})S^\dagger_{z_3}\big] \, +\, h.c. \right\rangle_{x_A}
\hspace{1cm}\nonumber\\
=-g_s^2\, (2\pi)^3\,N_c\frac{1}{4}\bigg[ \frac{1}{2}\delta^{ij}{\cal F}^{(2)}_{gg}(x_A,{\rm P})
-\frac{1}{2}
\bigg(\delta^{ij}-2\frac{\rm P^i\rm P^j}{\rm P^2}\bigg)
{\cal H}^{(2)}_{gg}(x_A,{\rm P})\bigg]\;,\hspace{-1cm}
\eeq
Eq. \eqref{barq_gamma_barq_gamma_34} can be (after target averaging) written as:
\beq
\label{barq_gamma_barq_gamma_34_F1F2}
&&
\Big\langle{\rm I}_{\bar q\gamma-\bar q\gamma}^{(3,4)}\Big\rangle_{x_A}={\cal M}_{\bar q\bar q}^{\bar\lambda\bar\lambda';\bar\eta\bar\eta'}\bigg(\xi_1,\frac{\xi_2}{\bar\xi_1}\bigg) (-2) g_s^2(2\pi)^3N_c
\nonumber\\
&&
\times \frac{1}{2}
\bigg\{
\bigg\lgroup \frac{\xi_2}{\bar\xi_1}\Pi^{\bar\eta';\bar\lambda'j}\big[q_1;b_0,{\rm Q}\big]-\xi_1\Pi^{\bar\lambda';\bar\eta'j}\big[{\rm Q}; b_0^{-1},q_1\big]\bigg\rgroup 
\bigg\lgroup \bar\xi_1\Pi^{\bar\lambda;\bar\eta i}\big[{\rm Q};b_0^{-1},q_1\big]\bigg\rgroup
\nonumber\\
&&
\hspace{0.5cm}
+
\bigg\lgroup \bar\xi_1\Pi^{\bar\lambda';\bar\eta'j}\big[{\rm Q}; b_0^{-1},q_1\big]\bigg\rgroup
\bigg\lgroup \frac{\xi_2}{\bar\xi_1}\Pi^{\bar\eta;\bar\lambda i}\big[q_1;b_0,{\rm Q}\big]-\xi_1\Pi^{\bar\lambda;\bar\eta i}\big[{\rm Q};b_0^{-1},q_1\big]\bigg\rgroup\bigg\}
\nonumber\\
&&
\times
\bigg[ 
\frac{1}{2}\delta^{ij}{\cal F}^{(2)}_{gg}(x_A,{\rm P})-\frac{1}{2}\bigg(\delta^{ij}-2\frac{\rm P^i\rm P^j}{\rm P^2}\bigg){\cal H}^{(2)}_{gg}(x_A,\rm P)\bigg]\;.
\eeq
As noticed in \cite{Marquet:2017xwy}, the two unpolarized TMDs that we introduced (${\cal F}^{(1)}_{gg}$ and ${\cal F}^{(2)}_{gg}$), and their polarized partners (${\cal H}^{(1)}_{gg}$ and ${\cal H}^{(2)}_{gg}$), are related to the gluon TMD that is build from a dipole in the adjoint representation: 
\beq
\label{def:F_adj}
{\cal F}_{adj}(x_A,{\rm P})=\frac{4C_F{\rm P}^2}{g_s^2}\frac{1}{(2\pi)^3}\int_{z_3y_3}e^{i{\rm P}\cdot(y_3-z_3)}d_A(z_3,y_3;x_A)\ ,
\eeq
where the adjoint dipole is defined in the standard way: 
\beq
d_A(z_3,y_3;x_A)=\frac{1}{N_c^2-1}\Big\langle\tr\big[S_A(z_3)S^\dagger_A(y_3)\big]\Big\rangle_{x_A}\;.
\eeq
Indeed, using the identity that relates the adjoint and fundamental representations of $SU(N_c)$ , given in Eq. \eqref{Adj_to_Fundamental}, it is straightforward to realize that
\beq
\label{F_in_adj}
{\cal F}^{(1)}_{gg}(x_A,{\rm P})+{\cal F}^{(1)}_{gg}(x_A,-{\rm P})-2{\cal F}^{(2)}_{gg}(x_A,{\rm P})=2{\cal F}_{adj}(x_A,{\rm P})\;,
\\
\label{H_in_adj}
{\cal H}^{(1)}_{gg}(x_A,{\rm P})+{\cal H}^{(1)}_{gg}(x_A,-{\rm P})-2{\cal H}^{(2)}_{gg}(x_A,{\rm P})=2{\cal F}_{adj}(x_A,{\rm P})\;.
\eeq
Thus, using Eqs. \eqref{F_in_adj} and \eqref{H_in_adj}, Eq. \eqref{barq_gamma_barq_gamma_34_F1F2} can be cast into the following expression in which the gluon TMDs  ${\cal F}^{(2)}_{gg}$ and ${\cal H}^{(2)}_{gg}$ are eliminated in favor of the adjoint dipole distribution ${\cal F}_{adj}(x_2,{\rm P})$: 
\beq
\label{barq_gamma_barq_gamma_34_Final}
&&
\Big\langle{\rm I}_{\bar q\gamma-\bar q\gamma}^{(3,4)}\Big\rangle_{x_A}={\cal M}_{\bar q\bar q}^{\bar\lambda\bar\lambda';\bar\eta\bar\eta'}\bigg(\xi_1,\frac{\xi_2}{\bar\xi_1}\bigg) g_s^2(2\pi)^3N_c
\nonumber\\
&&
\times \frac{1}{2}
\bigg\{
\bigg\lgroup \frac{\xi_2}{\bar\xi_1}\Pi^{\bar\eta';\bar\lambda'j}\big[q_1;b_0,{\rm Q}\big]-\xi_1\Pi^{\bar\lambda';\bar\eta'j}\big[{\rm Q}; b_0^{-1},q_1\big]\bigg\rgroup 
\bigg\lgroup \bar\xi_1\Pi^{\bar\lambda;\bar\eta i}\big[{\rm Q};b_0^{-1},q_1\big]\bigg\rgroup
\nonumber\\
&&
\hspace{0.5cm}
+
\bigg\lgroup \bar\xi_1\Pi^{\bar\lambda';\bar\eta'j}\big[{\rm Q}; b_0^{-1},q_1\big]\bigg\rgroup
\bigg\lgroup \frac{\xi_2}{\bar\xi_1}\Pi^{\bar\eta;\bar\lambda i}\big[q_1;b_0,{\rm Q}\big]-\xi_1\Pi^{\bar\lambda;\bar\eta i}\big[{\rm Q};b_0^{-1},q_1\big]\bigg\rgroup\bigg\}
\nonumber\\
&&
\times
\bigg\{
\frac{1}{2}\delta^{ij}
\bigg[2{\cal F}_{adj}(x_A,{\rm P})- {\cal F}^{(1)}_{gg}(x_A, {\rm P})- {\cal F}^{(1)}_{gg}(x_A, -{\rm P})\bigg]
\nonumber\\
&&
\hspace{0.4cm}
-\frac{1}{2}
\bigg(\delta^{ij}-2\frac{\rm P^i\rm P^j}{\rm P^2}\bigg)
\bigg[2{\cal F}_{adj}(x_A,{\rm P}) -{\cal H}^{(1)}_{gg}(x_A,{\rm P}) -{\cal H}^{(1)}_{gg}(x_A,-{\rm P}) \bigg] \bigg\}\;.
\eeq

Finally, we can consider the fifth term in Eq. \eqref{barq_gamma_barq_gamma_before_TMDs}. Using the definition of the last gluon TMD in the gluon channel (also known as the Weizs\"acker-Williams gluon distribution):
\beq
\label{def:TMD_3}
&&
\int_{z_3y_3}e^{i{\rm P}\cdot(y_3-z_3)}\left\langle\tr\big[(\partial^iS_{z_3})S^\dagger_{y_3}(\partial^jS_{y_3})S^\dagger_{z_3}\big]\right\rangle_{x_A}=
-g_s^2\, (2\pi)^3 \frac{1}{4}
\nonumber\\
&&
\hspace{3cm}
\times
\bigg[\frac{1}{2}\delta^{ij}{\cal F}^{(3)}_{gg}(x_A, {\rm P})
-\frac{1}{2}\bigg(\delta^{ij}-2\frac{\rm P^i\rm P^j}{\rm P^2}\bigg){\cal H}^{(3)}_{gg}(x_A, {\rm P})\bigg]\;,
\eeq
the integration over $z_3$ and $y_3$ in this term is performed, yielding:
\beq
\label{barq_gamma_barq_gamma_5_Final}
&&
\Big\langle{\rm I}^{(5)}_{\bar q\gamma-\bar q\gamma}\Big\rangle_{x_A}={\cal M}_{\bar q\bar q}^{\bar\lambda\bar\lambda';\bar\eta\bar\eta'}\bigg(\xi_1,\frac{\xi_2}{\bar\xi_1}\bigg) (-1) g_s^2 (2\pi)^3 \frac{1}{N_c}
\nonumber\\
&&
\times
\bigg\lgroup \frac{\xi_2}{\bar\xi_1}\Pi^{\bar\eta';\bar\lambda'j}\big[q_1;b_0,{\rm Q}\big]-\Pi^{\bar\lambda';\bar\eta'j}\big[{\rm Q}; b_0^{-1},q_1\big]\bigg\rgroup  \!\!
\bigg\lgroup \frac{\xi_2}{\bar\xi_1}\Pi^{\bar\eta;\bar\lambda i}\big[q_1;b_0,{\rm Q}\big]-\Pi^{\bar\lambda;\bar\eta i}\big[{\rm Q};b_0^{-1},q_1\big]\bigg\rgroup
\nonumber\\
&&
\times
\bigg[\frac{1}{2}\delta^{ij}{\cal F}^{(3)}_{gg}(x_A, {\rm P})
-\frac{1}{2}\bigg(\delta^{ij}-2\frac{\rm P^i\rm P^j}{\rm P^2}\bigg){\cal H}^{(3)}_{gg}(x_A, {\rm P})\bigg]\;.
\eeq
Eventually, putting everything together by adding Eqs. \eqref{barq_gamma_barq_gamma_12_Final}, \eqref{barq_gamma_barq_gamma_34_Final} and \eqref{barq_gamma_barq_gamma_5_Final}, we obtain: \beq
\label{final_barq_gamma_barq_gamma}
\Big\langle{\rm I}_{\bar q\gamma-\bar q\gamma}\Big\rangle_{x_A}&=&{\cal M}_{\bar q\bar q}^{\bar\lambda\bar\lambda';\bar\eta\bar\eta'}\bigg(\xi_1,\frac{\xi_2}{\bar\xi_1}\bigg)g_s^2(2\pi)^3N_c
\nonumber\\
&&
\times
\bigg\{
\Big[ {\rm H}^{(1)}_{gg}\Big]^{\bar\lambda\bar\lambda';\bar\eta\bar\eta';ij}_{\bar q\gamma-\bar q\gamma}
\bigg\lgroup \frac{1}{2}\delta^{ij}\bigg[ \frac{1}{2}{\cal F}^{(1)}_{gg}(x_A,{\rm P})+\frac{1}{2}{\cal F}^{(1)}_{gg}(x_A,-{\rm P})-\frac{1}{N_c^2}{\cal F}^{(3)}_{gg}(x_A,{\rm P})\bigg]
\nonumber\\
&&
\hspace{0.5cm}
-
\frac{1}{2}\bigg(\delta^{ij}-2\frac{\rm P^i\rm P^j}{\rm P^2}\bigg)
\bigg[ \frac{1}{2}{\cal H}^{(1)}_{gg}(x_A,{\rm P})+\frac{1}{2}{\cal H}^{(1)}_{gg}(x_A,-{\rm P})-\frac{1}{N_c^2}{\cal H}^{(3)}_{gg}(x_A,{\rm P})\bigg]\bigg\rgroup
\nonumber\\
&&
\hspace{0.4cm}
+\Big[ {\rm H}^{(adj)}_{gg}\Big]^{\bar\lambda\bar\lambda';\bar\eta\bar\eta';ij}_{\bar q\gamma-\bar q\gamma}\; \frac{\rm P^i\rm P^j}{\rm P^2} \; {\cal F}_{adj}(x_A,{\rm P})\bigg\}\;,
\eeq 
where the hard parts are defined as 
\beq
\label{hard_part_1_barq_gamma_barq_gamma}
\Big[ {\rm H}^{(1)}_{gg}\Big]^{\bar\lambda\bar\lambda';\bar\eta\bar\eta';ij}_{\bar q\gamma-\bar q\gamma}&=&
\bigg\lgroup \frac{\xi_2}{\bar\xi_1}\Pi^{\bar\eta';\bar\lambda'j}\big[q_1;b_0,{\rm Q}\big]-\Pi^{\bar\lambda';\bar\eta'j}\big[{\rm Q}; b_0^{-1},q_1\big]\bigg\rgroup 
\nonumber\\
&&
\hspace{-0.3cm}
\times
\bigg\lgroup \frac{\xi_2}{\bar\xi_1}\Pi^{\bar\eta;\bar\lambda i}\big[q_1;b_0,{\rm Q}\big]-\Pi^{\bar\lambda;\bar\eta i}\big[{\rm Q};b_0^{-1},q_1\big]\bigg\rgroup\;,
\eeq
and 
\beq
\label{hard_part_adj_barq_gamma_barq_gamma}
&&
\hspace{-0.5cm}
\Big[ {\rm H}^{(adj)}_{gg}\Big]^{\bar\lambda\bar\lambda';\bar\eta\bar\eta';ij}_{\bar q\gamma-\bar q\gamma}=
\bigg\lgroup \frac{\xi_2}{\bar\xi_1}\Pi^{\bar\eta';\bar\lambda'j}\big[q_1;b_0,{\rm Q}\big]-\xi_1\Pi^{\bar\lambda';\bar\eta'j}\big[{\rm Q}; b_0^{-1},q_1\big]\bigg\rgroup 
\bigg\lgroup \bar\xi_1\Pi^{\bar\lambda;\bar\eta i}\big[{\rm Q};b_0^{-1},q_1\big]\bigg\rgroup
\nonumber\\
&&
\hspace{1.5cm}
+
\bigg\lgroup \bar\xi_1\Pi^{\bar\lambda';\bar\eta'j}\big[{\rm Q}; b_0^{-1},q_1\big]\bigg\rgroup
\bigg\lgroup \frac{\xi_2}{\bar\xi_1}\Pi^{\bar\eta;\bar\lambda i}\big[q_1;b_0,{\rm Q}\big]-\xi_1\Pi^{\bar\lambda;\bar\eta i}\big[{\rm Q};b_0^{-1},q_1\big]\bigg\rgroup\;.
\eeq
Eq. \eqref{final_barq_gamma_barq_gamma} is the final result for the $\bar q\gamma-\bar q\gamma$ contribution to the correlation limit of the production cross section in the gluon channel. It is written in terms of two unpolarized TMDs: ${\cal F}^{(1)}_{gg}$, ${\cal F}^{(3)}_{gg}$ and their linearly polarized partners ${\cal H}^{(1)}_{gg}$ and ${\cal H}^{(3)}_{gg}$, as well as the adjoint dipole distribution ${\cal F}_{adj}$ for which the unpolarized and polarized versions coincide (as is the case for the fundamental dipole TMD ${\cal F}^{(1)}_{qg}$). This specific structure of the result is preserved for the other three contributions in the gluon channel, as we will see in the rest of the analysis. 

\subsubsection{$q\gamma-q\gamma$ contribution} 
Let us now study the contribution to the correlation limit of the production cross section in the gluon channel when the photon is emitted from the quark both in the amplitude and in the complex conjugate amplitude. In this contribution, the small parameters in which we expand are the transverse sizes of the dipole formed by the final quark and final photon ($r_q$) and the dipole formed by the intermediate quark (before the photon emission) and the final anti-quark ($\bar r_{\bar q}$) in the amplitude. $r'_q$ and $\bar r'_{\bar q}$ denote the transverse sizes of the same dipoles in the complex conjugate amplitude. These variables are defined as 
\beq
&&
r_q=z_1-z_2\; , \hspace{1cm} r'_q=y_1-y_2\;,\\
&&
\bar r_{\bar q}=v-z_3\; , \hspace{1.15cm} \bar r'_{\bar q}=v'-y_3\;.
\eeq
Performing this change of variables in Eq. \eqref{q_gamma_q_gamma_Cont} leads to the following expression for the $q\gamma-q\gamma$ contribution
\beq
\label{q_gamma_q_gamma_Before_expansion}
&&
{\rm I}_{q\gamma-q\gamma}=\int_{z_3y_3,r_qr'_q,\bar r_{\bar q}\bar r'_{\bar q} }
e^{i{\rm P}\cdot(y_1-z_1) -i{\rm K}\cdot(r'_q-r_q)-iq_3\cdot(\bar r'_{\bar q}-\bar r_{\bar q})} 8\, {\cal M}^{\bar\lambda\bar\lambda';\bar\eta\bar\eta'}_{qq}\bigg(\xi_1+\xi_2, \frac{\xi_2}{\xi_1+\xi_2}\bigg)
\nonumber\\
&&
\times
A^{\bar\lambda}(r_q)A^{\bar\lambda'}(r'_q)\, 
\tr\Big[
{\cal \bar{G}}^{\dagger c}_{q\gamma}(\xi_1,\xi_2; y_1,r'_q,\bar r'_{\bar q})\, {\cal \bar{G}}^{c}_{q\gamma}(\xi_1,\xi_2; z_1,r_q,\bar r_{\bar q})\Big]\;.
\eeq
Here, again we have performed the trivial integrations over $w$, $w'$, $v$ and $v'$. The newly introduced transverse momentum $\rm K$ is defined as 
\beq
\label{def:K}
{\rm K}=q_2+\frac{\xi_2}{\xi_1+\xi_2}q_3
\eeq
and the function ${\cal \bar{G}}^{c\bar \eta}_{q\gamma}$ encodes the Wilson line structure for the case when photon is emitted from the quark in the amplitude, and is obtained from ${\cal G}^{c\bar \eta}_{q\gamma}$ given in Eq.~\eqref{quark_photon}. After performing a leading-order Taylor expansion in powers of the small dipole sizes, it reads
\beq
\label{expanded_G_q_gamma}
{\cal \bar{G}}^{c\bar \eta}_{q\gamma}(\xi_1,\xi_2; z_1,r_q,\bar r_{\bar q})_{\alpha\beta}
&=& -\, (\xi_1+\xi_2) \, \bar r^i_{\bar q} \bigg[ {\cal A}^{\bar \eta}\bigg(1-\xi_1-\xi_2, \bar r_{\bar q}; \frac{\xi_2}{\xi_1+\xi_2},r_q\bigg)-A^{\bar\eta}(\bar r_{\bar q})\bigg]
\\
&&
\times\, 
 \big[(\partial^iS^\dagger_{z_1})t^cS_{z_1})\big]_{\alpha\beta}\nonumber\\
&&
+\, \bigg\{(1-\xi_1-\xi_2)\,\bar r^i_{\bar q} \, \bigg[{\cal A}^{\bar \eta}\bigg(1-\xi_1-\xi_2, \bar r_{\bar q}; \frac{\xi_2}{\xi_1+\xi_2},r_q\bigg)-A^{\bar\eta}(\bar r_{\bar q})\bigg] 
\nonumber\\
&&
\hspace{0.4cm}
+\, \frac{\xi_2}{\xi_1+\xi_2} \, r^i_q \, {\cal A}^{\bar \eta}\bigg(1-\xi_1-\xi_2, \bar r_{\bar q}; \frac{\xi_2}{\xi_1+\xi_2},r_q\bigg)
\bigg\} \big[S^\dagger_{z_1}t^c(\partial^iS_{z_1})\big]_{\alpha\beta}\;.\nonumber
\eeq
As in the case of the $\bar q\gamma-\bar q\gamma$ contribution, our next step is to plug in the above expansion into Eq. \eqref{q_gamma_q_gamma_Before_expansion} and trace over the fundamental color indices. After using the Fierz identity and performing some color algebra, the $q\gamma-q\gamma$ contribution reads
\beq
&&
{\rm I}_{q\gamma-q\gamma}=\int_{z_1y_1,r_qr'_q,\bar r_{\bar q}\bar r'_{\bar q}} e^{i{\rm P}\cdot(y_1-z_1)-i{\rm K}\cdot(r'_q-r_q)-iq_3\cdot(\bar r'_{\bar q}-\bar r_{\bar q})} \, 8\, {\cal M}^{\bar\lambda\bar\lambda';\bar\eta\bar\eta'}_{qq}\bigg(\xi_1+\xi_2,\frac{\xi_2}{\xi_1+\xi_2}\bigg)
\nonumber\\
&&
\times \,
A^{\bar\lambda}(r_q)A^{\bar\lambda'}(r'_q)
\bigg\{ 
\bigg\lgroup 
(\xi_1+\xi_2)\bar r'^j_{\bar q}
\bigg[ {\cal A}^{\bar\eta'}\bigg(1-\xi_1-\xi_2, \bar r'_{\bar q}; \frac{\xi_2}{\xi_1+\xi_2}, r'_q\bigg)-A^{\bar\eta'}(\bar r'_{\bar q})\bigg]
\bigg\rgroup
\nonumber\\
&&
\times
\bigg\lgroup 
(\xi_1+\xi_2)\bar r^i_{\bar q}\bigg[ {\cal A}^{\bar\eta}\bigg(1-\xi_1-\xi_2, \bar r_{\bar q}; \frac{\xi_2}{\xi_1+\xi_2}, r_q\bigg)-A^{\bar\eta}(\bar r_{\bar q})\bigg] \bigg\rgroup
\frac{1}{2}\tr\big[S_{z_1}S^{\dagger}_{y_1}\big] \tr\big[ (\partial^jS_{y_1})(\partial^iS^\dagger_{z_1})\big]
\nonumber
\\
&&
+
\bigg\lgroup 
(1-\xi_1-\xi_2) \, \bar r'^j_{\bar q} \,  \bigg[ {\cal A}^{\bar\eta'}\bigg(1-\xi_1-\xi_2, \bar r'_{\bar q}; \frac{\xi_2}{\xi_1+\xi_2},r'_q\bigg)-A^{\bar\eta'}(\bar r'_{\bar q})\bigg]
\nonumber
\\
&&
\hspace{0.5cm}
+ \, \frac{\xi_2}{\xi_1+\xi_2} \, r'^j_q \, {\cal A}^{\bar\eta'}\bigg(1-\xi_1-\xi_2,\bar r'_{\bar q}; \frac{\xi_2}{\xi_1+\xi_2}, r'_q\bigg)
\bigg\rgroup
\nonumber
\\
&&
\times
\bigg\lgroup
(1-\xi_1-\xi_2) \, \bar r^i_{\bar q} \bigg[ {\cal A}^{\bar\eta}\bigg(1-\xi_1-\xi_2, \bar r_{\bar q}; \frac{\xi_2}{\xi_1+\xi_2},r_q\bigg)-A^{\bar\eta}(\bar r_{\bar q})\bigg]
\nonumber
\\
&&
\hspace{0.5cm}
+ \, \frac{\xi_2}{\xi_1+\xi_2} \, r^i_q \, {\cal A}^{\bar\eta}\bigg(1-\xi_1-\xi_2,\bar r_{\bar q}; \frac{\xi_2}{\xi_1+\xi_2}, r_q\bigg)
\bigg\rgroup
\frac{1}{2}\tr\big[S^\dagger_{z_1}S_{y_1}\big] \tr\big[ (\partial^iS_{z_1})(\partial^jS^\dagger_{y_1})\big]
\nonumber
\\
&&
-\bigg\lgroup
(\xi_1+\xi_2)\, \bar r'^j_{\bar q}\, \bigg[ {\cal A}^{\bar\eta'}\bigg(1-\xi_1-\xi_2, \bar r'_{\bar q}; \frac{\xi_2}{\xi_1+\xi_2}, r'_q\bigg)-A^{\bar\eta'}(\bar r'_{\bar q})\bigg] \bigg\rgroup
\nonumber\\
&&
\times
\bigg\lgroup
(1-\xi_1-\xi_2) \, \bar r^i_{\bar q} \bigg[ {\cal A}^{\bar\eta}\bigg(1-\xi_1-\xi_2, \bar r_{\bar q}; \frac{\xi_2}{\xi_1+\xi_2},r_q\bigg)-A^{\bar\eta}(\bar r_{\bar q})\bigg]
\nonumber
\\
&&
\hspace{0.5cm}
+ \, \frac{\xi_2}{\xi_1+\xi_2} \, r^i_q \, {\cal A}^{\bar\eta}\bigg(1-\xi_1-\xi_2,\bar r_{\bar q}; \frac{\xi_2}{\xi_1+\xi_2}, r_q\bigg)
\bigg\rgroup
\frac{1}{2} \tr\big[(\partial^iS_{z_1})S^\dagger_{y_1}\big] \tr\big[(\partial^jS_{y_1})S^\dagger_{z_1}\big]
\nonumber\\
&&
-
\bigg\lgroup 
(1-\xi_1-\xi_2) \, \bar r'^j_{\bar q} \,  \bigg[ {\cal A}^{\bar\eta'}\bigg(1-\xi_1-\xi_2, \bar r'_{\bar q}; \frac{\xi_2}{\xi_1+\xi_2},r'_q\bigg)-A^{\bar\eta'}(\bar r'_{\bar q})\bigg]
\nonumber
\\
&&
\hspace{0.5cm}
+ \, \frac{\xi_2}{\xi_1+\xi_2} \, r'^j_q \, {\cal A}^{\bar\eta'}\bigg(1-\xi_1-\xi_2,\bar r'_{\bar q}; \frac{\xi_2}{\xi_1+\xi_2}, r'_q\bigg)
\bigg\rgroup
\nonumber\\
&&
\times
\bigg\lgroup 
(\xi_1+\xi_2)\bar r^i_{\bar q}\bigg[ {\cal A}^{\bar\eta}\bigg(1-\xi_1-\xi_2, \bar r_{\bar q}; \frac{\xi_2}{\xi_1+\xi_2}, r_q\bigg)-A^{\bar\eta}(\bar r_{\bar q})\bigg] \bigg\rgroup
\frac{1}{2} \tr\big[S_{z_1}(\partial^jS^\dagger_{y_1})\big] \tr\big[(\partial^iS^\dagger_{z_1})S_{y_1}\big]\nonumber\\
&&
+
\bigg\lgroup 
\bar r'^j_{\bar q} \bigg[ {\cal A}^{\bar\eta'}\bigg(1-\xi_1-\xi_2, \bar r'_{\bar q}; \frac{\xi_2}{\xi_1+\xi_2}, r'_q\bigg)-A^{\bar\eta'}(\bar r'_{\bar q})\bigg] 
\nonumber\\
&&
\hspace{0.5cm}
+\; 
\frac{\xi_2}{\xi_1+\xi_2} \, r'^j_q \, {\cal A}^{\bar\eta'}\bigg(1-\xi_1-\xi_2,\bar r'_{\bar q}; \frac{\xi_2}{\xi_1+\xi_2}, r'_q\bigg) 
\bigg\rgroup
\nonumber\\
&&
\times
\bigg\lgroup 
\bar r^i_{\bar q} \bigg[ {\cal A}^{\bar\eta}\bigg(1-\xi_1-\xi_2, \bar r_{\bar q}; \frac{\xi_2}{\xi_1+\xi_2}, r_q\bigg)-A^{\bar\eta}(\bar r_{\bar q})\bigg] 
\nonumber\\
&&
\hspace{0.5cm}
+\; 
\frac{\xi_2}{\xi_1+\xi_2} \, r^i_q \, {\cal A}^{\bar\eta}\bigg(1-\xi_1-\xi_2,\bar r_{\bar q}; \frac{\xi_2}{\xi_1+\xi_2}, r_q\bigg) 
\bigg\rgroup
\frac{1}{2N_c} \tr\big[(\partial^iS_{z_1})S^\dagger_{y_1}(\partial^jS_{y_1})S^\dagger_{z_1}\big]\bigg\}\;.
\eeq
The integrals over the small dipole sizes can be performed using the generic integrals given in Eq. \eqref{integral_1} and \eqref{integral_2}, and the resulting expression reads:
\beq
\label{q_gamma_q_gamma_before_TMDs}
&&
{\rm I}_{q\gamma-q\gamma}= 8\, {\cal M}^{\bar\lambda\bar\lambda';\bar\eta\bar\eta'}_{qq}\bigg(\xi_1+\xi_2, \frac{\xi_2}{\xi_1+\xi_2}\bigg) \int_{z_1y_1}e^{i{\rm P}\cdot(z_1-y_1)}
\nonumber\\
&&
\hspace{-0.3cm}
\times
\bigg\{\!\!
\bigg\lgroup (\xi_1+\xi_2) \Pi^{\bar\lambda';\bar\eta'j}\big[{\rm K}; b_1^{-1},q_3\big]\bigg\rgroup
\bigg\lgroup (\xi_1+\xi_2) \Pi^{\bar\lambda;\bar\eta i}\big[{\rm K}; b_1^{-1},q_3\big]\bigg\rgroup
\nonumber\\
&&
\times
\frac{1}{2} \tr\big[S_{z_1}S^\dagger_{y_1}\big] \tr\big[(\partial^jS_{y_1})(\partial^iS^\dagger_{z_1})\big]
\nonumber\\
&&
+
\bigg\lgroup 
\frac{\xi_2}{\xi_1+\xi_2}\Pi^{\bar\eta';\bar\lambda'j}\big[q_3; b_1, {\rm K}\big]
-(1-\xi_1-\xi_2)\Pi^{\bar\lambda';\bar\eta'j}\big[{\rm K}; b_1^{-1},q_3\big] 
\bigg\rgroup
\nonumber
\\
&&
\times
\bigg\lgroup 
\frac{\xi_2}{\xi_1+\xi_2} \Pi^{\bar\eta;\bar\lambda i}\big[q_3; b_1, {\rm K}\big]
-(1-\xi_1-\xi_2)\Pi^{\bar\lambda; \bar\eta i}\big[{\rm K}; b_1^{-1},q_3\big]
\bigg\rgroup
\nonumber\\
&&
\times
\frac{1}{2} \tr\big[S^\dagger_{z_1}S_{y_1}\big] \tr\big[(\partial^iS_{z_1})(\partial^jS^\dagger_{y_1})\big]
\nonumber\\
&&
+
\bigg\lgroup (\xi_1+\xi_2) \Pi^{\bar\lambda';\bar\eta'j}\big[{\rm K}; b_1^{-1},q_3\big]\bigg\rgroup\nonumber\\
&&\times
\bigg\lgroup 
\frac{\xi_2}{\xi_1+\xi_2} \Pi^{\bar\eta;\bar\lambda i}\big[q_3; b_1, {\rm K}\big]
-(1-\xi_1-\xi_2)\Pi^{\bar\lambda; \bar\eta i}\big[{\rm K}; b_1^{-1},q_3\big]
\bigg\rgroup
\nonumber\\
&&
\times
\frac{1}{2} \tr\big[(\partial^iS_{z_1})S^\dagger_{y_1}\big] \tr\big[(\partial^jS_{y_1})S^\dagger_{z_1}\big]
\nonumber\\
&&
+
\bigg\lgroup 
\frac{\xi_2}{\xi_1+\xi_2}\Pi^{\bar\eta';\bar\lambda'j}\big[q_3; b_1, {\rm K}\big]
-(1-\xi_1-\xi_2)\Pi^{\bar\lambda';\bar\eta'j}\big[{\rm K}; b_1^{-1},q_3\big] 
\bigg\rgroup
\nonumber\\
&&
\times
\bigg\lgroup (\xi_1+\xi_2) \Pi^{\bar\lambda;\bar\eta i}\big[{\rm K}; b_1^{-1},q_3\big]\bigg\rgroup\frac{1}{2} \tr\big[S_{z_1}(\partial^jS^\dagger_{y_1})\big] \tr\big[(\partial^iS^\dagger_{z_1}S_{y_1})\big]
\nonumber\\
&&
+
\bigg\lgroup 
\frac{\xi_2}{\xi_1+\xi_2}\Pi^{\bar\eta';\bar\lambda'j}\big[q_3; b_1, {\rm K}\big]
-\Pi^{\bar\lambda';\bar\eta'j}\big[{\rm K}; b_1^{-1},q_3\big] 
\bigg\rgroup
\\
&&
\times
\bigg\lgroup 
\frac{\xi_2}{\xi_1+\xi_2} \Pi^{\bar\eta;\bar\lambda i}\big[q_3; b_1, {\rm K}\big]
-\Pi^{\bar\lambda; \bar\eta i}\big[{\rm K}; b_1^{-1},q_3\big]
\bigg\rgroup
\frac{1}{2N_c} \tr\big[(\partial^iS_{z_1})S^\dagger_{y_1}(\partial^jS_{y_1})S^\dagger_{z_1}\big] \bigg\}\;,\nonumber
\eeq
with 
\beq
b_1=\frac{1}{(1-\xi_1-\xi_2)}\, \frac{\xi_2}{\xi_1+\xi_2} \, \bigg(1-\frac{\xi_2}{\xi_1+\xi_2}\bigg)\;.
\eeq
In order to integrate over the remaining transverse coordinates $y_1$ and $z_1$, we perform the same manipulations that we introduced for the calculation of the $\bar q \gamma-\bar q\gamma$ contribution. After target averaging, the first two terms of Eq. \eqref{q_gamma_q_gamma_before_TMDs} can then be written as 
\beq
\label{q_gamma_q_gamma_12_Final}
&&
\Big\langle{\rm I}_{q\gamma-q\gamma}^{(1,2)}\Big\rangle_{x_A}={\cal M}_{qq}^{\bar\lambda\bar\lambda';\bar\eta\bar\eta'}\bigg(\xi_1+\xi_2, \frac{\xi_2}{\xi_1+\xi_2}\bigg) g_s^2(2\pi)^3N_c
\nonumber\\
&&
\times
\frac{1}{2}
\bigg\{ 
\bigg\lgroup 
\frac{\xi_2}{\xi_1+\xi_2}\Pi^{\bar\eta';\bar\lambda'j}\big[q_3; b_1, {\rm K}\big]
-(1-\xi_1-\xi_2)\Pi^{\bar\lambda';\bar\eta'j}\big[{\rm K}; b_1^{-1},q_3\big] 
\bigg\rgroup
\nonumber
\\
&&
\times
\bigg\lgroup 
\frac{\xi_2}{\xi_1+\xi_2} \Pi^{\bar\eta;\bar\lambda i}\big[q_3; b_1, {\rm K}\big]
-(1-\xi_1-\xi_2)\Pi^{\bar\lambda; \bar\eta i}\big[{\rm K}; b_1^{-1},q_3\big]
\bigg\rgroup
\nonumber\\
&&
+
\bigg\lgroup (\xi_1+\xi_2) \Pi^{\bar\lambda';\bar\eta'j}\big[{\rm K}; b_1^{-1},q_3\big]\bigg\rgroup
\bigg\lgroup (\xi_1+\xi_2) \Pi^{\bar\lambda;\bar\eta i}\big[{\rm K}; b_1^{-1},q_3\big]\bigg\rgroup
\bigg\}
\nonumber\\
&&
\times
\bigg\{
\frac{1}{2}\delta^{ij}\bigg[{\cal F}^{(1)}_{gg}(x_A, {\rm P})+{\cal F}^{(1)}_{gg}(x_A, -{\rm P})\bigg]\nonumber\\
&&-\frac{1}{2}\bigg(\delta^{ij}-2\frac{\rm P^i\rm P^j}{\rm P^2}\bigg)
\bigg[{\cal H}^{(1)}_{gg}(x_A,{\rm P})+{\cal H}^{(1)}_{gg}(x_A,-{\rm P})\bigg]\bigg\}\;,
\eeq
where we have used the $i\leftrightarrow j$ symmetry of the momentum structures as well as the definition of the gluon TMD Eq. \eqref{def:TMD_1}. Similarly, the third and the fourth terms of Eq. \eqref{q_gamma_q_gamma_before_TMDs} give:
\beq
\label{q_gamma_q_gamma_34_Final}
&&
\Big\langle{\rm I}_{q\gamma-q\gamma}^{(3,4)}\Big\rangle_{x_A}={\cal M}_{qq}^{\bar\lambda\bar\lambda';\bar\eta\bar\eta'}\bigg(\xi_1+\xi_2, \frac{\xi_2}{\xi_1+\xi_2}\bigg) g_s^2 (2\pi)^3 N_c
\nonumber\\
&&
\times
\frac{1}{2}\bigg\{
\bigg\lgroup (\xi_1+\xi_2) \Pi^{\bar\lambda';\bar\eta'j}\big[{\rm K}; b_1^{-1},q_3\big]\bigg\rgroup
\nonumber\\
&&
\hspace{0.6cm}
\times
\bigg\lgroup 
\frac{\xi_2}{\xi_1+\xi_2} \Pi^{\bar\eta;\bar\lambda i}\big[q_3; b_1, {\rm K}\big]
-(1-\xi_1-\xi_2)\Pi^{\bar\lambda; \bar\eta i}\big[{\rm K}; b_1^{-1},q_3\big]
\bigg\rgroup
\nonumber\\
&&
\hspace{0.6cm}
+
\bigg\lgroup 
\frac{\xi_2}{\xi_1+\xi_2}\Pi^{\bar\eta';\bar\lambda'j}\big[q_3; b_1, {\rm K}\big]
-(1-\xi_1-\xi_2)\Pi^{\bar\lambda';\bar\eta'j}\big[{\rm K}; b_1^{-1},q_3\big] 
\bigg\rgroup
\nonumber\\
&&
\hspace{0.6cm}
\times
\bigg\lgroup (\xi_1+\xi_2) \Pi^{\bar\lambda;\bar\eta i}\big[{\rm K}; b_1^{-1},q_3\big]\bigg\rgroup
\bigg\}
\nonumber
\\
&&
\times
\bigg\{
\frac{1}{2}\delta^{ij}
\bigg[2{\cal F}_{adj}(x_A,{\rm P})- {\cal F}^{(1)}_{gg}(x_A, {\rm P})- {\cal F}^{(1)}_{gg}(x_A, -{\rm P})\bigg]
\nonumber\\
&&
\hspace{0.4cm}
-\frac{1}{2}
\bigg(\delta^{ij}-2\frac{\rm P^i\rm P^j}{\rm P^2}\bigg)
\bigg[2{\cal F}_{adj}(x_A,{\rm P}) -{\cal H}^{(1)}_{gg}(x_A,{\rm P}) -{\cal H}^{(1)}_{gg}(x_A,-{\rm P}) \bigg] \bigg\}\;.
\eeq
Again we have used the symmetry argument as in the case of similar terms in the $\bar q\gamma-\bar q\gamma$ contribution, and written the final result in terms of the adjoint dipole distribution using Eqs. \eqref{def:F_adj}, \eqref{F_in_adj} and \eqref{H_in_adj}. Finally, we can write the last term of Eq. \eqref{q_gamma_q_gamma_before_TMDs}, using the definition of the gluon TMD Eq. \eqref{def:TMD_3}, as
\beq
\label{q_gamma_q_gamma_5_Final}
\Big\langle{\rm I}_{q\gamma-q\gamma}^{(5)}\Big\rangle_{x_A}&=& {\cal M}_{qq}^{\bar\lambda\bar\lambda';\bar\eta\bar\eta'}\bigg(\xi_1+\xi_2, \frac{\xi_2}{\xi_1+\xi_2}\bigg) (-1) g_s^2 (2\pi)^3 \frac{1}{N_c}
\nonumber\\
&&
\times
\bigg\lgroup 
\frac{\xi_2}{\xi_1+\xi_2}\Pi^{\bar\eta';\bar\lambda'j}\big[q_3; b_1, {\rm K}\big]
-\Pi^{\bar\lambda';\bar\eta'j}\big[{\rm K}; b_1^{-1},q_3\big] 
\bigg\rgroup
\nonumber\\
&&
\times
\bigg\lgroup 
\frac{\xi_2}{\xi_1+\xi_2} \Pi^{\bar\eta;\bar\lambda i}\big[q_3; b_1, {\rm K}\big]
-\Pi^{\bar\lambda; \bar\eta i}\big[{\rm K}; b_1^{-1},q_3\big]
\bigg\rgroup
\nonumber\\
&&
\times
\bigg[\frac{1}{2}\delta^{ij}{\cal F}^{(3)}_{gg}(x_A, {\rm P})
-\frac{1}{2}\bigg(\delta^{ij}-2\frac{\rm P^i\rm P^j}{\rm P^2}\bigg){\cal H}^{(3)}_{gg}(x_A, {\rm P})\bigg]\;.
\eeq
We can now put Eqs. \eqref{q_gamma_q_gamma_12_Final}, \eqref{q_gamma_q_gamma_34_Final} and \eqref{q_gamma_q_gamma_5_Final} together, to write down the final expression for the $q\gamma-q\gamma$ contribution to the correlation limit of the production cross section in the gluon channel:
\beq
\label{final_q_gamma_q_gamma}
\Big\langle{\rm I}_{q\gamma-q\gamma}\Big\rangle_{x_A}&=&{\cal M}_{qq}^{\bar\lambda\bar\lambda';\bar\eta\bar\eta'}\bigg(\xi_1+\xi_2, \frac{\xi_2}{\xi_1+\xi_2}\bigg) g_s^2 (2\pi)^3 N_c
\nonumber\\
&&
\times
\bigg\{
\Big[ {\rm H}^{(1)}_{gg}\Big]^{\bar\lambda\bar\lambda';\bar\eta\bar\eta';ij}_{ q\gamma- q\gamma}
\bigg\lgroup \frac{1}{2}\delta^{ij}\bigg[ \frac{1}{2}{\cal F}^{(1)}_{gg}(x_A,{\rm P})+\frac{1}{2}{\cal F}^{(1)}_{gg}(x_A,-{\rm P})-\frac{1}{N_c^2}{\cal F}^{(3)}_{gg}(x_A,{\rm P})\bigg]
\nonumber\\
&&
\hspace{0.5cm}
-
\frac{1}{2}\bigg(\delta^{ij}-2\frac{\rm P^i\rm P^j}{\rm P^2}\bigg)
\bigg[ \frac{1}{2}{\cal H}^{(1)}_{gg}(x_A,{\rm P})+\frac{1}{2}{\cal H}^{(1)}_{gg}(x_A,-{\rm P})-\frac{1}{N_c^2}{\cal H}^{(3)}_{gg}(x_A,{\rm P})\bigg]\bigg\rgroup
\nonumber\\
&&
\hspace{0.4cm}
+\Big[ {\rm H}^{(adj)}_{gg}\Big]^{\bar\lambda\bar\lambda';\bar\eta\bar\eta';ij}_{ q\gamma- q\gamma}\; \frac{\rm P^i\rm P^j}{\rm P^2} \; {\cal F}_{adj}(x_A,{\rm P})\bigg\}\;,
\eeq
with the hard parts
\beq
\label{hard_part_1_q_gamma_q_gamma}
\Big[ {\rm H}^{(1)}_{gg}\Big]^{\bar\lambda\bar\lambda';\bar\eta\bar\eta';ij}_{ q\gamma- q\gamma}&=&
\bigg\lgroup 
\frac{\xi_2}{\xi_1+\xi_2}\Pi^{\bar\eta';\bar\lambda'j}\big[q_3; b_1, {\rm K}\big]
-\Pi^{\bar\lambda';\bar\eta'j}\big[{\rm K}; b_1^{-1},q_3\big] 
\bigg\rgroup
\nonumber\\
&
\times&
\bigg\lgroup 
\frac{\xi_2}{\xi_1+\xi_2} \Pi^{\bar\eta;\bar\lambda i}\big[q_3; b_1, {\rm K}\big]
-\Pi^{\bar\lambda; \bar\eta i}\big[{\rm K}; b_1^{-1},q_3\big]
\bigg\rgroup\;,
\eeq
and 
\beq
\label{hard_part_adj_q_gamma_q_gamma}
\Big[ {\rm H}^{(adj)}_{gg}\Big]^{\bar\lambda\bar\lambda';\bar\eta\bar\eta';ij}_{ q\gamma- q\gamma}=&&
\bigg\lgroup (\xi_1+\xi_2) \Pi^{\bar\lambda';\bar\eta'j}\big[{\rm K}; b_1^{-1},q_3\big]\bigg\rgroup
\nonumber\\
&&
\times
\bigg\lgroup 
\frac{\xi_2}{\xi_1+\xi_2} \Pi^{\bar\eta;\bar\lambda i}\big[q_3; b_1, {\rm K}\big]
-(1-\xi_1-\xi_2)\Pi^{\bar\lambda; \bar\eta i}\big[{\rm K}; b_1^{-1},q_3\big]
\bigg\rgroup
\nonumber\\
&&
+
\bigg\lgroup 
\frac{\xi_2}{\xi_1+\xi_2}\Pi^{\bar\eta';\bar\lambda'j}\big[q_3; b_1, {\rm K}\big]
-(1-\xi_1-\xi_2)\Pi^{\bar\lambda';\bar\eta'j}\big[{\rm K}; b_1^{-1},q_3\big] 
\bigg\rgroup
\nonumber\\
&&
\times
\bigg\lgroup (\xi_1+\xi_2) \Pi^{\bar\lambda;\bar\eta i}\big[{\rm K}; b_1^{-1},q_3\big]\bigg\rgroup\;.
\eeq
Before we conclude this subsection, we would like to emphasize that the TMD structures of the $\bar q\gamma-\bar q\gamma$ and $q\gamma-q\gamma$ contributions have exactly the same form. The difference between these two contributions only appears in the definition of the hard parts, and in the function that accounts for the product of the splitting amplitudes: ${\cal M}_g^{\bar\lambda\bar\lambda'; \bar\eta\bar\eta'}$. Moreover, the structure of hard parts in both contributions exhibit a similar pattern, the difference lying in the transverse momentum dependence of the pseudo-projectors and in the dependences on the longitudinal momentum ratios. This can be easily observed by comparing Eqs. \eqref{final_barq_gamma_barq_gamma}, \eqref{hard_part_1_barq_gamma_barq_gamma} and \eqref{hard_part_adj_barq_gamma_barq_gamma} with Eqs. \eqref{final_q_gamma_q_gamma}, \eqref{hard_part_1_q_gamma_q_gamma} and \eqref{hard_part_adj_q_gamma_q_gamma}.

\subsubsection{Crossed contributions \rm{(}${\bar q\gamma-q \gamma}$ and ${q\gamma-\bar q\gamma}${\rm )}} 
The remaining two contributions that we need to consider are the crossed ones: photon emission from the antiquark in the amplitude with photon emission from the quark in the complex conjugate amplitude, and vice versa. 

Let us start with the $\bar q\gamma-q\gamma$ contribution. As is obvious from the last two contributions, the small dipole sizes for this crossed contribution are
\beq
&&
r_q=z_1-z_2 \; , \hspace{1cm} r'_q=v'-y_1\;,\\
&&
\bar r_{\bar q}=v-z_3\; , \hspace{1.2cm} \bar r'_{\bar q}=y_3-y_2\;.
\eeq
After performing the above change of variables, Eq. \eqref{bar_q_gamma_q_gamma_Cont} can be written as 
\beq
&&
{\rm I}_{\bar q\gamma-q\gamma}=-\int_{z_1y_3, r_qr'_q, \bar r_{\bar q} \bar r'_{\bar q}} e^{i{\rm P}\cdot(y_3-z_1)-i\big[{\rm Q}\cdot \bar r'_{\bar q}-q_3\cdot\bar r_{\bar q}\big] -i\big[ q_1\cdot r'_q-{\rm K}\cdot r_q\big]} 
\, (-8)\, \widetilde{\cal M}^{\bar\lambda\bar\lambda';\bar\eta\bar\eta'}_g(\xi_1,\xi_2)
\nonumber\\
&&
\times
A^{\bar\lambda'}(\bar r'_{\bar q})A^{\bar\lambda}(r_q)
\tr\Big[ {\cal \bar{G}}^{\dagger c\bar\eta'}_{\bar q\gamma}(\xi_1,\xi_2; y_3,r'_q,\bar r'_{\bar q}) \,
{\cal \bar{G}}^{c\eta}_{q\gamma}(\xi_1,\xi_2;z_1,r_q,\bar r_{\bar q}) \Big]\;,
\eeq
where we performed the trivial integrations over $w, w', v$ and $v'$, and where the transverse momenta $\rm Q$ and $\rm K$ are defined in Eqs.~\eqref{def:Q} and \eqref{def:K}, respectively. The Taylor expansions for ${\cal \bar{G}}^{c\bar \eta}_{\bar q\gamma}$ and ${\cal \bar{G}}^{c\bar \eta}_{q\gamma}$ are already known (Eqs. \eqref{expanded_G_bar_q_gamma} and \eqref{expanded_G_q_gamma}), hence we can plug them into the above expression to calculate the first crossed contribution in the gluon channel, which reads after some color algebra:
\beq
&&
{\rm I}_{\bar q\gamma-q\gamma}=\int_{z_1y_3, r_qr'_q, \bar r_{\bar q} \bar r'_{\bar q}} e^{i{\rm P}\cdot(y_3-z_1)-i\big[{\rm Q}\cdot \bar r'_{\bar q}-q_3\cdot\bar r_{\bar q}\big] -i\big[ q_1\cdot r'_q-{\rm K}\cdot r_q\big]} 
\, (-8)\, \widetilde{\cal M}^{\bar\lambda\bar\lambda';\bar\eta\bar\eta'}_g(\xi_1,\xi_2)
A^{\bar\lambda'}(\bar r'_{\bar q})A^{\bar\lambda}(r_q)
\nonumber\\
&&
\hspace{-0.4cm}
\times
\bigg\{
\bigg\lgroup 
\frac{\xi_2}{\bar\xi_1}\bar r'^j_{\bar q} {\cal A}^{\bar\eta'}\bigg(\xi_1,r'_q;\frac{\xi_2}{\bar\xi_1},\bar r'_{\bar q}\bigg)
+\xi_1r'^j_q  \bigg[ {\cal A}^{\bar\eta'}\bigg(\xi_1,r'_q;\frac{\xi_2}{\bar\xi_1},\bar r'_{\bar q}\bigg) -A^{\bar\eta'}(r'_q)\bigg]
\bigg\rgroup
\nonumber\\
&&
\times
\bigg\lgroup
(\xi_1+\xi_2)\bar r^i_{\bar q}\bigg[ {\cal A}^{\bar\eta}\bigg(1-\xi_1-\xi_2,\bar r_{\bar q}; \frac{\xi_2}{\xi_1+\xi_2},r_q\bigg) -A^{\bar\eta}(\bar r_{\bar q})\bigg]
\bigg\rgroup
\nonumber\\
&&
\times
\frac{1}{2} \tr\big[S_{z_1}S^\dagger_{y_3}\big] \tr\big[(\partial^jS_{y_3})(\partial^iS^\dagger_{z_1})\big]
\nonumber\\
&&
+
\bigg\lgroup 
\bar \xi_1 r'^j_q \bigg[ {\cal A}^{\bar\eta'}\bigg(\xi_1,r'_q;\frac{\xi_2}{\bar\xi_1},\bar r'_{\bar q}\bigg) -A^{\bar\eta'}(r'_q)\bigg]
\bigg\rgroup
\nonumber\\
&&
\times
\bigg\lgroup
\frac{\xi_2}{\xi_1+\xi_2}r^i_q {\cal A}^{\bar \eta}\bigg(1-\xi_1-\xi_2,\bar r_{\bar q}; \frac{\xi_2}{\xi_1+\xi_2},r_q\bigg)
\nonumber\\
&&
\hspace{0.5cm}
+(1-\xi_1-\xi_2) \bar r^i_{\bar q}
\bigg[ {\cal A}^{\bar\eta}\bigg(1-\xi_1-\xi_2,\bar r_{\bar q}; \frac{\xi_2}{\xi_1+\xi_2},r_q\bigg) -A^{\bar\eta}(\bar r_{\bar q})\bigg]
\bigg\rgroup
\nonumber\\
&&
\times
\frac{1}{2} \tr\big[S_{y_3}S^\dagger_{z_1}\big] \tr\big[(\partial^iS_{z_1})(\partial^j S^\dagger_{y_3})\big]
\nonumber\\
&&
-
\bigg\lgroup 
\frac{\xi_2}{\bar\xi_1}\bar r'^j_{\bar q} {\cal A}^{\bar\eta'}\bigg(\xi_1,r'_q;\frac{\xi_2}{\bar\xi_1},\bar r'_{\bar q}\bigg)
+\xi_1r'^j_q  \bigg[ {\cal A}^{\bar\eta'}\bigg(\xi_1,r'_q;\frac{\xi_2}{\bar\xi_1},\bar r'_{\bar q}\bigg) -A^{\bar\eta'}(r'_q)\bigg]
\bigg\rgroup
\nonumber\\
&&
\times
\bigg\lgroup
\frac{\xi_2}{\xi_1+\xi_2}r^i_q {\cal A}^{\bar \eta}\bigg(1-\xi_1-\xi_2,\bar r_{\bar q}; \frac{\xi_2}{\xi_1+\xi_2},r_q\bigg)
\nonumber\\
&&
\hspace{0.5cm}
+(1-\xi_1-\xi_2) \bar r^i_{\bar q}
\bigg[ {\cal A}^{\bar\eta}\bigg(1-\xi_1-\xi_2,\bar r_{\bar q}; \frac{\xi_2}{\xi_1+\xi_2},r_q\bigg) -A^{\bar\eta}(\bar r_{\bar q})\bigg]
\bigg\rgroup
\nonumber\\
&&
\times
\frac{1}{2} \tr\big[(\partial^iS_{z_1})S^\dagger_{y_3}\big] \tr\big[(\partial^jS_{y_3})S^\dagger_{z_1}\big]
\nonumber\\
&&
-
\bigg\lgroup 
\bar \xi_1 r'^j_q \bigg[ {\cal A}^{\bar\eta'}\bigg(\xi_1,r'_q;\frac{\xi_2}{\bar\xi_1},\bar r'_{\bar q}\bigg) -A^{\bar\eta'}(r'_q)\bigg]
\bigg\rgroup
\nonumber\\
&&
\times
\bigg\lgroup
(\xi_1+\xi_2)\bar r^i_{\bar q}\bigg[ {\cal A}^{\bar\eta}\bigg(1-\xi_1-\xi_2,\bar r_{\bar q}; \frac{\xi_2}{\xi_1+\xi_2},r_q\bigg) -A^{\bar\eta}(\bar r_{\bar q})\bigg]
\bigg\rgroup
\nonumber\\
&&
\times
\frac{1}{2} \tr\big[S_{y_3}(\partial^iS^\dagger_{z_1})\big] \tr\big[S_{z_1}(\partial^jS^\dagger_{y_3})\big]
\nonumber\\
&&
+
\bigg\lgroup 
\frac{\xi_2}{\bar\xi_1}\bar r'^j_{\bar q} {\cal A}^{\bar\eta'}\bigg(\xi_1,r'_q;\frac{\xi_2}{\bar\xi_1},\bar r'_{\bar q}\bigg)
+r'^j_q  \bigg[ {\cal A}^{\bar\eta'}\bigg(\xi_1,r'_q;\frac{\xi_2}{\bar\xi_1},\bar r'_{\bar q}\bigg) -A^{\bar\eta'}(r'_q)\bigg]
\bigg\rgroup
\nonumber\\
&&
\times
\bigg\lgroup
\frac{\xi_2}{\xi_1+\xi_2}r^i_q {\cal A}^{\bar \eta}\bigg(1-\xi_1-\xi_2,\bar r_{\bar q}; \frac{\xi_2}{\xi_1+\xi_2},r_q\bigg)\nonumber\\
&&
+ 
\bar r^i_{\bar q}
\bigg[ {\cal A}^{\bar\eta}\bigg(1-\xi_1-\xi_2,\bar r_{\bar q}; \frac{\xi_2}{\xi_1+\xi_2},r_q\bigg) -A^{\bar\eta}(\bar r_{\bar q})\bigg]
\bigg\rgroup
\nonumber\\
&&
\times
\frac{1}{2N_c} \tr\big[(\partial^iS_{z_1})S^\dagger_{y_3}(\partial^jS_{y_3})S^\dagger_{z_1}\big]\bigg\}\;.
\eeq
We again use the generic integrals Eqs. \eqref{integral_1} and \eqref{integral_2} to integrate over the dipole sizes, which leads to
\beq
\label{bar_q_gamma_q_gamma_aft_integration}
&&
{\rm I}_{\bar q\gamma-q\gamma}=8\, \widetilde{\cal M}^{\bar\lambda\bar\lambda';\bar\eta\bar\eta'}_g(\xi_1,\xi_2)
\int_{z_1y_3}e^{i{\rm P}\cdot(y_3-z_1)}
\nonumber\\
&&
\times
\bigg\{
\bigg\lgroup
\frac{\xi_2}{\bar\xi_1}\Pi^{\bar\eta';\bar\lambda'j}\big[q_1;b_0,{\rm Q}\big]-\xi_1\Pi^{\bar\lambda';\bar\eta'j}\big[{\rm Q}; b_0^{-1},q_1\big]
\bigg\rgroup
\bigg\lgroup
(\xi_1+\xi_2)\Pi^{\bar\lambda;\bar\eta i}\big[{\rm K}; b_1^{-1},q_3\big]
\bigg\rgroup
\nonumber\\
&&
\times
\frac{1}{2} \tr\big[S_{z_1}S^\dagger_{y_3}\big] \tr\big[(\partial^jS_{y_3})(\partial^iS^\dagger_{z_1})\big]
\nonumber\\
&&
+
\bigg\lgroup
\bar\xi_1\Pi^{\bar\lambda';\bar\eta'j}\big[{\rm Q};b_0^{-1},q_1\big] 
\bigg\rgroup
\bigg\lgroup
\frac{\xi_2}{\xi_1+\xi_2}\Pi^{\bar\eta;\bar\lambda i}\big[q_3;b_1,{\rm K}\big]-(1-\xi_1-\xi_2)\Pi^{\bar\lambda;\bar\eta i}\big[{\rm K};b_1^{-1},q_3\big]
\bigg\rgroup
\nonumber\\
&&
\times
\frac{1}{2} \tr\big[S_{y_3}S^\dagger_{z_1}\big] \tr\big[(\partial^iS_{z_1})(\partial S^\dagger_{y_3})\big]
\nonumber\\
&&
+
\bigg\lgroup
\frac{\xi_2}{\bar\xi_1}\Pi^{\bar\eta';\bar\lambda'j}\big[q_1;b_0,{\rm Q}\big]-\xi_1\Pi^{\bar\lambda';\bar\eta'j}\big[{\rm Q}; b_0^{-1},q_1\big]
\bigg\rgroup
\nonumber\\
&&
\times
\bigg\lgroup
\frac{\xi_2}{\xi_1+\xi_2}\Pi^{\bar\eta;\bar\lambda i}\big[q_3;b_1,{\rm K}\big]-(1-\xi_1-\xi_2)\Pi^{\bar\lambda;\bar\eta i}\big[{\rm K};b_1^{-1},q_3\big]
\bigg\rgroup\nonumber\\
&& 
\times
\frac{1}{2} \tr\big[(\partial^iS_{z_1})S^\dagger_{y_3}\big] \tr\big[(\partial^jS_{y_3})S^\dagger_{z_1}\big]
\nonumber\\
&&
+
\bigg\lgroup
\bar\xi_1\Pi^{\bar\lambda';\bar\eta'j}\big[{\rm Q};b_0^{-1},q_1\big] 
\bigg\rgroup
\bigg\lgroup
(\xi_1+\xi_2)\Pi^{\bar\lambda;\bar\eta i}\big[{\rm K}; b_1^{-1},q_3\big]
\bigg\rgroup
\frac{1}{2} \tr\big[S_{y_3}(\partial^iS^\dagger_{z_1})\big] \tr\big[S_{z_1}(\partial^jS^\dagger_{y_3})\big]
\nonumber\\
&&
-
\bigg\lgroup
\frac{\xi_2}{\bar\xi_1}\Pi^{\bar\eta';\bar\lambda'j}\big[q_1;b_0,{\rm Q}\big]-\Pi^{\bar\lambda';\bar\eta'j}\big[{\rm Q}; b_0^{-1},q_1\big]
\bigg\rgroup\nonumber\\
&&
\times
\bigg\lgroup
\frac{\xi_2}{\xi_1+\xi_2}\Pi^{\bar\eta;\bar\lambda i}\big[q_3;b_1,{\rm K}\big]-\Pi^{\bar\lambda;\bar\eta i}\big[{\rm K};b_1^{-1},q_3\big]
\bigg\rgroup
\nonumber\\
&&
\times
\frac{1}{2N_c} \tr\big[(\partial^iS_{z_1})S^\dagger_{y_3}(\partial^jS_{y_3})S^\dagger_{z_1}\big]
\bigg\}\;.
\eeq

At this point we would like to mention that above expression does not have the $i\leftrightarrow j$ symmetry. It is obvious that this symmetry will be restored once we combine the $\bar q\gamma-q\gamma$ with the $q\gamma-\bar q\gamma$ contributions. Therefore, we postpone introducing the gluon TMDs, and first consider the second crossed contribution, namely $q\gamma-\bar q\gamma$. The small dipole sizes for this contribution are defined as 
\beq
&&
r_q=v-z_1\; , \hspace{1.1cm} r'_q=y_1-y_2\;,\\
&&
\bar r_{\bar q}=z_3-z_2\; , \hspace{1cm} \bar r'_{\bar q}=v'-y_3\;.
\eeq 
Expressed in these variables, Eq. \eqref{q_gamma_bar_q_gamma_Cont} can be written as 
\beq
&&
{\rm I}_{q\gamma-\bar q\gamma}=-\int_{z_3y_1,r_qr'_q,\bar r_{\bar q}\bar r'_{\bar q}}
e^{i{\rm P}\cdot(y_1-z_3)-i\big[q_3\cdot\bar r'_{\bar q}-{\rm Q}\cdot\bar r_{\bar q}\big]-i\big[{\rm K}\cdot r'_q-q_1\cdot r_q\big]}
\, (-8) \, \widetilde{\cal M}^{\bar\lambda\bar\lambda';\bar\eta\bar\eta'}_g(\xi_1,\xi_2)
\nonumber\\
&&
\times
A^{\bar\lambda}(\bar r_{\bar q})A^{\bar\lambda'}(r'_q) \, 
\tr\Big[ {\cal \bar{G}}^{\dagger c\bar\eta'}_{q\gamma}(\xi_1,\xi_2;y_1,r'_q,\bar r'_{\bar q}) \, {\cal \bar{G}}^{c\bar\eta}_{\bar q\gamma}(\xi_1,\xi_2;z_3,r_q,\bar r_{\bar q})\Big]\;,
\eeq
and using the Taylor expansions of the functions ${\cal \bar{G}}^{c\bar \eta}_{q\gamma}$ (Eq. \eqref{expanded_G_q_gamma}) and ${\cal \bar{G}}^{c\bar \eta}_{q\gamma}$ (Eq. \eqref{expanded_G_bar_q_gamma}), we get:
\beq
&&
{\rm I}_{q\gamma-\bar q\gamma}=\int_{z_3y_1,r_qr'_q,\bar r_{\bar q}\bar r'_{\bar q}}
e^{i{\rm P}\cdot(y_1-z_3)-i\big[q_3\cdot\bar r'_{\bar q}-{\rm Q}\cdot\bar r_{\bar q}\big]-i\big[{\rm K}\cdot r'_q-q_1\cdot r_q\big]}
\, (-8) \, \widetilde{\cal M}^{\bar\lambda\bar\lambda';\bar\eta\bar\eta'}_g(\xi_1,\xi_2)
A^{\bar\lambda'}( r'_{ q})A^{\bar\lambda}(\bar r_{\bar q})\nonumber\\
&&
\hspace{-0.4cm}
\times
\bigg\{
\bigg\lgroup
\frac{\xi_2}{\xi_1+\xi_2}r'^j_q{\cal A}^{\bar\eta'}\bigg(1-\xi_1-\xi_2,\bar r'_{\bar q}; \frac{\xi_2}{\xi_1+\xi_2},r'_q\bigg)
\nonumber\\
&&
\hspace{0.5cm}
+
(1-\xi_1-\xi_2)\bar r'^j_{\bar q}\bigg[ {\cal A}^{\bar\eta'}\bigg(1-\xi_1-\xi_2,\bar r'_{\bar q}; \frac{\xi_2}{\xi_1+\xi_2},r'_q\bigg)-A^{\bar\eta'}(\bar r'_{\bar q})\bigg]
\bigg\rgroup
\nonumber\\
&&
\times
\bigg\lgroup
\bar\xi_1r^i_q\bigg[ {\cal A}^{\bar\eta}\bigg(\xi_1,r_q;\frac{\xi_2}{\bar\xi_1},\bar r_{\bar q}\bigg)-A^{\bar\eta}(r_q)\bigg]
\bigg\rgroup
\frac{1}{2} \tr\big[S_{y_1}S^\dagger_{z_3}\big] \tr\big[(\partial^iS_{z_3})(\partial^jS^\dagger_{y_3})\big]
\nonumber\\
&&
+
\bigg\lgroup
(\xi_1+\xi_2)\bar r'^j_{\bar q}\bigg[ {\cal A}^{\bar\eta'}\bigg(1-\xi_1-\xi_2,\bar r'_{\bar q}; \frac{\xi_2}{\xi_1+\xi_2},r'_q\bigg)-A^{\bar\eta'}(\bar r'_{\bar q})\bigg]
\bigg\rgroup
\nonumber\\
&&
\times
\bigg\lgroup
\frac{\xi_2}{\bar\xi_1}\bar r^i_{\bar q} {\cal A}^{\bar\eta}\bigg( \xi_1,r_q;\frac{\xi_2}{\bar\xi_1},\bar r_{\bar q}\bigg)
+\xi_1 r^i_q\bigg[ {\cal A}^{\bar\eta}\bigg( \xi_1,r_q;\frac{\xi_2}{\bar\xi_1},\bar r_{\bar q}\bigg) -A^{\bar\eta}(r_q)\bigg]
\bigg\rgroup
\nonumber\\
&&
\times
\frac{1}{2} \tr\big[S_{z_3}S^\dagger_{y_1}\big] \tr\big[(\partial^jS_{y_1})(\partial^iS^\dagger_{z_3})\big]
\nonumber\\
&&
-
\bigg\lgroup
\frac{\xi_2}{\xi_1+\xi_2}r'^j_q{\cal A}^{\bar\eta'}\bigg(1-\xi_1-\xi_2,\bar r'_{\bar q}; \frac{\xi_2}{\xi_1+\xi_2},r'_q\bigg)
\nonumber\\
&&
\hspace{0.5cm}
+
(1-\xi_1-\xi_2)\bar r'^j_{\bar q}\bigg[ {\cal A}^{\bar\eta'}\bigg(1-\xi_1-\xi_2,\bar r'_{\bar q}; \frac{\xi_2}{\xi_1+\xi_2},r'_q\bigg)-A^{\bar\eta'}(\bar r'_{\bar q})\bigg]
\bigg\rgroup\nonumber\\
&&
\times
\bigg\lgroup
\frac{\xi_2}{\bar\xi_1}\bar r^i_{\bar q} {\cal A}^{\bar\eta}\bigg( \xi_1,r_q;\frac{\xi_2}{\bar\xi_1},\bar r_{\bar q}\bigg)
+\xi_1 r^i_q\bigg[ {\cal A}^{\bar\eta}\bigg( \xi_1,r_q;\frac{\xi_2}{\bar\xi_1},\bar r_{\bar q}\bigg) -A^{\bar\eta}(r_q)\bigg]
\bigg\rgroup
\nonumber\\
&&
\times
\frac{1}{2} \tr\big[S_{y_1}(\partial^iS^\dagger_{z_3})\big] \tr\big[S_{z_3}(\partial^jS^\dagger_{y_1})\big]
\nonumber\\
&&
-
\bigg\lgroup
(\xi_1+\xi_2)\bar r'^j_{\bar q}\bigg[ {\cal A}^{\bar\eta'}\bigg(1-\xi_1-\xi_2,\bar r'_{\bar q}; \frac{\xi_2}{\xi_1+\xi_2},r'_q\bigg)-A^{\bar\eta'}(\bar r'_{\bar q})\bigg]
\bigg\rgroup
\nonumber\\
&&
\times
\bigg\lgroup
\bar\xi_1r^i_q\bigg[ {\cal A}^{\bar\eta}\bigg(\xi_1,r_q;\frac{\xi_2}{\bar\xi_1},\bar r_{\bar q}\bigg)-A^{\bar\eta}(r_q)\bigg]
\bigg\rgroup
\frac{1}{2} \tr\big[(\partial^jS_{y_1})S^\dagger_{z_3}\big] \tr\big[(\partial^iS_{z_3})S^\dagger_{y_1}\big]
\nonumber\\
&&
+
\bigg\lgroup
\frac{\xi_2}{\xi_1+\xi_2}r'^j_q{\cal A}^{\bar\eta'}\bigg(1-\xi_1-\xi_2,\bar r'_{\bar q}; \frac{\xi_2}{\xi_1+\xi_2},r'_q\bigg)
\nonumber\\
&&
\hspace{0.5cm}
+
\bar r'^j_{\bar q}\bigg[ {\cal A}^{\bar\eta'}\bigg(1-\xi_1-\xi_2,\bar r'_{\bar q}; \frac{\xi_2}{\xi_1+\xi_2},r'_q\bigg)-A^{\bar\eta'}(\bar r'_{\bar q})\bigg]
\bigg\rgroup
\nonumber\\
&&
\times
\bigg\lgroup
\frac{\xi_2}{\bar\xi_1}\bar r^i_{\bar q} {\cal A}^{\bar\eta}\bigg( \xi_1,r_q;\frac{\xi_2}{\bar\xi_1},\bar r_{\bar q}\bigg)
+ r^i_q\bigg[ {\cal A}^{\bar\eta}\bigg( \xi_1,r_q;\frac{\xi_2}{\bar\xi_1},\bar r_{\bar q}\bigg) -A^{\bar\eta}(r_q)\bigg]
\bigg\rgroup
\nonumber\\
&&
\times
\frac{1}{2N_c} \tr\big[(\partial^jS_{y_1})S^\dagger_{z_3}(\partial^iS_{z_3})S^\dagger_{y_1}\big]\bigg\}\;.
\eeq
Once again, we use the generic integrals introduced in Eqs. \eqref{integral_1} and \eqref{integral_2} to perform the integrations over the dipole sizes, after which the result reads:
\beq
\label{q_gamma_bar_q_gamma_aft_integration}
&&
{\rm I}_{q\gamma-\bar q\gamma}=8\, \widetilde{\cal M}^{\bar\lambda\bar\lambda';\bar\eta\bar\eta'}_g(\xi_1,\xi_2)
\int_{z_3y_1}e^{i{\rm P}\cdot(y_1-z_3)}
\nonumber\\
&&
\hspace{-0.4cm}
\times
\bigg\{
\bigg\lgroup
\frac{\xi_2}{\xi_1+\xi_2}\Pi^{\bar\eta';\bar\lambda'j}\big[q_3;b_1,{\rm K}\big]-(1-\xi_1-\xi_2)\Pi^{\bar\lambda';\bar\eta'j}\big[{\rm K};b_1^{-1},q_3\big]
\bigg\rgroup
\bigg\lgroup 
\bar\xi_1\Pi^{\bar\lambda;\bar\eta i}\big[{\rm Q}; b_0^{-1},q_1\big]
\bigg\rgroup
\nonumber\\
&&
\times
\frac{1}{2} \tr\big[S_{y_1}S^\dagger_{z_3}\big] \tr\big[(\partial^iS_{z_3})(\partial^jS^\dagger_{y_1})\big]
\nonumber\\
&&
+
\bigg\lgroup
(\xi_1+\xi_2)\Pi^{\bar\lambda';\bar\eta'j}\big[{\rm K};b_1^{-1},q_3\big]
\bigg\rgroup
\bigg\lgroup
\frac{\xi_2}{\bar\xi_1}\Pi^{\bar\eta;\bar\lambda i}\big[q_1;b_0,{\rm Q}\big] -\xi_1\Pi^{\bar\lambda;\bar\eta i}\big[{\rm Q};b_0^{-1},q_1\big]
\bigg\rgroup
\nonumber\\
&&
\times
\frac{1}{2} \tr\big[S_{z_3}S^\dagger_{y_1}\big] \tr\big[(\partial^jS_{y_1})(\partial^iS^\dagger_{z_3})\big]
\nonumber\\
&&
+
\bigg\lgroup
\frac{\xi_2}{\xi_1+\xi_2}\Pi^{\bar\eta';\bar\lambda'j}\big[q_3;b_1,{\rm K}\big]-(1-\xi_1-\xi_2)\Pi^{\bar\lambda';\bar\eta'j}\big[{\rm K};b_1^{-1},q_3\big]
\bigg\rgroup
\nonumber\\
&&
\times
\bigg\lgroup
\frac{\xi_2}{\bar\xi_1}\Pi^{\bar\eta;\bar\lambda i}\big[q_1;b_0,{\rm Q}\big] -\xi_1\Pi^{\bar\lambda;\bar\eta i}\big[{\rm Q};b_0^{-1},q_1\big]
\bigg\rgroup
\frac{1}{2} \tr\big[S_{y_1}(\partial^iS^\dagger_{z_3})\big] \tr\big[ S_{z_3}(\partial^jS^\dagger_{y_1})\big]
\nonumber\\
&&
+
\bigg\lgroup
(\xi_1+\xi_2)\Pi^{\bar\lambda';\bar\eta'j}\big[{\rm K};b_1^{-1},q_3\big]
\bigg\rgroup
\bigg\lgroup 
\bar\xi_1\Pi^{\bar\lambda;\bar\eta i}\big[{\rm Q}; b_0^{-1},q_1\big]
\bigg\rgroup
\frac{1}{2} \tr\big[ (\partial^jS_{y_1})S^\dagger_{z_3}\big] \tr\big[(\partial^iS_{z_3})S^\dagger_{y_1}\big]
\nonumber\\
&&
-
\bigg\lgroup
\frac{\xi_2}{\xi_1+\xi_2}\Pi^{\bar\eta';\bar\lambda'j}\big[q_3;b_1,{\rm K}\big]-\Pi^{\bar\lambda';\bar\eta'j}\big[{\rm K};b_1^{-1},q_3\big]
\bigg\rgroup
\nonumber\\
&&
\times
\bigg\lgroup
\frac{\xi_2}{\bar\xi_1}\Pi^{\bar\eta;\bar\lambda i}\big[q_1;b_0,{\rm Q}\big] -\Pi^{\bar\lambda;\bar\eta i}\big[{\rm Q};b_0^{-1},q_1\big]
\bigg\rgroup
\frac{1}{2N_c} \tr\big[(\partial^jS_{y_1})S^\dagger_{z_3}(\partial^iS_{z_3})S^\dagger_{y_1}\big]
\bigg\}\;.
\eeq
Comparing Eqs. \eqref{bar_q_gamma_q_gamma_aft_integration} and \eqref{q_gamma_bar_q_gamma_aft_integration}, it is obvious that the $i\leftrightarrow j$ symmetry is restored when the $\bar q\gamma-q\gamma$ and $q\gamma-\bar q\gamma$ contributions are added. Thus, after combining these two crossed contributions, we can integrate over the remaining two coordinates by using the definitions of the TMDs, and the final result of the crossed contribution reads:
\beq
\label{crossed_final}
&&
\Big\langle{\rm I}_{\bar q\gamma-q\gamma}+{\rm I}_{q\gamma-\bar q\gamma}\Big\rangle_{x_A}= -\widetilde{\cal M}_g^{\bar\lambda\bar\lambda'; \bar\eta\bar\eta'}(\xi_1,\xi_2) g_s^2 (2\pi)^3 N_c 
\nonumber\\
&&
\times
\bigg\{
\Big[ {\rm H}^{(1)}_{gg}\Big]^{\bar\lambda\bar\lambda';\bar\eta\bar\eta';ij}_{cross}
\bigg\lgroup \frac{1}{2}\delta^{ij}\bigg[ \frac{1}{2}{\cal F}^{(1)}_{gg}(x_A,{\rm P})+\frac{1}{2}{\cal F}^{(1)}_{gg}(x_A,-{\rm P})-\frac{1}{N_c^2}{\cal F}^{(3)}_{gg}(x_A,{\rm P})\bigg]
\nonumber\\
&&
-
\frac{1}{2}\bigg(\delta^{ij}-2\frac{\rm P^i\rm P^j}{\rm P^2}\bigg)
\bigg[ \frac{1}{2}{\cal H}^{(1)}_{gg}(x_A,{\rm P})+\frac{1}{2}{\cal H}^{(1)}_{gg}(x_A,-{\rm P})-\frac{1}{N_c^2}{\cal H}^{(3)}_{gg}(x_A,{\rm P})\bigg]\bigg\rgroup
\nonumber\\
&&
+\Big[ {\rm H}^{(adj)}_{gg}\Big]^{\bar\lambda\bar\lambda';\bar\eta\bar\eta';ij}_{ cross}\; \frac{\rm P^i\rm P^j}{\rm P^2} \; {\cal F}_{adj}(x_A,{\rm P})\bigg\}\;,
\eeq
with the hard parts
\beq
\label{hard_part_1_cross}
\Big[ {\rm H}^{(1)}_{gg}\Big]^{\bar\lambda\bar\lambda';\bar\eta\bar\eta';ij}_{cross}&=&\bigg\lgroup
\frac{\xi_2}{\bar\xi_1}\Pi^{\bar\eta';\bar\lambda'j}\big[q_1;b_0,{\rm Q}\big]-\Pi^{\bar\lambda';\bar\eta'j}\big[{\rm Q}; b_0^{-1},q_1\big]
\bigg\rgroup
\nonumber\\
&&
\times
\bigg\lgroup
\frac{\xi_2}{\xi_1+\xi_2}\Pi^{\bar\eta;\bar\lambda i}\big[q_3;b_1,{\rm K}\big]-\Pi^{\bar\lambda;\bar\eta i}\big[{\rm K};b_1^{-1},q_3\big]
\bigg\rgroup
\nonumber\\
&&
+
\bigg\lgroup
\frac{\xi_2}{\xi_1+\xi_2}\Pi^{\bar\eta';\bar\lambda'j}\big[q_3;b_1,{\rm K}\big]-\Pi^{\bar\lambda';\bar\eta'j}\big[{\rm K};b_1^{-1},q_3\big]
\bigg\rgroup
\nonumber\\
&&
\times
\bigg\lgroup
\frac{\xi_2}{\bar\xi_1}\Pi^{\bar\eta;\bar\lambda i}\big[q_1;b_0,{\rm Q}\big] -\Pi^{\bar\lambda;\bar\eta i}\big[{\rm Q};b_0^{-1},q_1\big]
\bigg\rgroup\;,
\eeq
and 
\beq
\label{hard_part_adj_cross}
\Big[ {\rm H}^{(adj)}_{gg}\Big]^{\bar\lambda\bar\lambda';\bar\eta\bar\eta';ij}_{ cross}&=&-
\bigg\lgroup
\frac{\xi_2}{\bar\xi_1}\Pi^{\bar\eta';\bar\lambda'j}\big[q_1;b_0,{\rm Q}\big]-\xi_1\Pi^{\bar\lambda';\bar\eta'j}\big[{\rm Q}; b_0^{-1},q_1\big]
\bigg\rgroup
\nonumber\\
&&
\hspace{0.3cm}
\times
\bigg\lgroup
\frac{\xi_2}{\xi_1+\xi_2}\Pi^{\bar\eta;\bar\lambda i}\big[q_3;b_1,{\rm K}\big]-(1-\xi_1-\xi_2)\Pi^{\bar\lambda;\bar\eta i}\big[{\rm K};b_1^{-1},q_3\big]
\bigg\rgroup\nonumber\\
&&
\hspace{0.3cm}
-
\bigg\lgroup
\frac{\xi_2}{\xi_1+\xi_2}\Pi^{\bar\eta';\bar\lambda'j}\big[q_3;b_1,{\rm K}\big]-(1-\xi_1-\xi_2)\Pi^{\bar\lambda';\bar\eta'j}\big[{\rm K};b_1^{-1},q_3\big]
\bigg\rgroup
\nonumber\\
&&
\hspace{0.3cm}
\times
\bigg\lgroup
\frac{\xi_2}{\bar\xi_1}\Pi^{\bar\eta;\bar\lambda i}\big[q_1;b_0,{\rm Q}\big] -\xi_1\Pi^{\bar\lambda;\bar\eta i}\big[{\rm Q};b_0^{-1},q_1\big]
\bigg\rgroup
\nonumber\\
&&
\hspace{0.3cm}
-
\bigg\lgroup
\bar\xi_1\Pi^{\bar\lambda';\bar\eta'j}\big[{\rm Q};b_0^{-1},q_1\big] 
\bigg\rgroup
\bigg\lgroup
(\xi_1+\xi_2)\Pi^{\bar\lambda;\bar\eta i}\big[{\rm K}; b_1^{-1},q_3\big]
\bigg\rgroup
\nonumber\\
&&
\hspace{0.3cm}
-
\bigg\lgroup
(\xi_1+\xi_2)\Pi^{\bar\lambda';\bar\eta'j}\big[{\rm K};b_1^{-1},q_3\big]
\bigg\rgroup
\bigg\lgroup 
\bar\xi_1\Pi^{\bar\lambda;\bar\eta i}\big[{\rm Q}; b_0^{-1},q_1\big]
\bigg\rgroup\;.
\eeq
Eq. \eqref{crossed_final} is the final result for the crossed contributions $\bar q\gamma-q\gamma$ and $q\gamma-\bar q\gamma$, written in terms of the unpolarized gluon TMDs ${\cal F}^{(1)}_{gg}$ and ${\cal F}^{(3)}_{gg}$, their linearly polarized partners (${\cal H}^{(1)}_{gg}$ and ${\cal H}^{(3)}_{gg}$), together with the adjoint dipole distribution ${\cal F}_{adj}$.  

The full result of the correlation limit of the partonic cross section in the gluon initiated channel is given by Eq. \eqref{gluon_channel_partonic}, where each contribution ${\rm I}_{\bar q\gamma-\bar q\gamma}$, ${\rm I}_{q\gamma- q\gamma}$, and ${\rm I}_{\bar q\gamma- q\gamma}+{\rm I}_{ q\gamma-\bar q\gamma}$ is given in Eqs. \eqref{final_barq_gamma_barq_gamma},  \eqref{final_q_gamma_q_gamma} and \eqref{crossed_final}, respectively.  Each of these contributions exhibit the same TMD structure, the only difference being in the hard parts and in the functions ${\cal M}_g$  and $\widetilde{\cal M}_g$.

\subsection{Final factorized expressions}
\label{TMD-summary}

Combining the expressions together as described in the end of the two previous subsections, the full cross sections can be cast into TMD factorized expressions of the form
\begin{eqnarray}
\frac{d\sigma^{qA\to \gamma gq+X}}{d^3\uq_1d^3\uq_2d^3\uq_3}&=&
2\pi\delta(p^+\!-\!q_1^+\!-\!q_2^+\!-\!q_3^+)
\bigg\{
\left[{\rm H}^{(1)}_{qg}\right]^{ij} \frac{{\rm P}^i{\rm P}^j}{\rm P^2}{\cal F}^{(1)}_{qg}(x_A, {\rm P})
\nonumber\\
&+&\; \left[{\rm H}^{(2)}_{qg}\right]^{ij} \bigg[\frac{1}{2}\delta^{ij}{\cal F}^{(2)}_{qg}(x_A, {\rm P})
\!-\!\frac{1}{2}\bigg(\delta^{ij}\!-\!2\frac{{\rm P}^i{\rm P}^j}{\rm P^2}\bigg){\cal H}^{(2)}_{qg}(x_A, {\rm P})\bigg]\bigg\}
\label{final-quark}\\
\frac{d\sigma^{gA\to q\gamma\bar q+X}}{d^3\uq_1d^3\uq_2d^3\uq_3}&=&
2\pi\delta(p^+\!-\!q_1^+\!-\!q_2^+\!-\!q_3^+)
\bigg\{\Big[ {\rm H}^{(adj)}_{gg}\Big]^{ij}\frac{\rm P^i\rm P^j}{\rm P^2} \; {\cal F}_{adj}(x_A,{\rm P})\nonumber\\
&+&\Big[ {\rm H}^{(1)}_{gg}\Big]^{ij}
\bigg\lgroup \frac{1}{2}\delta^{ij}\bigg[ \frac{1}{2}{\cal F}^{(1)}_{gg}(x_A,{\rm P})\!+\!\frac{1}{2}{\cal F}^{(1)}_{gg}(x_A,-{\rm P})\!-\!\frac{1}{N_c^2}{\cal F}^{(3)}_{gg}(x_A,{\rm P})\bigg]\\
&-&\frac{1}{2}\bigg(\delta^{ij}\!-\!2\frac{\rm P^i\rm P^j}{\rm P^2}\bigg)
\bigg[ \frac{1}{2}{\cal H}^{(1)}_{gg}(x_A,{\rm P})\!+\!\frac{1}{2}{\cal H}^{(1)}_{gg}(x_A,-{\rm P})\!-\!\frac{1}{N_c^2}{\cal H}^{(3)}_{gg}(x_A,{\rm P})\bigg]\bigg\rgroup\bigg\}\ ,\nonumber
\label{final-gluon}
\end{eqnarray}
where the fundamental-dipole gluon TMD ${\cal F}^{(1)}_{qg}$, the adjoint-dipole gluon TMD ${\cal F}_{adj}$ and the Weizs\"acker-Williams 
(unpolarized and linearly-polarized) distributions ${\cal F}^{(3)}_{gg}$ and ${\cal H}^{(3)}_{gg}$ are defined in Eqs. \eqref{Fq_1}, \eqref{def:F_adj} and \eqref{def:TMD_3}, respectively. Those TMDs are $C$-even: ${\cal F}(x_A,{\rm P})={\cal F}(x_A,{-\rm P})$ (since the action of the charge conjugation operator on the gauge fields is $CA_\mu C=-A_\mu^T $ \cite{Hatta2005}, it is easy to see from the definitions of the gluon TMDs that for each of them $C$-parity amounts to $C \mathcal{F}(x_2,\mathrm{P}) C=\mathcal{F}(x_2,\mathrm{-P})$). The other four gluon distributions $\mathcal{F}_{qg}^{(2)}$, $\mathcal{H}_{qg}^{(2)}$, $\mathcal{F}_{gg}^{(1)}$ and $\mathcal{H}_{gg}^{(1)}$, defined in Eqs. \eqref{Fq_2} and \eqref{def:TMD_1}, are not $C$-even, although in the gluon channel Eq. \eqref{final-gluon}, $\mathcal{F}_{gg}^{(1)}$ and $\mathcal{H}_{gg}^{(1)}$ appears in a $C$-even combinations. In the case of the quark channel Eq. \eqref{final-quark}, $\mathcal{F}_{qg}^{(2)}(x_A,-{\rm P})$ and $\mathcal{H}_{qg}^{(2)}(x_A,-{\rm P})$ would appear if the incoming quark was replaced by an antiquark.

The overall hard factors given by:
\begin{eqnarray}
\left[{\rm H}^{(i)}_{qg}\right]^{ij}=\frac{1}{(2\pi)^6}\frac{g_s^4g_e^2}{8q_1^+ q_2^+}\sum_{\bar\lambda\bar\lambda';\bar\eta\bar\eta'}
&&\bigg\{{\cal M}_q^{\bar \lambda\bar\lambda'; \bar \eta\bar\eta'}\bigg(\xi_1, \frac{\xi_2}{\bar\xi_1}\bigg)
{\left[{\rm H}^{(i)}_{qg}\right]}_{\it bef-bef}^{\bar\lambda\bar\lambda';\bar\eta\bar\eta'; ij} \\
&+&{\cal M}_q^{\bar \lambda\bar\lambda'; \bar \eta\bar\eta'}\bigg(\xi_2, \frac{\xi_1}{\bar\xi_2}\bigg)
{\left[{\rm H}^{(i)}_{qg}\right]}_{\it aft-aft}^{\bar\lambda\bar\lambda';\bar\eta\bar\eta'; ij}\nonumber\\
&-&\widetilde{\cal M}_q^{\bar \lambda\bar\lambda'; \bar \eta\bar\eta'}(\xi_1,\xi_2)\left(
{\left[{\rm H}^{(i)}_{qg}\right]}_{\it bef-aft}^{\bar\lambda\bar\lambda';\bar\eta\bar\eta'; ij}
+{\left[{\rm H}^{(i)}_{qg}\right]}_{\it bef-aft}^{\bar\lambda\bar\lambda';\bar\eta\bar\eta'; ij}
\right)\bigg\}
\nonumber
\end{eqnarray}
and
\begin{eqnarray}
\left[{\rm H}^{(i)}_{gg}\right]^{ij}=\frac{1}{(2\pi)^6}\frac{g_s^4g_e^2}{8 p^+ q_2^+}\frac{1}{2C_F}
\sum_{\bar\lambda\bar\lambda';\bar\eta\bar\eta'}&&\bigg\{
{\cal M}_{\bar q\bar q}^{\bar\lambda\bar\lambda';\bar\eta\bar\eta'}\bigg(\xi_1,\frac{\xi_2}{\bar\xi_1}\bigg)
 \Big[ {\rm H}^{(i)}_{gg}\Big]^{\bar\lambda\bar\lambda';\bar\eta\bar\eta';ij}_{\bar q\gamma-\bar q\gamma}\nonumber\\
&+&{\cal M}_{qq}^{\bar\lambda\bar\lambda';\bar\eta\bar\eta'}\bigg(\xi_1+\xi_2, \frac{\xi_2}{\xi_1+\xi_2}\bigg)
\Big[ {\rm H}^{(i)}_{gg}\Big]^{\bar\lambda\bar\lambda';\bar\eta\bar\eta';ij}_{ q\gamma- q\gamma}\nonumber\\
&-&\widetilde{\cal M}_g^{\bar\lambda\bar\lambda'; \bar\eta\bar\eta'}(\xi_1,\xi_2)
\Big[ {\rm H}^{(i)}_{gg}\Big]^{\bar\lambda\bar\lambda';\bar\eta\bar\eta';ij}_{cross}\bigg\}\ .
\end{eqnarray}
where the various pieces of the quark-initiated channel are given in Eqs.
\eqref{hard_part_1_bef_bef}-\eqref{hard_part_2_bef_bef}, 
\eqref{hard_part_1_aft_aft}-\eqref{hard_part_2_aft_aft},
\eqref{hard_part_1_bef_aft}-\eqref{hard_part_2_bef_aft} and Eqs.
\eqref{hard_part_1_aft_bef}-\eqref{hard_part_2_aft_bef}, while those of the gluon-initiated channel are Eqs. 
\eqref{hard_part_1_barq_gamma_barq_gamma}-\eqref{hard_part_adj_barq_gamma_barq_gamma}, 
\eqref{hard_part_1_q_gamma_q_gamma}-\eqref{hard_part_adj_q_gamma_q_gamma}, and 
\eqref{hard_part_1_cross}-\eqref{hard_part_adj_cross}.

We therefore demonstrated, for the first time, that the matching of the CGC with TMD factorization, in the regime where the validity region of the two formalisms overlaps (at low-$x$ and in the presence two ordered scales) which was first established in \cite{firstlowxTMDs} for two-particle final states, may also hold for three-particle final states. We have explicitly worked out the case of dijet + photon production, in order to keep the color flow simple, and plan to investigate the trijet case in the future. As expected, the gluon TMD content is the same as for the $q\to qg $ and $g\to q\bar q$ dijet channels, with the difference that linearly-polarized distributions appear for $q\to qg\gamma$ and $g\to q\bar q\gamma$ even with massless quarks. This is due the non-zero virtuality of the intermediate (anti-)quark states, which effectively acts as a mass.

\section{Conclusion and outlook}

In conclusion, we have computed the production cross section of a hard photon and two hard jets in forward $pA$ collisions. The computation is performed adopting the hybrid formalism which is suitable for forward collisions. More precisely, we have considered two different channels: the quark initiated channel and the gluon initiated one. In the former, the quark coming form the dilute projectile emits a gluon and a photon which then scatter off the target via eikonal interactions, producing a photon together with a quark jet and a gluon jet. For this channel, we have taken into account the two possible cases, depending on whether the photon is emitted before or after the gluon in the amplitude and in the complex conjugate amplitude. The results for the ensuing four different contributions - {\it bef-bef} (Eq. \eqref{bef-bef-final}), {\it aft-aft} (Eq. \eqref{aft-aft-final}), {\it aft-bef} (Eq. \eqref{aft-bef-final}) and {\it aft-bef} (Eq. \eqref{bef-aft-final}) - are calculated separately and the final result for the partonic level production cross section in this channel is given as a sum of each of these contributions in Eq. \eqref{Q_initiated_full}. In order to get the full cross section, this result should be convolved with the quark PDF as stated in Eq. \eqref{Q_in_full_X_section}.

In the gluon initiated channel, the gluon coming from the dilute projectile splits into a quark-antiquark pair and the photon is emitted either from the quark or from the antiquark forming three particles which then scatter through the target. In this channel, the partonic level production cross section is also calculated separately for the four possible cases: photon emission from the antiquark both in the amplitude and in the complex conjugate amplitude ($\bar q\gamma-\bar q\gamma$ contribution, Eq. \eqref{barq_gamma_squared_final}), photon emission from the quark both in the amplitude and in the complex conjugate amplitude ($q\gamma-q\gamma$ contribution, Eq. \eqref{q_gamma_squared_final}) and the two crossed contributions where the photon is radiated from the quark (or from the antiquark) in the amplitude and from the antiquark (or from the quark) in the complex conjugate amplitude ($\bar q\gamma-q\gamma$ contribution Eq. \eqref{crossed_1_fin} and $q\gamma-\bar q\gamma$ Eq. \eqref{crossed_2_fin}). The partonic level cross section is given as a sum of each of these four contributions (Eq. \eqref{gluon_channel_partonic}), and one needs to convolve the partonic level cross section with the gluon PDF in order to get the full cross section as stated in Eq. \eqref{G_in_full_X_section}. In both channels, the cross sections are written in terms of the standard dipole and quadrupole amplitudes in the fundamental representation. 

In the correlation limit, the transverse momenta of the three final state particles are much larger than the saturation scale of the target, whereas their
total transverse momentum $P$ is small, parametrically of the order of saturation scale. We have shown that in this limit, the production cross section can be simplified significantly, and cast in a factorized form involving transverse-momentum-dependent (TMD) gluon distributions. This is summarized in section \ref{TMD-summary}. We demonstrated by an explicit calculation that the correspondence between the CGC on the one hand, and TMD factorization on the other hand, in the region of overlap of both theories, remains valid beyond the simplest $2\to2$ processes that were considered previously. Moreover, our calculation provides an important subleading ($\alpha_s^2 \alpha_{em}$ vs. $\alpha_s^3$) contribution to the forward three-jet cross section, a part of which has been published recently \cite{Iancu:2018aa}, since forward photons are experimentally indistinguishable from forward jets. Finally, the computation in this work constitutes a first step towards photon-jet production at NLO, and eventually a complete NLO calculation of photon production in the hybrid formalism, which we are presently pursuing.

\section*{Acknowledgments}
TA thanks Nestor Armesto for many discussions related to the subject of this work. TA  gratefully acknowledges the support from {\it Bourses du Gouvernement Fran\c{c}ais (BGF)- S\'{e}jour de recherche}, and expresses his gratitude to CPHT, Ecole Polytechnique, and to the Institute of Nuclear Physics, Polish Academy of Sciences, for hospitality when part of this work was done. PT thanks Krzysztof Kutak for useful discussions, and for his hospitality in The Henryk Niewodniczanski Institute of Nuclear Physics (IFJ PAN), which was supported by the FWO-PAS grant VS.070.16N. RB and PT thank the National Centre for Nuclear Research (NCBJ) for hospitality and support when part of this work was done. The work of TA is supported by Grant No. 2017/26/M/ST2/01074 of the National Science Centre, Poland. The work of CM is supported in part by the Agence Nationale de la Recherche under the project ANR-16-CE31-0019-02. PT is funded from the European Research Council (ERC) under the European Union's Horizon 2020 research and innovation programme (grant agreement No. 647981, 3DSPIN). The work of RB is supported by the National Science Center, Poland, grant No. 2015/17/B/ST2/01838,  by the U.S. Department of Energy, Office of Nuclear Physics, under Contracts No. DE-SC0012704 and by an LDRD grant from Brookhaven Science Associates. This work has been performed in the framework of COST Action CA15213 ``Theory of hot matter and relativistic heavy-ion collisions'' (THOR).
\appendix

\section{Derivation of the outgoing gluon wave function}

\subsection{Derivation of the splitting functions in the gluon initiated channel}
\label{App:Derivation}
In this appendix, our primary aim is to derive the momentum space expressions of the splitting functions, $F^{(2)}_{(\rm\bf  q\bar q-\bar q\gamma)}$ and $F^{(2)}_{(\rm \bf q\bar q-q\gamma)}$, which account for two successive emissions, and appear at $O(g_eg_s)$ when computing the dressed gluon state. 

The interaction Hamiltonian for a gluon with three-momentum $(\underline{q} +\underline{p})$, color $c$ and polarization $j$, splitting into quark with three-momentum $\underline{p}$, color $\beta$ and spin $s$ and an antiquark with three-momentum $\underline{q}$, color $\alpha$ and spin $s'$ reads \cite{Elastic_vs_Inelastic}
\beq
&&
H_I^{g\to q\bar q}=g_st\, ^c_{\alpha\beta} \int \frac{d^3\uq}{(2\pi)^3} \frac{d^3\underline{p}}{(2\pi)^3}
\frac{1}{2\sqrt{2(p^++q^+)}}\bigg[ \frac{p^i}{p^+}-\frac{q^i}{q^+}\bigg]\bigg\{\frac{q^+-p^+}{q^++p^+}\delta^{ij}\delta_{s,-s'}-i\epsilon^{ij}\sigma^3_{s,-s'}\bigg\} 
\nonumber\\
&&
\hspace{5cm}
\times\; 
b^{\dagger\beta}_s(p^+,p)  d^{\dagger\alpha}_{s'}(q^+,q)  a_j^c(p^++q^+,p+q)\;,
\eeq
where $b^\dagger_s$ ($d^\dagger_{s'}$) are the quark (antiquark) creation operators and $a_j$ is the gluon annihilation operator. The function $F^{(1)}_{g\to q\bar q}$, which accounts for the momentum structure of the splitting, as well as the spin and polarization structure, is defined as
\beq
&&
\label{F1_gqq}
F^{(1)}_{g\to q\bar q}\big[ ({\bf q})[p^+, p], ({\bf \bar q})[q^+,q]\Big]^j_{s's}=\frac{1}{2\sqrt{2(p^++q^+)}}\bigg[ \frac{p^i}{p^+}-\frac{q^i}{q^+}\bigg] \nonumber\\
&&
\hspace{3cm}
\times\: \bigg\{\frac{q^+-p^+}{q^++p^+}\delta^{ij}\delta_{s,-s'}-i\epsilon^{ij}\sigma^3_{s,-s'}\bigg\}\frac{1}{w_{p+q}-w_p-w_q}\;,
\eeq  
with 
\beq
\label{energy}
w_p=\frac{p^2}{2p^+}\;.
\eeq
Remember that, in order to compute the dressed gluon state at order $g_eg_s$, we need to consider two cases, which differ in whether the photon is emitted from the quark or from the antiquark, after the splitting of the gluon in the quark-antiquark pair. We would like to keep the momenta of the final state particles (quark, antiquark and photon) fixed. Thus, we examine the two cases separately. Moreover, even though in this work the incoming gluon has vanishing transverse momentum, in this appendix we will keep it more general and only set it to zero at the very end. 
\\
\\
$\bullet$ {\it Photon emission from the antiquark}\\
We consider the incoming gluon with three-momentum $\underline{p}$, which splits into a quark with three-momentum $\underline{k_1}$ and antiquark with three-momentum $\underline{p}-\underline{k_1}$. Then, a photon with three-momentum $\underline{k_2}$ is emitted from the antiquark, and the momentum of the final state antiquark is $\underline{p}-\underline{k_1}-\underline{k_2}$. In this setup, using Eq. \eqref{F1_gqq}, the first splitting function can simply be written as 
\beq
&&
F^{(1)}_{g\to q\bar q}\big[ ({\bf q})[k_1^+, k_1], ({\bf \bar q})[p^+-k_1^+,p-k_1]\Big]^{\eta}_{ss'}=\frac{1}{2\sqrt{2(p^+)}}\bigg[ \frac{(p-k_1)^{\bar\eta}}{p^+-k_1^+}-\frac{k_1^{\bar\eta}}{k_1^+}\bigg]
\nonumber\\
&&
\times\; 
\bigg[\frac{p^+-k_1^+-k_1^+}{p^+}\delta^{\eta\bar\eta}\delta_{s,-s'}-i\epsilon^{\eta\bar\eta}\sigma^3_{s,-s'}\bigg] 
\frac{1}{w_{p}-w_{k_1}-w_{p-k_1}}\;.
\eeq
Defining the momentum ratio $\xi_1=k_1^+/p^+$ and using Eq. \eqref{energy}, the above formula can be further simplified, yielding
\beq
\label{F1gqq_mom}
F^{(1)}_{g\to q\bar q}\big[ ({\bf q})[k_1^+, k_1], ({\bf \bar q})[p^+-k_1^+,p-k_1]\Big]^{\eta}_{s's}&=&
\frac{-1}{\sqrt{2p^+}}\frac{(\xi_1p-k_1)^{\bar\eta}}{(\xi_1p-k_1)^2}\Psi^{\eta\bar\eta}_{ss'}(\xi_1)\;,
\eeq
where $\Psi^{\eta\bar\eta}_{ss'}(\xi_1)$ is defined in Eq. \eqref{split_qbarq}.

In order to calculate the splitting function for two successive emissions, we use standard second order perturbation theory. For the second splitting, we need the interaction Hamiltonian for a quark (or antiquark) emitting a photon, which reads \cite{Elastic_vs_Inelastic}
\beq
\label{H_qgamma}
&&
H_I^{q\to q\gamma}=g_s\int \frac{d^3\underline{p}}{(2\pi)^3}\frac{d^3\underline{q}}{(2\pi)^3}\frac{1}{2\sqrt{2q^+}}
\bigg(\frac{q^i}{q^+}-\frac{p^i}{p^+}\bigg)
\bigg[ \frac{2p^++q^+}{p^++q^+}\delta^{ij}\delta_{s's}-i\epsilon^{ij}\sigma^3_{s's}\frac{q^+}{p^++q^+}\bigg]
\nonumber\\
&&
\hspace{5cm}
\times\;
b^\beta_{s'}(p^++q^+,p+q)b^{\dagger\beta}_s(p^+,p)\gamma^\dagger_j(q^+,q)\;,
\eeq
with $b_s$ being the annihilation operator for the quark and $\gamma^\dagger_j$ the creation operator for the photon. Using the above interaction Hamiltonian and second order perturbation theory, after renaming the momentum of the final state particles, the function $F^{(2)}_{(q\bar q-\bar q\gamma)}$ reads
\beq
&&
F^{(2)}_{(q\bar q-\bar q\gamma)}\Big[ ({\bf q})[k_1^+,k_1], (\gamma) [k_2^+,k_2], ({\bf \bar q})[p^+-k_1^+-k_2^+,p-k_1-k_2]\Big]^{\lambda\eta}_{ss''}\nonumber
\\
&&
=
 \Bigg\{\frac{-1}{\sqrt{2p^+}}\frac{(\xi_1p-k_1)^{\bar\eta}}{(\xi_1p-k_1)^2}  \Psi^{\eta\bar\eta}_{s's}(\xi_1)\Bigg\}
 \Bigg\{ 
 \frac{1}{2\sqrt{2k_2^+}}
\bigg[ \frac{(p-k_1-k_2)^{\bar\lambda}}{(p^+-k_1^+-k_2^+)}-\frac{k_2^{\bar\lambda}}{k_2^+}\bigg]
\nonumber\\
&&
\times\; 
\bigg[ \frac{(p^+-k_1^+)+(p^+-k_1^+-k_2^+)}{p^+-k_1^+}\delta^{\lambda\bar\lambda}\delta_{s's''}-i\epsilon^{\bar\lambda\lambda}\sigma^3_{s's''}\frac{k_2^+}{p^+-k_1^+}\bigg]\nonumber\\
&&
\times
\frac{1}{w_p-w_{p-k_1-k_2}-w_{k_1}-w_{k_2}}
\Bigg\}.
\eeq
Defining the momentum ratios $\xi_2=k_2^+/p^+$, $\bar\xi_1=(1-\xi_1)$, $\bar\xi_2=(1-\xi_2)$ and using Eq. \eqref{energy} to calculate the energy denominator, the splitting function $F^{(2)}_{(q\bar q- \bar q\gamma)}$ yields Eq. \eqref{F2_antiq-photon}.
\\
$\bullet$ {\it Photon emission from the quark}\\
As mentioned earlier, we would like to keep the momenta of the final state particles in the incoming gluon wave function fixed, to have a quark with three-momentum $\underline{k_1}$, a photon with three-momentum $\underline{k_2}$, and an antiquark with three-momentum $\underline{p}-\underline{k_1}-\underline{k_2}$. Since in this case the photon is emitted from the quark, this means that the incoming gluon with three-momentum $\underline{p}$ splits into a quark with three-momentum $\underline{k_1}+\underline{k_2}$ and an antiquark with $\underline{p}-\underline{k_1}-\underline{k_2}$. In this setup, the function that accounts for the momentum, spin, and polarization structure of the gluon splitting into a quark-antiquark pair reads
\beq
&&
F^{(1)}_{g\to q\bar q}\big[({\bf q})[\underline{k}_1+\underline{k}_2], ({\bf \bar q})[\underline{p}-\underline{k}_1-\underline{k}_2]\big]^{\eta}_{s's''}
=
\frac{1}{2\sqrt{2p^+}}\bigg[ \frac{(p-k_1-k_2)^{\bar\eta}}{p^+-k_1^+-k_2^+}-\frac{(k_1+k_2)^{\bar\eta}}{k_1^++k_2^+}\bigg]
\nonumber\\
&&
\times\, 
\bigg[ \frac{p^+-2(k_1^+-k_2^+)}{p^+}\delta^{\eta\bar\eta}\delta_{s',-s''}-i\epsilon^{\eta\bar\eta}\sigma^3_{s',-s''}\bigg]
\frac{1}{w_p-w_{k_1+k_2}-w_{p-k_1-k_2}}\;.
\eeq
After using Eq. \eqref{energy}, the above formula can be simplified and written in a more compact way: 
\beq
F^{(1)}_{g\to q\bar q}\big[({\bf q})[\underline{k}_1+\underline{k}_2], ({\bf \bar q})[\underline{p}-\underline{k}_1-\underline{k}_2]\big]^{\eta}_{s's''}
=
\frac{-1}{\sqrt{2p^+}}\frac{\big[(\xi_1+\xi_2)p-(k_1+k_2)\big]^{\bar\eta}}{\big[(\xi_1+\xi_2)p-(k_1+k_2)\big]^2}
\Psi^{\eta\bar\eta}_{s's''}(\xi_1+\xi_2)\,.
\nonumber\\
\eeq 
In the second step, the quark with three-momentum $\underline{k_1}+\underline{k}_2$ splits into a photon with $\underline{k}_2$ and a quark with $\underline{k}_1$. In this case, the function that accounts for the two successive splittings can be written as 
\beq
&&
F^{(2)}_{(q\bar q-q\gamma)}\big[({\bf q})[k_1^+,k_1], (\gamma)[k_2^+,k_2], ({\bf \bar q})[p^+-k_1^+-k_2^+, p-k_1-k_2] \big]^{\lambda\eta}_{ss''}\nonumber\\
&&
=\frac{-1}{\sqrt{2p^+}}\frac{\big[(\xi_1+\xi_2)p-(k_1+k_2)\big]^{\bar\eta}}{\big[(\xi_1+\xi_2)p-(k_1+k_2)\big]^2}
\Psi^{\eta\bar\eta}_{s's''}(\xi_1+\xi_2) \, 
\frac{1}{2\sqrt{2k_2^+}}\bigg[\frac{k_2^{\bar\lambda}}{k_2^+}-\frac{k_1^{\bar\lambda}}{k_1^+}\bigg]
\nonumber\\
&&
\times\; 
\bigg[\frac{2k_1^++k_2^+}{k_1^++k_2^+}\delta^{\lambda\bar\lambda}_{ss'}-i\epsilon^{\lambda\bar\lambda}\sigma^3_{ss'}\frac{k_2^+}{k_1^++k_2^+}\bigg]
\frac{1}{w_p-w_{k_1}-w_{k_2}-w_{p-k_1-k_2}}\;,
\eeq
which simplifies to the expression given in Eq. \eqref{F2_q-photon} after using Eq. \eqref{energy}. 
 
\subsection{Dressed gluon state in the mixed space }
\label{App:Dressed_in_mixed_space}
Even though it is fairly straightforward to go from the full momentum space expression to the mixed space expression of the dressed gluon state, for the sake of the completeness we provide the details of the calculation in this appendix.  Our starting point is the perturbative expression of the dressed gluon state in full momentum space given in Eq. \eqref{dressed_gluon_mom}. The explicit expressions of the splitting functions in full momentum space are given in Eqs. \eqref{F1}, \eqref{F2_antiq-photon} and \eqref{F2_q-photon}. We would like to emphasize that these expressions are given for the incoming gluon with nonvanishing transverse momentum, even though in our setup it is assumed to be zero. In this appendix, we stick to the most general case, i.e. we keep the transverse momentum of the incoming gluon nonzero, and only set it to zero at the end.
\\
\\
(i) {\it bare gluon component} The two dimensional Fourier transform of the bare gluon component of the dressed gluon state can be be written in a completely trivial way: 
\beq
\big|({\bf g})[p^+,p]^c_{\eta}\big\rangle_0=\int_we^{ip\cdot w}\big|({\bf g})[p^+,w]^c_\eta\big\rangle_0\;.
\eeq 
\\
(ii) {\it bare quark-antiquark component}  The bare quark-antiquark component of the dressed gluon state can be written as 
\beq
&&
g_s \, t^c_{\alpha\beta} 
\int \frac{dk_1^+}{2\pi} \frac{d^2k_1}{2\pi}
F^{(1)}_{(g\to q\bar q)}\Big[ {(\bf q)} [k_1^+,k_1], ({\bf \bar q})[p^+-k_1^+, p-k_1]\Big]^\eta_{ss'}
\nonumber\\
&&
\hspace{5cm}
\times\; 
\big| ({\bf q}) [k_1^+,k_1]^\alpha_s; ({\bf \bar q})[p^+-k_1^+,p-k_1]^\beta_{s'}\big\rangle_0
\nonumber
\\
&&
= \; 
g_s\, t^c_{\alpha\beta}\int \frac{dk_1^+}{2\pi}\frac{dk_2^+}{(2\pi)^2} \int_{vx_1z_1z_2} e^{-ik_1\cdot(x_1+z_1)-i(p-k_1)\cdot(z_2+v)}
\nonumber\\
&&
\times\; 
F^{(1)}_{(g\to q\bar q)}\Big[ ({\bf q}) [k_1^+,z_1], ({\bf \bar q}) [p^+-k_1^+,z_2]\Big]^\eta_{ss'} 
\big| ({\bf q})[k_1^+,x_1]^\alpha_s; ({\bf \bar q})[p^+-k_1^+,v]^\beta_{s'}\big\rangle_0\;.
\eeq
Integration over $k_1$ gives $\delta^{(2)}\big[z_1-(z_2+v-x_1)\big]$. After renaming $z_2=w-v$ and performing the trivial integral over $z_1$ by realizing the delta function, we get
\beq
&&
g_s \, t^c_{\alpha\beta} 
\int \frac{dk_1^+}{2\pi} \frac{d^2k_1}{2\pi}
F^{(1)}_{(g\to q\bar q)}\Big[ {(\bf q)} [k_1^+,k_1], ({\bf \bar q})[p^+-k_1^+, p-k_1]\Big]^\eta_{ss'}
\nonumber\\
&&
\hspace{5cm}
\times\; 
\big| ({\bf q}) [k_1^+,k_1]^\alpha_s; ({\bf \bar q})[p^+-k_1^+,p-k_1]^\beta_{s'}\big\rangle_0
\nonumber
\\
&&
= \; 
g_s\, t^c_{\alpha\beta}\int \frac{dk_1^+}{2\pi} \int_{wvx_1}e^{-ip\cdot w} F^{(1)}_{(g\to q\bar q)}\big[ ({\bf q}) [k_1^+,w-x_1], ({\bf \bar q}) [p^+-k_1^+, w-v]\Big]^\eta_{ss'}
\nonumber\\
&&
\hspace{5cm}
\times\; 
\Big| {(\bf q)}[k_1^+,x_1]^\alpha_s; ({\bf \bar q}) [p^+-k_1^+,v]^\beta_{s'}\big\rangle_0\;.
\eeq 
Now, we need the mixed space expression of the splitting function: 
\beq
&&
F^{(1)}_{(g\to q\bar q)}\Big[ ({\bf q})[k_1^+,w-x_1], ({\bf \bar q}) [p^+-k_1^+,w-v]\Big]^\eta_{ss'}=\int \frac{d^2q_1}{(2\pi)^2}\frac{d^2q}{(2\pi)^2}e^{iq_1\cdot (w-x_1)+i(q-q_1)\cdot(w-v)}
\nonumber\\
&&
\hspace{8cm}
\times\; 
\frac{(\xi_1q-q_1)^{\bar\eta}}{(\xi_1q-q_1)^2}\frac{-1}{\sqrt{2p^+}}\Psi^{\eta\bar\eta}_{ss'}(\xi_1)\;.
\eeq
After performing the following change of variables 
\beq
\xi_1q-q_1=\xi_1P\; , \;\;\;\;\;\; q_1=K\;, 
\eeq
the integration over $K$ and $P$ factorizes and both integrals can be performed easily, yielding
\beq
F^{(1)}_{(g\to q\bar q)}\Big[ ({\bf q})[k_1^+,w-x_1], ({\bf \bar q}) [p^+-k_1^+,w-v]\Big]^\eta_{ss'}&=&\delta^{(2)}\big[ w-(\xi_1x_1+\bar\xi_1v)\big]
\nonumber\\
&&
\times\frac{(-i)}{\sqrt{2p^+}}\Psi^{\eta\bar\eta}_{ss'}(\xi_1)A^{\bar\eta}(v-x_1)\;,\quad
\eeq
where $A^{\bar\eta}(v-x_1)$ is the standard Weizs\"acker-Williams field in coordinate space whose expression is given in Eq. \eqref{Standard_WW}.  Finally, we can write the bare quark-antiquark component of the dressed gluon state in the mixed space as
\beq
&&
g_s\, t^c_{\alpha\beta} \int\frac{dk_1^+}{2\pi} \int_{wvx_1}e^{ip\cdot w}\delta^{(2)}\big[w-(\xi_1x_1+\bar\xi_1v)\big]
\frac{(-i)}{\sqrt{2p^+}}\Psi^{\eta\bar\eta}_{ss'}(\xi_1)
A^{\bar\eta}(v-x_1)
\nonumber\\
&&
\hspace{4cm}
\times\; 
\big| ({\bf q})[k_1^+,x_1]^\alpha_s; ({\bf \bar q})[p^+-k_1^+,v]^\beta_{s'}\big\rangle_0\;.
\eeq
(iii) {\it bare quark-antiquark-photon component} As discussed earlier, there are two different contributions to the quark-antiquark-photon component of the dressed gluon wave function, depending on whether the photon is emitted from the antiquark or from the quark. We consider each case separately and start from the emission of the photon from the antiquark: 
\beq
\label{derv_F2_antiq_1}
&&
g_sg_e \, t^c_{\alpha\beta}\int\frac{dk_1^+}{2\pi}\frac{d^2k_1}{(2\pi)^2}\frac{dk_2^+}{2\pi}\frac{d^2k_2}{(2\pi)^2} 
\nonumber\\
&&
\times\; 
F^{(2)}_{(q\bar q-\bar q\gamma)}\Big[({\bf q})[k_1^+,k_1], (\gamma)[k_2^+,k_2], ({\bf \bar q})[p^+-k_1^+-k_2^+, p-k_1-k_2]\Big]^{\lambda\eta}_{ss''} 
\nonumber\\
&&
\times\; 
\big| ({\bf q})[k_1^+,k_1]^\alpha_s; (\gamma)[k_2^+,k_2]^\lambda; ({\bf \bar q})[p^+-k_1^+-k_2^+, p-k_1-k_2]^\beta_{s''}\big\rangle_0
\nonumber\\
&&
=\;
g_sg_e t^c_{\alpha\beta} 
\int\frac{dk_1^+}{2\pi}\frac{d^2k_1}{(2\pi)^2}\frac{dk_2^+}{2\pi}\frac{d^2k_2}{(2\pi)^2} 
\int_{z_1,z_2,z_3,x_1,x_2,x_3}
e^{ik_1\cdot(x_1+z_1)-ik_2\cdot(x_2+z_2)-i(p-k_1-k_2)\cdot(x_3+z_3)}
\nonumber\\
&&
\times\; 
F^{(2)}_{(q\bar q-\bar q\gamma)}\Big[ ({\bf q})[k_1^+,z_1], (\gamma)[k_2^+,z_2], ({\bf \bar q})[p^+-k_1^+-k_2^+, z_3]\Big]^{\lambda\eta}_{ss''}
\nonumber\\
&&
\times\; 
\big| ({\bf q}) [k_1^+,x_1]_s^\alpha; (\gamma)[k_2^+,x_2]^\lambda; ({\bf \bar q})[p^+-k_1^+-k_2^+, x_3]^\beta_{s''}\big\rangle_0\;. 
\eeq 
After renaming $z_3=w-x_3$, we can first integrate over $k_1$ and $k_2$, and then over $z_1$ and $z_2$. The right hand side of Eq. \eqref{derv_F2_antiq_1} then reads
\beq
\label{F2_antiq_mid}
&&
g_sg_e\, t^c_{\alpha\beta}\int \frac{dk_1^+}{2\pi} \frac{dk_2^+}{2\pi} \int_{wx_1x_2x_3}e^{-ip\cdot w} 
\nonumber\\
&&
\times\; 
F^{(2)}_{(q\bar q-\bar q\gamma)}\Big[ ({\bf q})[k_1^+,w-x_1]^\alpha_s; (\gamma)[k_2^+,w-x_2]; ({\bf \bar q}) [p^+-k_1^+-k_2^+,w-x_3]\Big]^{\lambda\eta}_{ss''}
\nonumber\\
&&
\times\; 
\big| ({\bf q})[k_1^+,x_1]^\alpha_s ; (\gamma)[k_2^+,x_2]; ({\bf \bar q})[p^+-k_1^+-k_2^+,x_3]\big\rangle_0\;.
\eeq 
Now, let us calculate the Fourier transform of the function $F^{(2)}$: 
\beq
\label{derv_F2_antIq_2}
&&
F^{(2)}_{(q\bar q-\bar q\gamma)}\Big[ ({\bf q})[k_1^+,w-x_1], (\gamma)[k_2^+,w-x_2], ({\bf \bar q})[p^+-k_1^+-k_2^+,w-x_3]\Big]^{\lambda\eta}_{ss''}
\\
&&
=\; 
\int\frac{d^2q_1}{(2\pi)^2}\frac{d^2q_2}{(2\pi)^2}\frac{d^2q}{(2\pi)^2} 
e^{iq_1\cdot(w-x_1)+iq_2\cdot(w-x_2)+i(q-q_1-q_2)\cdot(w-x_3)}
\frac{(-1)}{\sqrt{2p^+}}\Psi^{\eta\bar\eta}_{ss'}(\xi_1)
\nonumber\\
&&
\times\; 
\frac{1}{\sqrt{2\xi_2p^+}}\phi^{\lambda\bar\lambda}_{s's''}\bigg(\frac{\xi_2}{\bar\xi_1}\bigg)
\frac{(\xi_1q-q_1)^{\bar\eta}}{(\xi_1q-q_1)^2} 
\frac{ \xi_1 \big[\xi_2(q-q_1)-\bar\xi_1q_2\big]^{\bar\lambda}}{\xi_2 (\xi_1 q-q_1)^2+\xi_1(\xi_2 q-q_2)^2-(\xi_2q_1-\xi_1q_2)^2}\;.
\nonumber
\eeq
After performing the following change of variables 
\beq
\xi_1q-q_1&=&P\;,\\
\xi_2(q-q_1)-\bar\xi_1q_2&=&\bar\xi_1K\;,
\eeq
the integrations over $P$, $K$, and $q$ factorize, and the integral over the latter results in a delta function. The right hand side of Eq. \eqref{derv_F2_antIq_2} can then be written as 
\beq
&&
\frac{(-1)}{\sqrt{2p^+}}\Psi^{\eta\bar\eta}_{ss'}(\xi_1)
\frac{1}{\sqrt{2\xi_2p^+}}\phi^{\lambda\bar\lambda}_{s's''}\bigg(\frac{\xi_2}{\bar\xi_1}\bigg)
\delta^{(2)}\big[w-\big\{\xi_1x_1+\xi_2x_2+(1-\xi_1-\xi_2)x_3\big\}\big]
\nonumber\\
&&
\times\; 
\int\frac{d^2P}{(2\pi)^2}\frac{d^2K}{(2\pi)^2}e^{-iP\cdot\frac{(w-x_1)}{\bar\xi_1} -iK\cdot(x_3-x_2)} \, 
\frac{P^{\bar\eta}}{P^2} \,  \frac{K^{\bar\lambda}}{K^2+c_0P^2}\;,
\eeq  
with $c_0=\frac{\xi_2(1-\xi_1-\xi_2)}{\xi_1\bar\xi_1^2}$. The integral over $P$ and $K$ can be performed easily (see for example \cite{Altinoluk:2018uax}):
\beq
\label{Generic_F2_Int}
\int \frac{d^2P}{(2\pi)^2}\frac{d^2K}{(2\pi)^2}e^{iP\cdot r+iK\cdot r'} 
\frac{P^{\bar\eta}}{P^2} \, \frac{K^{\bar \lambda}}{K^2+c_0P^2}=-\frac{1}{(2\pi)^2}\frac{r'^{\bar\lambda}}{r'^2}\, \frac{r^{\bar\eta}}{r^2+c_0r'^2}\;.
\eeq
Using the above formula, and realizing that 
\beq
&&
\delta^{(2)}\big[ w-\big\{ \xi_1x_1+\xi_2x_2+(1-\xi_1-\xi_2)x_3\big\}\big]=\int_v\delta^{(2)}\big[ w-(\xi_1x_1+\bar\xi_1v)\big]\
\nonumber\\
&&
\hspace{5cm}
\times\; 
\delta^{(2)}\bigg[v-\bigg\{\bigg(1-\frac{\xi_2}{\bar\xi_1}\bigg)x_3+\frac{\xi_2}{\bar\xi_1}x_2\bigg\}\bigg]\;,
\eeq
we can write the mixed space expression of the splitting function as
\beq
&&
F^{(2)}_{(q\bar q-\bar q\gamma)}\Big[ ({\bf q})[k_1^+,w-x_1], (\gamma)[k_2^+,w-x_2], ({\bf \bar q})[p^+-k_1^+-k_2^+,w-x_3]\Big]^{\lambda\eta}_{ss''}
\nonumber\\
&&
=\int_v \delta^{(2)}\big[ w-(\xi_1x_1+\bar\xi_1v)\big]\delta^{(2)}\bigg[v-\bigg\{\bigg(1-\frac{\xi_2}{\bar\xi_1}\bigg)x_3+\frac{\xi_2}{\bar\xi_1}x_2\bigg\}\bigg]
\nonumber\\
&&
\times\; 
\frac{(-i)}{\sqrt{2p^+}}\Psi^{\eta\bar\eta}_{ss'}(\xi_1)\frac{(+i)}{\sqrt{2p^+\xi_2}}\phi^{\lambda\bar\lambda}_{s's''}\bigg( \frac{\xi_2}{\bar\xi_1}\bigg) A^{\bar\lambda}(x_3-x_2) 
{\cal A}^{\bar\eta}\bigg( \xi_1, v-x_1; \frac{\xi_2}{\bar\xi_1}x_3-x_2\bigg)\;,\;
\eeq
where ${\cal A}^{\bar\eta}$ is the modified Weizs\"acker-Williams field in the coordinate space, whose definition is given in Eq. \eqref{Modified_WW}.  

We can now consider the second case, in which the photon is emitted from the quark. Eqs. \eqref{derv_F2_antiq_1} and \eqref{F2_antiq_mid} hold for this case as well. The difference between the two cases appears when we calculate the Fourier transform of the splitting function, since their coordinate space expressions are different: 
\beq
&&
F^{(2)}_{(q\bar q- q\gamma)}\Big[ ({\bf q})[k_1^+,w-x_1], (\gamma)[k_2^+,w-x_2], ({\bf \bar q})[p^+-k_1^+-k_2^+,w-x_3]\Big]^{\lambda\eta}_{ss''}
\nonumber\\
&&
=\; 
\int \frac{d^2q}{(2\pi)^2}\frac{d^2q_1}{(2\pi)^2}\frac{d^2q_2}{(2\pi)^2}
e^{iq_1\cdot(w-x_1)+iq_2\cdot(w-x_2)+i(q-q_1-q_2)\cdot(w-x_3)}
\frac{(-1)}{\sqrt{2p^+}}\Psi^{\eta\bar\eta}_{s's''}(\xi_1+\xi_2)
\nonumber\\
&&
\times\; 
\frac{1}{\sqrt{2\xi_2p^+}}\phi^{\lambda\bar\lambda}_{s's}\bigg(\frac{\xi_2}{\xi_1+\xi_2}\bigg) 
\frac{\big[ (\xi_1+\xi_2)q-(q_1+q_2)\big]^{\bar\eta}}{\big[(\xi_1+\xi_2)q-(q_1+q_2)\big]^2}
\nonumber\\
&&
\times\; 
\frac{(1-\xi_1-\xi_2)(\xi_2q_1-\xi_1q_2)^{\bar\lambda}}{\xi_2(\xi_1q-q_1)^2+\xi_1(\xi_2q-q_2)^2-(\xi_2q_1-\xi_1q_2)^2}\;.
\eeq
We perform the following change of variables to factorize the transverse momentum integrals 
\beq
(\xi_1+\xi_2)q-(q_1+q_2)=P\;, \\
\xi_2q_1-\xi_1q_2=K\;.
\eeq
After this, the integration over $q$ gives a delta function, and the result reads:
\beq
&&
F^{(2)}_{(q\bar q- q\gamma)}\Big[ ({\bf q})[k_1^+,w-x_1], (\gamma)[k_2^+,w-x_2], ({\bf \bar q})[p^+-k_1^+-k_2^+,w-x_3]\Big]^{\lambda\eta}_{ss''}
\nonumber\\
&&
=\; 
\frac{(-1)}{\sqrt{2p^+}}\Psi^{\eta\bar\eta}_{s's''}(\xi_1+\xi_2)
\frac{1}{\sqrt{2\xi_2p^+}}\phi^{\lambda\bar\lambda}_{ss'}\bigg(\frac{\xi_2}{\xi_1+\xi_2}\bigg) 
\delta^{(2)}\big[ w-\big\{ \xi_1x_1+\xi_2x_2+(1-\xi_1-\xi_2)x_3\big\}\big]
\nonumber\\
&&
\times\; 
\int \frac{d^2P}{(2\pi)^2}\frac{d^2K}{(2\pi)^2}
e^{iK\cdot \frac{(x_2-x_1)}{\xi_1+\xi_2}+iP\cdot\frac{(w-x_3)}{\xi_1+\xi_2}}
\frac{P^{\bar\eta}}{P^2}\frac{1}{\xi_1+\xi_2}\frac{K^{\bar\lambda}}{K^2+c_0P^2}\;,
\eeq
with $c_0=\frac{\xi_1\xi_2}{(1-\xi_1-\xi_2)(\xi_1+\xi_2)^2}$. The integrations over $P$ and $K$ can be performed by the generic integral given in  Eq. \eqref{Generic_F2_Int}. After noting  that 
\beq
\delta^{(2)}\big[ w-\big\{ \xi_1x_1+\xi_2x_2+(1-\xi_1-\xi_2)x_3\big\} \big]&=&\int_v \delta^{(2)}\big[ w-\big\{ (\xi_1+\xi_2)v+(1-\xi_1-\xi_2)x_3\big\}\big]
\nonumber
\\
&&
\hspace{-2cm}
\times\; 
\delta^{(2)}
\bigg[ v-\bigg\{\bigg(1-\frac{\xi_2}{\xi_1+\xi_2}\bigg)x_1+\frac{\xi_2}{\xi_1+\xi_2}x_2\bigg\}\bigg]\;,
\eeq
and 
\beq
(w-x_3)^{\bar\eta}=(\xi_1+\xi_2)(v-x_3)^{\bar\eta}\;,
\eeq
the mixed space expression for $F^{(2)}_{(q\bar q-q\gamma)}$ reads
\beq
&&
F^{(2)}_{(q\bar q-q\gamma)}\Big[ 
({\bf q})[k_1^+,w-x_1], (\gamma)[k_2^+, w-x_2], ({\bf \bar q})[p^+-k_1^+-k_2^+, w-x_3]
\Big]^{\lambda\eta}_{ss''}
\nonumber\\
&&
= \int_v\delta^{(2)}\big[ w-\big\{ (\xi_1+\xi_2)v+(1-\xi_1-\xi_2)x_3\big] \delta^{(2)}\bigg[ v-\bigg\{\bigg(1-\frac{\xi_2}{\xi_1+\xi_2}\bigg)x_1+\frac{\xi_2}{\xi_1+\xi_2}x_2\bigg\}\bigg]
\nonumber\\
&&
\times\; 
\frac{(-i)}{\sqrt{2p^+}} \Psi^{\eta\bar\eta}_{s's''}(\xi_1+\xi_2)
\frac{(+i)}{\sqrt{2\xi_2p^+}}\phi^{\lambda\bar\lambda}_{ss'}\bigg(\frac{\xi_2}{\xi_1+\xi_2}\bigg) A^{\bar\lambda}(x_2-x_1)
\nonumber\\
&&
\times\; 
{\cal A}^{\bar\eta}\bigg(1-\xi_1-\xi_2, v-x_3; \frac{\xi_2}{\xi_1+\xi_2}, x_2-x_1\bigg)\;.
\eeq
Putting everything together, one can get the mixed space expression of the dressed gluon wave function as given in Eq. \eqref{dressed_gluon_mixed_sp}. 
\subsection{From bare to dressed components }
\label{App:bare2dressed}
In order to calculate the production cross section, one needs to calculate the expectation value of the number operator in the outgoing gluon wave function. The number operator is defined in terms of the creation/annihilation operators of the dressed photon, dressed quark and dressed antiquark. Therefore, the outgoing gluon wave function has to be written in terms of the dressed components instead of the bare ones. Let us explain this procedure in a schematic way.  The dressed states can be written in terms of the bare ones up to perturbative $O(g_eg_s)$ corrections as 
\beq
\label{dressed_in_terms_of_bares}
|g\rangle_D&\simeq &|g\rangle_0+g_sF^{(1)}_{(q\bar q)}|q\bar q\rangle_0+g_sg_e\Big[ F^{(2)}_{(q\bar q-\bar q\gamma)}+F^{(2)}_{(q\bar q-q\gamma)}\Big]|q\bar q\gamma\rangle_0\;,
\nonumber\\
|q\bar q\rangle_D&\simeq&|q\bar q\rangle_0+g_e\Big[ F^{(1)}_{(q\gamma)}+ F^{(1)}_{(\bar q\gamma)}\Big] |q\bar q\gamma\rangle_0\;,\\
|q\bar q\gamma\rangle_D&\simeq&|q\bar q\gamma\rangle_0\;. \nonumber
\eeq
The outgoing wave function is given by the dressed gluon state after eikonally interacting with the target. Written schematically 
(c.f.r. Eq. \eqref{out_gluon_bare}), it reads
\beq
\label{out_with_bares}
|g\rangle_{\rm out}&=&S_A(w)|g\rangle_0+g_sF^{(1)}_{q\bar q}S_F^\dagger(v)S_F(x_1) |q\bar q\rangle_0
\nonumber\\
&&
+g_sg_e\Big[  F^{(2)}_{(q\bar q-\bar q\gamma)}+F^{(2)}_{(q\bar q-q\gamma)}\Big]S^\dagger_F(x_3)S_F(x_1) |q\bar q\gamma\rangle_0\;.
\eeq
We can now use Eq. \eqref{dressed_in_terms_of_bares} to rewrite Eq. \eqref{out_with_bares} and group the dressed components. The schematic result for the outgoing wave funcion  in terms of the dressed components reads
\beq
\label{out_with_dressed}
&&
\hspace{-1cm}
|g\rangle_{\rm out}= S_A(w)|g\rangle_D+g_sF^{(1)}_{q\bar q}\Big[ S^\dagger_F(v)S_F(x_1)-S_A(w)\Big] |q\bar q\rangle_D\nonumber\\
&&
\hspace{-0.95cm}
+g_sg_e\Big\{ \Big[ S^\dagger_F(x_3)S_F(x_1)-S_A(w)\Big]F_{(q\bar q-\bar q\gamma)}^{(2)}-\big[ S^\dagger_F(v)S_F(x_1)-S_A(w)\Big]F^{(1)}_{q\bar q} F^{(1)}_{\bar q\gamma}
\nonumber\\
&&
\hspace{0cm}
+
\Big[ S^\dagger_F(x_3)S_F(x_1)-S_A(w)\Big]F_{(q\bar q- q\gamma)}^{(2)}-
\Big[S^\dagger_F(x_3)S_F(v)-S_A(w)\Big] F^{(1)}_{q\bar q}F^{(1)}_{q\gamma}\Big\}\;,
\eeq
which, when written in full detail, is the same as Eq. \eqref{gluon_out_final}. 
\subsection{Product of splitting amplitudes}
\label{App:splittings}
In this appendix, we present the details of the calculation of the product of splitting amplitudes. We start with the quark initiated channel, for this channel we have three different structures for the product of splitting amplitudes that are listed below.\\
(i) Photon emission before the gluon emission both in the amplitude and complex conjugate amplitude:
\begin{equation}
\begin{aligned} & \phi_{ss'}^{\lambda\bar{\lambda}}(\xi_{1})\phi_{s's''}^{\eta\bar{\eta}}\biggl(\frac{\xi_{2}}{\bar{\xi}_{1}}\biggr)\phi_{s\bar{s}'}^{\lambda\bar{\lambda}'*}(\xi_{1})\phi_{\bar{s}'s''}^{\eta\bar{\eta}'*}\biggl(\frac{\xi_{2}}{\bar{\xi}_{1}}\biggr)\\
 & =\Bigl[(2-\xi_{1})\delta^{\lambda\bar{\lambda}}\delta_{ss'}-i\xi_{1}\epsilon^{\lambda\bar{\lambda}}\sigma_{ss'}^{3}\Bigr]\Bigl[(2-\xi_{1})\delta^{\lambda\bar{\lambda}'}\delta_{s\bar{s}'}+i\xi_{1}\epsilon^{\lambda\bar{\lambda}'}\sigma_{s\bar{s}'}^{3}\Bigr]\\
 & \times\biggl[\biggl(2-\frac{\xi_{2}}{\bar{\xi}_{1}}\biggr)\delta^{\eta\bar{\eta}}\delta_{s's''}-i\frac{\xi_{2}}{\bar{\xi}_{1}}\epsilon^{\eta\bar{\eta}}\sigma_{s's''}^{3}\biggr]\biggl[\biggl(2-\frac{\xi_{2}}{\bar{\xi}_{1}}\biggr)\delta^{\eta\bar{\eta}'}\delta_{\bar{s}'s''}+i\frac{\xi_{2}}{\bar{\xi}_{1}}\epsilon^{\eta\bar{\eta}'}\sigma_{\bar{s}'s''}^{3}\biggr]\;,\\
 & =\Bigl\{\Bigl[(2-\xi_{1})^{2}+\xi_{1}^{2}\Bigr]\delta^{\bar{\lambda}\bar{\lambda}'}\delta_{s'\bar{s}'}+2i\xi_{1}(2-\xi_{1})\epsilon^{\bar{\lambda}\bar{\lambda}'}\sigma_{s'\bar{s}'}^{3}\Bigr\}\\
 & \times\biggl\{\biggl[\biggl(2-\frac{\xi_{2}}{\bar{\xi}_{1}}\biggr)^{2}+\frac{\xi_{2}^{2}}{\bar{\xi}_{1}^{2}}\biggr]\delta^{\bar{\eta}\bar{\eta}'}\delta_{s'\bar{s}'}+2i\frac{\xi_{2}}{\bar{\xi}_{1}}\biggl(2-\frac{\xi_{2}}{\bar{\xi}_{1}}\biggr)\epsilon^{\bar{\eta}\bar{\eta}'}\sigma_{s'\bar{s}'}^{3}\biggr\}\;,\\
 & =8\, \mathcal{M}_{q}^{\bar{\lambda}\bar{\lambda}';\bar{\eta}\bar{\eta}'}\biggl(\xi_{1},\frac{\xi_{2}}{\bar{\xi}_{1}}\biggr)\;,
\end{aligned}
\end{equation}
where $\mathcal{M}_{q}^{\bar{\lambda}\bar{\lambda}';\bar{\eta}\bar{\eta}'}\biggl(\xi_{1},\frac{\xi_{2}}{\bar{\xi}_{1}}\biggr)$ is defined in Eq. \eqref{F_bef-bef}.\\

(ii) Photon emission after the gluon emission both in the amplitude and in the complex conjugate amplitude: The product of splitting amplitudes for this contribution can be obtained from Eq. \eqref{F_bef-bef} by simply exchanging $\xi_1 \leftrightarrow \xi_2$.

(iii) Photon emission before the gluon emission in the amplitude and after the gluon emission in the complex conjugate amplitude:
\begin{equation}
\begin{aligned} & \phi_{ss'}^{\eta\bar{\eta}}(\xi_{2})\phi_{s's''}^{\lambda\bar{\lambda}}\biggl(\frac{\xi_{1}}{\bar{\xi}_{2}}\biggr)\phi_{s\bar{s}'}^{\lambda\bar{\lambda}'*}(\xi_{1})\phi_{\bar{s}'s''}^{\eta\bar{\eta}'*}\biggl(\frac{\xi_{2}}{\bar{\xi}_{1}}\biggr)\\
 & =\Bigl[(2-\xi_{2})\delta^{\eta\bar{\eta}}\delta_{ss'}-i\xi_{2}\epsilon^{\eta\bar{\eta}}\sigma_{ss'}^{3}\Bigr]\Bigl[(2-\xi_{1})\delta^{\lambda\bar{\lambda}'}\delta_{s\bar{s}'}+i\xi_{1}\epsilon^{\lambda\bar{\lambda}'}\sigma_{s\bar{s}'}^{3}\Bigr]\\
 & \times\biggl[\biggl(2-\frac{\xi_{1}}{\bar{\xi}_{2}}\biggr)\delta^{\lambda\bar{\lambda}}\delta_{s's''}-i\frac{\xi_{1}}{\bar{\xi}_{2}}\epsilon^{\lambda\bar{\lambda}}\sigma_{s's''}^{3}\biggr]\biggl[\biggl(2-\frac{\xi_{2}}{\bar{\xi}_{1}}\biggr)\delta^{\eta\bar{\eta}'}\delta_{\bar{s}'s''}+i\frac{\xi_{2}}{\bar{\xi}_{1}}\epsilon^{\eta\bar{\eta}'}\sigma_{\bar{s}'s''}^{3}\biggr]\;,\\
 & =\Bigl[(2-\xi_{2})(2-\xi_{1})\delta^{\eta\bar{\eta}}\delta^{\lambda\bar{\lambda}'}\delta_{s'\bar{s}'}+\xi_{1}\xi_{2}\epsilon^{\eta\bar{\eta}}\epsilon^{\lambda\bar{\lambda}'}\delta_{s'\bar{s}'}\\
 & -i\xi_{2}(2-\xi_{1})\epsilon^{\eta\bar{\eta}}\delta^{\lambda\bar{\lambda}'}\sigma_{s'\bar{s}'}^{3}+i\xi_{1}(2-\xi_{2})\delta^{\eta\bar{\eta}}\epsilon^{\lambda\bar{\lambda}'}\sigma_{s'\bar{s}'}^{3}\Bigr]\\
 & \times\biggl[\biggl(2-\frac{\xi_{1}}{\bar{\xi}_{2}}\biggr)\biggl(2-\frac{\xi_{2}}{\bar{\xi}_{1}}\biggr)\delta^{\eta\bar{\eta}'}\delta^{\lambda\bar{\lambda}}\delta_{s'\bar{s}'}+\frac{\xi_{1}\xi_{2}}{\bar{\xi}_{1}\bar{\xi}_{2}}\epsilon^{\eta\bar{\eta}'}\epsilon^{\lambda\bar{\lambda}}\delta_{s'\bar{s}'}\\
 & -i\frac{\xi_{1}}{\bar{\xi}_{2}}\biggl(2-\frac{\xi_{2}}{\bar{\xi}_{1}}\biggr)\delta^{\eta\bar{\eta}'}\epsilon^{\lambda\bar{\lambda}}\sigma_{s'\bar{s}'}^{3}+i\frac{\xi_{2}}{\bar{\xi}_{1}}\biggl(2-\frac{\xi_{1}}{\bar{\xi}_{2}}\biggr)\epsilon^{\eta\bar{\eta}'}\delta^{\lambda\bar{\lambda}}\sigma_{s'\bar{s}'}^{3}\biggr]\;,\\
 & =8\, \widetilde{\mathcal{M}}_{q}^{\bar{\lambda}\bar{\lambda}';\bar{\eta}\bar{\eta}'}(\xi_{1},\xi_{2})\;,
\end{aligned}
\end{equation}
where $\widetilde{\mathcal{M}}_{q}^{\bar{\lambda}\bar{\lambda}';\bar{\eta}\bar{\eta}'}(\xi_{1},\xi_{2})$ is defined in Eq. \eqref{F_cross}.

In the gluon initiated channel, we have three different structures for the product of splitting amplitudes: \\
(iv) Photon emission from the antiquark both in the amplitude and in the complex conjugate amplitude:
\begin{equation}
\begin{alignedat}{1} & \Psi_{s\tilde{s}}^{\eta\bar{\eta}'*}(\xi_1)\phi_{\tilde{s}s'}^{\lambda\bar{\lambda}'*}\biggl(\frac{\xi_2}{\bar{\xi_1}}\biggr)\Psi_{s\bar{s}}^{\eta\bar{\eta}}(\xi_1)\phi_{\bar{s}s'}^{\lambda\bar{\lambda}}\biggl(\frac{\xi_2}{\bar{\xi_1}}\biggr)\\
 & =\bigl[(1-2\xi_1)\delta^{\eta\bar{\eta}'}\delta_{s,-\tilde{s}}+i\epsilon^{\eta\bar{\eta}'}\sigma_{s,-\tilde{s}}^{3}\bigr]\bigl[(1-2\xi_1)\delta^{\eta\bar{\eta}}\delta_{s,-\bar{s}}-i\epsilon^{\eta\bar{\eta}}\sigma_{s,-\bar{s}}^{3}\bigr]\\
 & \times\biggl[\Bigl(2-\frac{\xi_2}{\bar{\xi_1}}\Bigr)\delta^{\lambda\bar{\lambda}'}\delta_{\tilde{s}s'}+i\frac{\xi_2}{\bar{\xi_1}}\epsilon^{\lambda\bar{\lambda}'}\sigma_{\tilde{s}s'}^{3}\biggr]\biggl[\Bigl(2-\frac{\xi_2}{\bar{\xi_1}}\Bigr)\delta^{\lambda\bar{\lambda}}\delta_{\bar{s}s'}-i\frac{\xi_2}{\bar{\xi_1}}\epsilon^{\lambda\bar{\lambda}}\sigma_{\bar{s}s'}^{3}\biggr]\;,\\
 & =\Bigl\{\bigl[(1-2\xi_1)^{2}+1\bigr]\delta^{\bar{\eta}\bar{\eta}'}\delta_{\bar{s}\tilde{s}}-2i(1-2\xi_1)\epsilon^{\bar{\eta}\bar{\eta}'}\sigma_{\bar{s},\tilde{s}}^{3}\Bigr\}\\
 & \times\biggl\{\biggl[\Bigl(2-\frac{\xi_2}{\bar{\xi_1}}\Bigr)^{2}+\Bigl(\frac{\xi_2}{\bar{\xi_1}}\Bigr)^{2}\biggr]\delta^{\bar{\lambda}\bar{\lambda}'}\delta_{\bar{s}\tilde{s}}+2i\frac{\xi_2}{\bar{\xi_1}}\Bigl(2-\frac{\xi_2}{\bar{\xi_1}}\Bigr)\epsilon^{\bar{\lambda}\bar{\lambda}'}\sigma_{\tilde{s}\bar{s}}^{3}\biggr\}\;,\\
 & =8\, \mathcal{M}_{\bar{q}\bar{q}}^{\bar{\lambda}\bar{\lambda}';\bar \eta\bar{\eta}'}\biggl(\xi_1,\frac{\xi_2}{\bar{\xi_1}}\biggr)\;
\end{alignedat}
\end{equation}
where $\mathcal{M}_{\bar{q}\bar{q}}^{\bar{\lambda}\bar{\lambda}';\bar\eta\bar{\eta}'}\biggl(\xi_1,\frac{\xi_2}{\bar{\xi_1}}\biggr)$ is defined in Eq. \eqref{product_splitt_squared}.
\\
(v) Photon emission from the quark both in the amplitude and in the complex conjugate amplitude:
\begin{equation}
\begin{aligned} & \Psi_{\tilde{s}s'}^{\eta\bar{\eta}'*}(\xi_1+\xi_2)\phi_{s\tilde{s}}^{\lambda\bar{\lambda}'*}\biggl(\frac{\xi_2}{\xi_1+\xi_2}\biggr)\Psi_{\bar{s}s'}^{\eta\bar{\eta}}(\xi_1+\xi_2)\phi_{s\bar{s}}^{\lambda\bar{\lambda}}\biggl(\frac{\xi_2}{\xi_1+\xi_2}\biggr)\\
 & =\bigl[\bigl(1-2\xi_1-2\xi_2\bigr)\delta^{\eta\bar{\eta}'}\delta_{\tilde{s},-s'}+i\epsilon^{\eta\bar{\eta}'}\sigma_{\tilde{s},-s'}^{3}\bigr]\bigl[\bigl(1-2\xi_1-2\xi_2\bigr)\delta^{\eta\bar{\eta}}\delta_{\bar{s},-s'}-i\epsilon^{\eta\bar{\eta}}\sigma_{\bar{s},-s'}^{3}\bigr]\\
 & \times\biggl[\bigg(2-\frac{\xi_2}{\xi_1+\xi_2}\bigg)\delta^{\lambda\bar{\lambda}'}\delta_{s\tilde{s}}+i\frac{\xi_2}{\xi_1+\xi_2}\epsilon^{\lambda\bar{\lambda}'}\sigma_{s\tilde{s}}^{3}\biggr]\biggl[\bigg(2-\frac{\xi_2}{\xi_1+\xi_2}\bigg)\delta^{\lambda\bar{\lambda}}\delta_{s\bar{s}}-i\frac{\xi_2}{\xi_1+\xi_2}\epsilon^{\lambda\bar{\lambda}}\sigma_{s\bar{s}}^{3}\biggr]\;,\\
 & =\biggl\{\Bigl[\bigl(1-2\xi_1-2\xi_2\bigr)^{2}+1\Bigr]\delta^{\bar{\eta}\bar{\eta}'}\delta_{\tilde{s}\bar{s}}+2i\bigl(1-2\xi_1-2\xi_2\bigr)\epsilon^{\bar{\eta}\bar{\eta}'}\sigma_{\tilde{s}\bar{s}}^{3}\biggr\}\\
 & \times\biggl\{\biggl[\bigg(2-\frac{\xi_2}{\xi_1+\xi_2}\bigg)^{2}+\biggl(\frac{\xi_2}{\xi_1+\xi_2}\biggr)^{2}\biggr]\delta^{\bar{\lambda}\bar{\lambda}'}\delta_{\tilde{s}\bar{s}}+2i\frac{\xi_2}{\xi_1+\xi_2}\bigg(2-\frac{\xi_2}{\xi_1+\xi_2}\bigg)\epsilon^{\bar{\lambda}\bar{\lambda}'}\sigma_{\bar{s}\tilde{s}}^{3}\biggr\}\;,\\
 & =8\, \mathcal{M}_{qq}^{\bar{\lambda}\bar{\lambda}';\bar\eta\bar{\eta}'}\biggl(\xi_1+\xi_2,\frac{\xi_2}{\xi_1+\xi_2}\biggr)\;
\end{aligned}
\end{equation}
where $\mathcal{M}_{qq}^{\bar{\lambda}\bar{\lambda}';\bar\eta\bar{\eta}'}\biggl(\xi_1+\xi_2,\frac{\xi_2}{\xi_1+\xi_2}\biggr)$ is defined in Eq. \eqref{M_qq}. 
\\
(vi) Photon emission from the quark in the amplitude and from the antiquark in the complex conjugate amplitude:
\begin{equation}
\begin{aligned} & \Psi_{\tilde{s}s'}^{\eta\bar{\eta}'*}(\xi_1+\xi_2)\phi_{s\tilde{s}}^{\lambda\bar{\lambda}'*}\biggl(\frac{\xi_2}{\xi_1+\xi_2}\biggr)\Psi_{s\bar{s}}^{\eta\bar{\eta}}(\xi_1)\phi_{\bar{s}s'}^{\lambda\bar{\lambda}}\biggl(\frac{\xi_2}{\bar{\xi_1}}\biggr)\\
 & =\Bigl\{\Bigl[1-2(\xi_1+\xi_2)\Bigr]\delta^{\eta\bar{\eta}'}\delta_{\tilde{s},-s'}+i\epsilon^{\eta\bar{\eta}'}\sigma_{\tilde{s},-s'}^{3}\Bigr\}\biggl[\bigg(2-\frac{\xi_2}{\xi_1+\xi_2}\bigg)\delta^{\lambda\bar{\lambda}'}\delta_{s\tilde{s}}+i\frac{\xi_2}{\xi_1+\xi_2}\epsilon^{\lambda\bar{\lambda}'}\sigma_{s\tilde{s}}^{3}\biggr]\\
 & \times\Bigl[(1-2\xi_1)\delta^{\eta\bar{\eta}}\delta_{s,-\bar{s}}-i\epsilon^{\eta\bar{\eta}}\sigma_{s,-\bar{s}}^{3}\Bigr]\biggl[\biggl(\frac{\xi_2}{\bar{\xi_1}}-2\biggr)\delta^{\lambda\bar{\lambda}}\delta_{\bar{s}s'}+i\frac{\xi_2}{\bar{\xi_1}}\epsilon^{\lambda\bar{\lambda}}\sigma_{\bar{s}s'}^{3}\biggr]\\
 & =\biggl\{\Bigl[1-2(\xi_1+\xi_2)\Bigr]\bigg(2-\frac{\xi_2}{\xi_1+\xi_2}\bigg)\delta^{\eta\bar{\eta}'}\delta^{\lambda\bar{\lambda}'}\delta_{s,-s'}-\frac{\xi_2}{\xi_1+\xi_2}\epsilon^{\lambda\bar{\lambda}'}\epsilon^{\eta\bar{\eta}'}\delta_{s,-s'}\\
 & +i\bigg(2-\frac{\xi_2}{\xi_1+\xi_2}\bigg)\delta^{\lambda\bar{\lambda}'}\epsilon^{\eta\bar{\eta}'}\sigma_{s,-s'}^{3}+i\frac{\xi_2}{\xi_1+\xi_2}\Bigl[1-2(\xi_1+\xi_2)\Bigr]\epsilon^{\lambda\bar{\lambda}'}\delta^{\eta\bar{\eta}'}\sigma_{s,-s'}^{3}\biggr\}\\
 & \times\biggl[(1-2\xi_1)\biggl(\frac{\xi_2}{\bar{\xi_1}}-2\biggr)\delta^{\eta\bar{\eta}}\delta^{\lambda\bar{\lambda}}\delta_{s,-s'}-\frac{\xi_2}{\bar{\xi_1}}\epsilon^{\lambda\bar{\lambda}}\epsilon^{\eta\bar{\eta}}\delta_{s,-s'}\\
 & -i\biggl(\frac{\xi_2}{\bar{\xi_1}}-2\biggr)\delta^{\lambda\bar{\lambda}}\epsilon^{\eta\bar{\eta}}\sigma_{s,-s'}^{3}+i\frac{\xi_2}{\bar{\xi_1}}(1-2\xi_1)\delta^{\eta\bar{\eta}}\epsilon^{\lambda\bar{\lambda}}\sigma_{-s,s'}^{3}\biggr]\;,\\
 & =-8\,\widetilde{\mathcal{M}}_g^{\bar{\lambda}\bar{\lambda}';\bar\eta\bar{\eta}'}\bigl(\xi_1,\xi_2\bigr)
\end{aligned}
\end{equation}
where $\widetilde{\mathcal{M}}_g^{\bar{\lambda}\bar{\lambda}';\bar\eta\bar{\eta}'}\bigl(\xi_1,\xi_2\bigr)$ is defined in Eq. \eqref{splitting_mixed}.

\section{Integrals}
\label{App:Integrals}
In this appendix, we present the details of the calculation of the two nontrivial integrals that one needs to perform in the correlation limit. The first integral is
\beq
I_1^{\eta i, \lambda}=\int d^2r_g d^2r_\gamma e^{i{\rm K}\cdot r_\gamma+i{\rm Q}\cdot r_g}
A^\eta(r_g)r^i_g\A^\lambda\bigg(\xi_1,r_\gamma; \frac{\xi_2}{\bar\xi_1},r_g\bigg)\;.
\eeq
Using the explicit expressions of the standard and modified Weizs\"acker-Williams fields given in Eqs. \eqref{Standard_WW} and \eqref{Modified_WW}, one can rewrite the above integral as 
\beq
\label{integral_1_form}
I_1^{\eta i,\lambda}=\frac{1}{(2\pi)^2}\int d^2r_g d^2r_\gamma e^{i{\rm K}\cdot r_\gamma+i{\rm Q}\cdot r_g}\,  \frac{r_g^\eta r_g^i}{r_g^2} \, \frac{r_\gamma^\lambda}{r_\gamma^2+c_0r_g^2}\;.
\eeq
For convenience, let us define  
\beq
I_1^{\eta i, \lambda}=\frac{1}{(2\pi)^2}\tilde{I}_1^{\eta i, \lambda}\;,
\eeq
where 
\beq
\tilde{I}_1^{\eta i,\lambda}=\int d^2r \, d^2\bar{r} \, e^{ip\cdot r+i \bar{p}\cdot \bar{r}} \, \frac{\bar{r}^\eta}{\bar{r}^2}\bar{r}^i\frac{r^\lambda}{r^2+c_0\bar{r}^2}\;.
\eeq
Here, $c_0=\frac{1}{\xi_1}\frac{\xi_2}{\bar\xi_1}\bigg(1-\frac{\xi_2}{\bar\xi_1}\bigg)$, and for practical reasons we have renamed $r_\gamma$ and $r_g$ as $r$ and $\bar{r}$, respectively. Likewise, the conjugate momenta ${\rm K}$ and $\rm Q$ are renamed as $p$ and $\bar p$. Again for convenience, we introduce a constant $a_0$ and defined the integral that needs to be calculated as 
\beq
\tilde{I}_{1,a_0}^{\eta i, \lambda}=\int d^2r d^2{\bar r}\, e^{ip\cdot r+i\bar{p}\cdot\bar{r}}\frac{\bar r^\eta \bar r^i}{[\bar r^2+a_0^2]}\; \frac{r^\lambda}{[r^2+c_0(\bar r^2+a_0^2)]}\;,
\eeq
such that 
\beq
\tilde{I}_1^{\eta i, \lambda}=\lim_{a_0\to0} \tilde{I}_{1,a_0}^{\eta i, \lambda}\;.
\eeq
As the next step, we introduce Schwinger parameters and rewrite the integral as 
\beq
\tilde{I}_{1,a_0}^{\eta i, \lambda}=\int d^2r d^2{\bar r} \int_0^{+\infty}d\sigma\,  dt  \, \bar r^\eta\bar r^i r^\lambda e^{ip\cdot r+i\bar p\cdot \bar r}\, e^{-\sigma[r^2+c_0(\bar r^2+a_0^2)]}e^{-t[\bar r^2+a_0^2]}\;.
\eeq
After completing the squares in the exponent, and renaming the integration variables as 
\beq
\label{cov_4_int}
r'=r-\frac{ip}{2\sigma}\; , \hspace{1cm} \bar r'=\bar r-\frac{i\bar p}{2(t+c_0\sigma)}\;,
\eeq
the integral becomes:
\beq
\label{Int_mid_1}
\tilde{I}_{1,a_0}^{\eta i, \lambda}&=&\int d^2r' \bigg( \frac{ip^\lambda}{2\sigma}\bigg)e^{-\sigma r'^2}
\int d^2\bar r' \bigg( \bar r'^\eta \bar r'^i-\frac{\bar p^\eta\bar p^i}{4(t+c_0\sigma)^2}\bigg) e^{-(t+c_0\sigma)\bar r'^2}
\nonumber\\
&&
\times\; 
\int_0^{+\infty} d\sigma dt \; e^{-(t+c_0\sigma)a_0^2}\, e^{-p^2/4\sigma}\, e^{-\bar p^2/4(t+c_0\sigma)}\;.
\eeq
Now, the integrations over $r'$ and $\bar r'$ can be performed trivially, and the result reads
\beq
\tilde{I}_{1,a_0}^{\eta i, \lambda}&=&\int_0^{+\infty} d\sigma \int_0^{+\infty} dt \bigg(\frac{i\pi}{2}\frac{p^\lambda}{\sigma^2}\bigg)\frac{\pi}{2}
\bigg[ \frac{\delta^{\eta i}}{(t+c_0\sigma)^2}-\frac{\bar p^\eta\bar p^i}{2(t+c_0\sigma)^3}\bigg] e^{-(t+c_0\sigma)a_0^2} 
\nonumber\\
&&
\times\; 
e^{-p^2/4\sigma} \, e^{-{\bar p}^2/4(t+c_0\sigma)}\;. 
\eeq
Before we calculate the integrations over the Schwinger parameters $\sigma$ and $t$, we first perform a change of variables and define $\sigma'=p^2/4\sigma$, after which we perform another change of variables where we define $t'=t+c_0p^2/4\sigma'$. The integral now reads
\beq
\tilde{I}_{1,a_0}^{\eta i, \lambda}&=&\int_0^{+\infty} d\sigma' \int_{c_0p^2/4\sigma'}^{+\infty} dt' \bigg(i2\pi \frac{p^\lambda}{p^2}\bigg) \frac{\pi}{2}\bigg[ \frac{\delta^{\eta i}}{t'^2}-\frac{\bar p^\eta \bar p^i}{2t'^3}\bigg] e^{-t'a_0^2}\, e^{-\bar p^2/4t'}\, e^{-\sigma'}\;.
\eeq
After changing the order of the two integrations, the integration over $\sigma'$ can be performed in a trivial manner: 
\beq
\tilde{I}_{1,a_0}^{\eta i, \lambda}&=&i \pi^2 \frac{p^\lambda}{p^2}\int_0^{+\infty}dt'\bigg[ \frac{\delta^{\eta i}}{t'^2}-\frac{\bar p^\eta \bar p^i}{2t'^3}\bigg] e^{-t'a_0^2} e^{-(\bar p^2+c_0p^2)/4t'}
\eeq
The integration over $t'$ can be performed thanks to the following formula
\beq
\label{generic_int}
\int_0^{+\infty}dt\;  t^{\nu-1}\; e^{-B/t}\; e^{-Ct}= 2\bigg(\frac{B}{C}\bigg)^{\nu/2} {\rm K}_{-\nu}\left( 2\sqrt{BC}\right)\;,
\eeq
where ${\rm K}_{-\nu}$ is the Modified Bessel function of the second kind, and the final result reads
\beq
\tilde{I}_{1,a_0}^{\eta i, \lambda}&=&i 4\pi^2 \frac{p^\lambda}{p^2}
\bigg[\delta^{\eta i}\frac{a_0}{\sqrt{\bar p^2+c_0p^2}} \; {\rm K}_1\left(|a_0|\sqrt{\bar p^2+c_0p^2}\right)\nonumber\\ 
&&-\frac{\bar p^\eta\bar p^i}{\bar p^2+c_0p^2}a_0^2\; {\rm K}_2\left( |a_0|\sqrt{\bar p^2+c_0p^2}\right) \bigg]\;.
\eeq
From this, we obtain the expression for $\tilde{I}_1^{\eta i, \lambda}$ by taking the limit $a_0\to 0$:
\beq
\label{int_1_final}
\tilde{I}_1^{\eta i, \lambda}=\lim_{a_0\to 0}\tilde{I}_{1,a_0}^{\eta i, \lambda}= i4\pi^2\frac{p^\lambda}{p^2}\frac{1}{\bar p^2+c_0p^2}\bigg[\delta^{\eta i}-2\frac{\bar p^\eta \bar p^i}{\bar p^2+c_0p^2}\bigg]\;.
\eeq
Finally, after renaming the momenta $p$ and $\bar p$ back into $\rm K$ and $\rm Q$, we arrive at Eq. \eqref{integral_1}. 

The second integral that we encounter in our calculation of the correlation limit is
\beq
I_2^{\eta,i\lambda}=\int d^2r_g d^2r_\gamma \, e^{i{\rm K}\cdot r_\gamma+i{\rm Q}\cdot r_g} A^\eta(r_g)\, r^i_\gamma \bigg[ \A^\lambda\bigg(\xi_1,r_\gamma; \frac{\xi_2}{\bar \xi_1},r_g\bigg) -A^{\lambda}(r_\gamma) \bigg]\;.
\eeq
In order to evaluate this integral, we use the result of the first integral. First, let us write explicitly the difference between the modified and standard Weizs\"acker-Williams fields, defined in  Eqs. \eqref{Standard_WW} and \eqref{Modified_WW}: 
\beq
{\cal A}^{\lambda}\bigg(\xi_1,r_\gamma; \frac{\xi_2}{\bar\xi_1},r_g\bigg)-A^{\lambda}(r_\gamma)=\frac{1}{2\pi}\frac{r_\gamma^\lambda}{r_\gamma^2}\, r_g^2\, \frac{1}{r_g^2+c_0^{-1}r_\gamma^2}\;.
\eeq
 After plugging this expression into the expression for $I_2^{\eta, i\lambda}$, we get:
 \beq
I_2^{\eta,i\lambda}= -\frac{1}{(2\pi)^2} \int d^2r_g d^2r_\gamma \, e^{i{\rm K}\cdot r_\gamma+i{\rm Q}\cdot r_g}\frac{r_\gamma^ir_\gamma^{\lambda}}{r_\gamma^2} \frac{r_g^\eta}{r_g^2+c_0^{-1}r_\gamma^2}\;.
\eeq
 Note that the form of the second integral is the same as the first integral given in Eq. \eqref{integral_1_form}, with the exchange of $r_g\leftrightarrow r_\gamma$. Therefore, we can simply read off the result from Eq. \eqref{int_1_final}: 
 \beq
 I_2^{\eta, i\lambda}=-i\frac{\bar p^\eta}{\bar p^2}\frac{1}{p^2+c_0^{-1}\bar p^2}\bigg[ \delta^{\lambda i}-2\frac{p^\lambda p^i}{p^2+c_0^{-1}\bar p^2}\bigg]\;.
 \eeq
Finally, renaming $p$ into ${\rm K}$ and $\bar p$ into ${\rm Q}$, we arrive at Eq. \eqref{integral_2}.



\end{document}